\documentclass[aps,prd,twocolumn,groupedaddress, showkeys, showpacs]{revtex4}
\usepackage{graphicx,subfigure,enumerate,pstricks,latexsym}
\usepackage[english]{babel}

\providecommand{\dif}{\mathrm{d}} 
\def\beq{\begin{equation}}
\def\eeq{\end{equation}}
\def\bea{\begin{eqnarray}}
\def\eea{\end{eqnarray}}

\def\d{\dif}
\def\lightarrow{\rightarrow}

\def\SS{\tilde{\Sigma}}
\def\cosm{{\lambda}}

\def\xx{\tilde{x}}
\def\yy{\tilde{y}}
\def\JJ{\tilde{J}}
\def\EE{\tilde{E}}
\def\rr{\tilde{r}}
\def\tt{\theta}
\def\tit{\tilde{\tau}}

\def\XXB{X}

\newcommand{\nnd}{\discretionary{--}{--}{--}}
\newcommand{\Schw}{Schwarzschild}
\newcommand{\dS}{de~Sitter}

\begin{document}

\title{Current-carrying string loops in black-hole spacetimes with a repulsive cosmological constant}
\author{M. Kolo\v{s}}
\author{Z. Stuchl\'{\i}k}
\affiliation{Institute of Physics, Faculty of Philosophy \& Science, Silesian University in Opava, Bezru\v{c}ovo n\'{a}m.13, CZ-74601 Opava, CzechRepublic}

\begin{abstract}
Current-carrying string-loop dynamics in Schwarzschild-de Sitter spacetimes characterized by the cosmological parameter $\lambda=\frac{1}{3} \Lambda M^2$ is investigated. With attention concentrated to the axisymmetric motion of string loops it is shown that the resulting motion is governed by the presence of an outer tension barrier and an inner angular momentum barrier that are influenced by the black hole gravitational field given by the mass $M$ and the cosmic repulsion given by the cosmological constant $\Lambda$. The gravitational attraction could cause capturing of the string having low energy by the black hole or trapping in its vicinity; with energy high enough, the string can escape (scatter) to infinity. The role of the cosmic repulsion becomes important in vicinity of the so called static radius where the gravitational attraction is balanced by the cosmic repulsion - it is demonstrated both in terms of the effective potential of the string motion and the basin boundary method reflecting its chaotic character that a potential barriere exists along the static radius behind which no trapped oscillations may exist. The trapped states of the string loops, governed by the interplay of the gravitating mass $M$ and the cosmic repulsion, are allowed only in Schwarzschild-de Sitter spacetimes with the cosmological parameter $\lambda < \lambda_{\rm trap} \sim 0.00497$. The trapped oscillations can extend close to the radius of photon circular orbit, down to $r_{\rm mt} \sim 3.3 M$.
\end{abstract}

 
  
\maketitle

\section{Introduction}\label{intro}

Recent cosmological tests indicate presence of dark energy with properties close to those of nonzero (but very small) repulsive cosmological constant ($\Lambda > 0$) responsible for the observed present acceleration of the expansion of our universe \cite{Rie-etal:2004:ASTRJ2:}. More precisely, these cosmological tests indicate that the dark energy represents about $74.5\%$ of the energy content of the observable universe that is very close to the critical energy density $\rho_{\rm crit}$ corresponding to the almost flat universe predicted by the inflationary scenario \cite{Spe-etal:2007:ASTJS:}. Further, there 
are strong indications that the dark energy equation of state is very close to those corresponding to the vacuum 
energy, i.e., to the cosmological constant \cite{Cald-Kami:2009:NATURE:}. Therefore, it is quite important 
to study the cosmological and astrophysical consequences of the effect of the observed cosmological constant implied 
by the cosmological tests to be $\Lambda \approx 1.3\times 10^{-56}{\rm cm}^{-2}$. 

The relevance of the repulsive cosmological constant in the cosmological models was discussed in detail by \cite{Mis-Tho-Whe:1973:Gra:}. Its role in the vacuola models of mass concentrations immersed in the expanding universe is treated in \cite{Stu:1983:BULAI:, Stu:1984:BULAI:,Uzan-Ellis-Larena:2010:arXiv:1005.1809v1:,Grenon-Lake:2010:PHYSR4:} and its relevance is even shown for astrophysical situations related to active galactic nuclei and their central supermassive black holes \cite{Stu:2005:MODPLA:}. The black hole spacetimes with the $\Lambda$ term included are described in spherically symmetric case by the \Schw\nnd\dS{} (SdS) geometry \cite{Kot:1918:,Stu-Hle:1999:PHYSR4:,Stu:2000:APS:,Boh:2004:GENRG:} and in axially symmetric, rotating case by the Kerr-de Sitter (KdS) geometry \cite{Car:1973:BlaHol:}. In the spacetimes with the repulsive cosmological term, motion of photons is treated in a series of papers \cite{Stu:1990:BULAI:,Stu-Cal:1991:GENRG2:,Stu-Hle:2000:CLAQG:,Lake:2002:PHYSR4:,Bak-etal:2007:CEJP:,Ser:2008:,Ser:2009:,Mul:2008:,Scha-Zai:2008:}, while motion of test particles and perfect fluid was studied in \cite{Stu:1983:BULAI:,Stu-Sla:2004:PHYSR4:,Kra:2004:CLAQG:,Kra:2005:CLAQG:,Kra:2007:CLAQG:,Cru-Oli-Vil:2005:CLAQG:,Kag-Kun-Lam:2006:PHYSR4:,Ior:2009:NewA:}. Equilibrium positions of spinning test particles were investigated in \cite{Stu:1999:ACTPS2:,Stu-Hle:2001:PHYSR4:,Stu-Kov:2006:CLAQG:,Mor-Moh:2009:GRG:}. The cosmological constant can be relevant in both the geometrically thin \cite{Stu:2005:MODPLA:,Stu-Hle:1999:PHYSR4:,Stu-Sla:2004:PHYSR4:,Mul-Asch:2007:CLAQG:NonMonoVel,Sla-Stu:2008:CLAQG:} and thick accretion discs \cite{Stu-Sla-Hle:2000:ASTRA:,Sla-Stu:2005:CLAQG:,Rez-Zan-Fon:2003:AA:,Asch:2008:,Stu-Sla-Tor-Abr:2005:PHYSR4:} orbiting around supermassive black holes in the center of giant galaxies. Moreover, it was recently shown that in spherically symmetric spacetimes such disc structures can be described with high precision by an appropriately chosen Pseudo-Newtonian potential \cite{Stu-Kov:2008:INTJMD:, Stu-Sla-Kov:2009:CLAQG:} that appears to be useful in studies of motion of interacting galaxies \cite{Stu-Schee:2010:}. 

\begin{figure}
\includegraphics[]{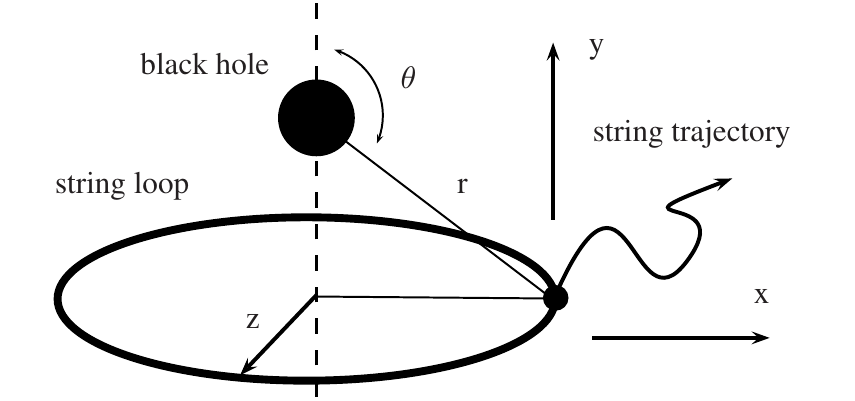}
\caption{Schematic picture of a string loop moving around a black hole. Assumed axial symmetry of the string loop allows to investigate only one point on the loop; one point path can represent whole string movement. Trajectory of the loop is then represented by the black curve on the picture, given in 2D $x$-$y$ plot.  \label{schema}}
\end{figure}

Quite recently, an interesting study of relativistic current carrying strings moving axisymmetrically along the axis of a Kerr black hole appeared \cite{Jac-Sot:2009:PHYSR4:}. Tension of such a loop string prevents its expansion beyond some radius, while its worldsheet current introduces an angular momentum barrier preventing the loop from collapsing into the black hole. There is an important possible astrophysical relevance of the current carrying strings \cite{Jac-Sot:2009:PHYSR4:} as they could in a simplified way represent plasma that exhibits associated string-like behavior via dynamics of the magnetic field lines in the plasma \cite{Chri-Hin:1999:PhRvD:,Sem-Dya-Pun:2004:Sci:} or due to thin isolated flux tubes of plasma that could be described by an one-dimensional string \cite{Spr:1981:AA:}. Such a configuration was also studied in  \citep{Lar:1994:CLAQG:,Fro-Lar:1999:CLAQG:}. It has been proposed in \citep{Jac-Sot:2009:PHYSR4:} that this current configuration can be used as a model for jet formation. 

Here we investigate dynamics of a current carrying string loop, characterized by its tension and angular momentum, in the field of a \Schw\nnd\dS{} black hole, generalizing thus the previous works \citep{Fro-Lar:1999:CLAQG:, Gu-Cheng:2007:GRG:} related to the string loops determined by tension only. Considering capture, trapping and scattering of circular string loops by the black hole we identify the role of the repulsive cosmological constant in the string dynamics. It is well known that even such a very simple axisymmetric two-body problem demonstrates apparent chaotic behavior reflected by the so called strange repeller due to the presence of the tension-terms in the motion equations, see \citep{Fro-Lar:1999:CLAQG:}. In order to reflect the chaotic character of the string motion, we use the method of Poincare surfaces and we determine the basin-boundary separating the space of initial condition due to different asymptotic outcomes of the string motion.

We study a string loop threaded on to an axis of the black hole chosen to be the $y$-axis - see Fig.\ref{schema}. The string loop can oscillate, changing its radius in $x$-$z$ plane, while propagating in $y$ direction. The string loop tension and worldsheet current form barriers governing its dynamics. These barriers are modified by the gravitational attraction of the black hole characterized by the mass $M$ and the cosmic repulsion determined by the cosmological constant $\Lambda$. We focus our attention on the interplay of the gravitational attraction and cosmic repulsion influencing the string-loop motion.

%
%
%
%

\section{Relativistic current carrying strings loop in spherically symmetric spacetimes}
The relativistic description of the string motion can be given in terms of a properly chosen action reflecting both the string and spacetime properties and enabling derivation of the equations motion. We summarize the equations of string motion in the standard form discussed in \cite{Jac-Sot:2009:PHYSR4:}. 

The string worldsheet is described by the spacetime coordinates $X^{\alpha}(\sigma^{a})$ with $\alpha = 0,1,2,3$ given as functions of two worldsheet coordinates $\sigma^{a}$ with $a = 0,1$ that imply induced metric on the worldsheet in the form
\beq
      h_{ab}= g_{\alpha\beta}X^\alpha_{,a}X^\beta_{,b}.
\eeq 
The string current localized on the worldsheet is described by a scalar field $\phi({\sigma^a})$. Dynamics of the string, inspired by an effective description of superconducting strings representing topological defects occurring in the theory with multiple scalar fields undergoing spontaneous symmetry breaking \cite{Wit:1985:NuclPhysB:}, is described by the action
\beq
 S = \int \dif^2 \sigma \sqrt{-h} (\mu/c + h^{ab} \varphi_{,a}\varphi_{,b}).\label{katolicka_akce}
\eeq
where $ \varphi_{,a} = j_a $ determines current of the string and $\mu > 0$ reflects the string tension. For $j_a = 0, \mu = 0$ we have null strings. 

Varying the action with respect to the induced metric $h_{ab}$ yields the worldsheet stress-energy tensor density (being of density weight one with respect to worldsheet coordinate transformations)
\beq
\SS^{ab}= \sqrt{-h} \left( 2 j^a j^b - (\mu + j^2) h^{ab}\right),
\eeq
where
\beq
   j^a = h^{ab}j_{b} , \quad j^2 = h^{ab}j_{a}j_{b}.
\eeq
The contribution from the string tension with $\mu > 0$ has a positive energy density and a negative pressure (tension). The current contribution is traceless, due to the conformal invariance of the action - it can be considered as a $1+1$ dimensional massless radiation fluid with positive energy density and equal pressure \cite{Jac-Sot:2009:PHYSR4:}. 

Varying the action with respect to $X^{\alpha}$ we arrive to equations of motion
\beq
 (\SS^{ab} g_{\alpha\lambda} X^\alpha_{,a})_{,b} - \frac{1}{2} \SS^{ab} g_{\alpha\beta,\lambda} X^\alpha_{,a} X^\beta_{,b} = 0 , \label{EqOfMotion}
\eeq
while varying the action with respect to $\varphi$ yields the $1+1$ dimensional wave equation
\beq
         (\sqrt{-h} h^{ab} \varphi_{,a})_{,b} = 0 ,
\eeq
i.e., the current is divergenceless. Similarly, $\Sigma^{ab}_{;b} = 0$, i.e., the worldsheet stress-energy tensor is divergenceless with respect to $h_{ab}$ covariant derivative \cite{Jac-Sot:2009:PHYSR4:}.

Any two-dimensional metric is conformally flat metric, i.e., there is
\beq
            h_{ab} = \Omega^2 \eta_{ab},
\eeq
where $\eta_{ab}$ is locally flat metric and $\Omega$ is a worldsheet scalar function. Adopting coordinates $\sigma^a = (\tau, \sigma)$ such that $\eta_{\tau \sigma} = 0$ and $\eta_{\tau \tau} = -\eta_{\sigma \sigma} = -1$, the conformally flat gauge is equivalent to the conditions
\beq
    h_{\tau \sigma} = 0, \, h_{\tau \tau} + h_{\sigma \sigma} = 0, \, \sqrt{-h} h^{ab} = \eta^{ab}. \label{gCondition}
\eeq
Then the conformal factor is given by
\beq
           h_{\sigma \sigma} = \Omega^2 \eta_{\sigma \sigma} = g_{\phi \phi}.
\eeq
In the conformal gauge, the equation of motion of the scalar field reads
\beq
          \varphi_{,\tau \tau} - \varphi_{, \sigma \sigma} = 0.
\eeq

In spherically symmetric spacetimes, the axisymmetric string loops can be characterized by coordinates
\beq
 X^\alpha(\tau,\sigma) = (t(\tau),r(\tau),\theta(\tau),\sigma).
\eeq
(Notice that in axially symmetric spacetimes the conditions has to be more complex \cite{Jac-Sot:2009:PHYSR4:}.)
The assumption of axisymmetry implies that the current is independent of $\sigma$ and $j_{a,\sigma} = 0$. Using the scalar field equation of motion we can conclude that the scalar field can be expressed in linear form with constants $j_{\sigma}$ and $j_{\tau}$
\beq
          \varphi = j_{\sigma}\sigma + j_{\tau}\tau.
\eeq
Introducing new variables 
\beq
          J^2 \equiv j_\sigma^2 + j_\tau^2, \quad \omega \equiv -j_\sigma / j_\tau ,
\eeq
we express the components of the worldsheet stress-energy density $\SS^{ab}$ in the form
\bea
&& \SS^{\tau\tau} = \frac{J^2}{g_{\phi\phi}} + \mu , \quad \SS^{\sigma\sigma} = \frac{J^2}{g_{\phi\phi}} - \mu ,\\ 
&& \SS^{\sigma\tau} = \frac{-2 \omega J^2}{g_{\phi\phi} (1+\omega^2)}.
\eea

The string dynamics depends on the current through the worldsheet stress-energy tensor. The dependence is expressed using the parameters $J^2$ and $\omega$. The minus sign in the definition of $\omega$ is chosen in order to obtain correspondence of positive angular momentum and positive $\omega$.

\subsection{Equations of the string motion}

Using the gauge and string axisymmetry conditions, the equations of string motion (\ref{EqOfMotion}) take the form
\bea
  \left( \SS^{\tau\tau} g_{\alpha\lambda} X^\alpha \right)_{,\tau}  - \frac{1}{2} \left( \SS^{\tau\tau} g_{\alpha\beta,\lambda} X^\alpha X^\beta + \SS^{\sigma\sigma} g_{\phi\phi,\lambda} \right) \nonumber \\ = 0.  \label{EqOfMotion02}
\eea
For $\lambda = \phi$ the equation is satisfied identically. For $\lambda = t$ it yields the energy conservation  condition 
\beq
 \SS^{\tau\tau} g_{tt} \dot{t} = - E, \label{konzerva}
\eeq
where $E$  is a constant that has to be identified with total string energy related to the Killing vector field ($\frac{\partial}{\partial \phi}$) divided by $2\pi$. For $\lambda = r$ and $\lambda = \theta$ the motion equations read
\bea
\left(\SS^{\tau\tau} g_{rr}\dot{r} \right)_{,\tau} & = & \frac{1}{2}\left(\SS^{\tau\tau} g_{\alpha\beta,r} X^\alpha X^\beta + \SS^{\sigma\sigma} g_{\phi\phi,r} \right) \label{EqOfMotion03-01}\\
\left(\SS^{\tau\tau} g_{\theta\theta} \dot{\theta} \right)_{,\tau} & = & \frac{1}{2} \left(\SS^{\tau\tau}g_{\alpha\beta,\theta} X^\alpha X^\beta + \SS^{\sigma\sigma} g_{\phi\phi,\theta} \right). \label{EqOfMotion03-02}
\eea
%
%

In the spherically symmetric spacetimes the equations of motion of the string loop are significantly simplified due to the symmetry properties. The motion is independent of the parameter $\omega$ and depends only on the parameter $J^2$ representing the angular momentum. 

\subsection{Integrals of the motion}
For a Killing vector field $\xi^{\alpha}$ a conserved Killing current must exist \cite{Jac-Sot:2009:PHYSR4:}. Writing the Killing equation in the form $g_{\alpha \beta, \lambda} \xi^{\lambda} = 0$, the worldsheet vector density that is the contraction of the worldsheet energy-density with the pullback of the Killing one-form to the worldsheet
\beq
      {\cal J}_\xi^b = \SS^{ab} X^{\alpha}_{,a} g_{\alpha\lambda} \xi^\lambda
\eeq
has to satisfy the conservation law
\beq
       {\cal J}_\xi^b{}_{,b}=0.
\eeq
Integrating over any closed worldsheet cross section determines the conserved quantity related to the Killing vector field
\beq
         Q_\xi=\int {\cal J}_\xi^b \d S_b.
\eeq
Taking the integral over a surface of constant $\tau$ we obtain, because of the axisymmetry of the string loop, the relation
\beq
         -Q_{\xi} = {\cal E} = 2\pi E,
\eeq
where $E$ is the constant introduced in Eq.(\ref{konzerva}) having the meaning of the Killing energy of the string divided by $2\pi$. For the axial Killing vector we arrive at
\beq
         Q_{\xi}  = L = -4\pi j_{\sigma} j_{\tau}
\eeq
which is clearly a constant for the solution we consider. It should be stressed that without the current the string carries no angular momentum.

\subsection{Motion in spherically symmetric spacetimes and its effective potential}

We can write any vacuum spherically symmetric spacetime in the form
\beq
 \d s^2 = -A(r) \d t^2 + A^{-1}(r) \d r^2 + r^2 (\d \theta^2 + \sin^2 \d \phi^2), \label{SfSymMetrika}
\eeq
The characteristic function $A(r)$ describes a particular spacetime. 

The gauge condition (\ref{gCondition}), takes the form
\beq
       \dot{t}^2 = A(r)^{-2} \dot{r}^2 + A(r)^{-1} r^2 \dot{\theta}^2 + A(r)^{-1} r^2 \sin^2 \theta. \label{GaugeCond01}
\eeq

The components of the string equations of motion (\ref{EqOfMotion02}) take relatively simple form for the spherically symmetric metric (\ref{SfSymMetrika}). For $\lambda = t$ the conservation law reads (\ref{konzerva})
\beq
        \dot{t} A(r) \SS^{\tau\tau} = E . \label{Conservation_Law}
\eeq
Considering (\ref{GaugeCond01}), we arrive for $\lambda = r$ and $\lambda = \theta$ to the formulae
\bea
 \ddot{r} &=& {\dot{\theta}}^2 \left(A(r) r - \frac{ \partial_{r} A(r)}{2} r^2 \right) -\dot{r} \frac{\partial_{\tau}\Sigma^{\tau\tau}{}}{\Sigma^{\tau \tau}} \nonumber \\
 && + \sin^2\theta \left( \frac{\Sigma^{\sigma\sigma}}{\Sigma^{\tau \tau}} A(r) r - \frac{\partial_{r}A(r)}{2} r^2 \right), \label{MovingEq01} \\
 \ddot{\theta} &=& -\frac{2}{r} \dot{r} \dot{\theta} - \frac{\partial_{\tau}\Sigma^{\tau\tau}{}}{\Sigma^{\tau \tau}} \dot{\theta} + \frac{\Sigma^{\sigma\sigma}}{\Sigma^{\tau \tau}} \sin\theta \cos\theta, \label{MovingEq02}
\eea
where
\beq
\partial_{\tau}\Sigma^{\tau\tau} = -\frac{2 J^2}{g_{\phi\phi}^2} (r \sin^2\theta \dot{r} +
   r^2 \sin\theta \cos\theta \dot{\theta}).
\eeq
Component $\SS^{\sigma\tau}$ of the worldsheet stress-energy tensor is not present in string equations of motion (\ref{MovingEq01}-\ref{MovingEq02}). Further, the string motion does not depend on $\omega$, which is obvious for spherically symmetric spacetimes.

Considering the gauge condition (\ref{GaugeCond01}) and the conservation law (\ref{Conservation_Law}), the equations of motion can be expressed in the ``integrated'' form
\beq
 A(r) (\SS^{\tau\tau})^2 (A(r)^{-1} \dot{r}^2 + r^2 \dot{\theta}^2 ) + V(r,\theta) = 0,
\eeq
where
\beq
 V(r,\theta) = -E^2 + A(r) r^2 \sin^2 \theta (\SS^{\tau\tau})^2
\eeq
plays the role of an ``effective potential''. Since the first term of the equations of motion is always positive in the static parts of the spherically symmetric spacetimes, the string motion is confined to the region where 
\beq
            V(r,\theta) \leq 0 ,
\eeq
being bounded by the relation fulfilled by the energy parameter
\bea
 E = E_{\rm b}(r,\theta) &\equiv& \sqrt{A(r)} \, r \sin \theta \, \SS^{\tau\tau} \\
 &=& \sqrt{A(r)} \left( \frac{J^2}{r \sin \theta}  +  \mu r \sin \theta  \right). \label{BRelation}
\eea
We find directly that $E_{\rm b}(r,\theta) = 0$ just where $A(r)=0$, i.e., at the spacetime horizons; it diverges at infinity and at $r=0$ when $J^2>0$.
Assuming that the string loop will start its motion from rest, i.e., assuming $\dot{r}(0) = 0$ and $\dot{\theta}(0)=0$, the initial position of the string will be located at some point of the energy boundary $E_{\rm b}$ of its motion.

It is useful to use the Cartesian coordinates that are introduced in the form 
\beq
 x = r \sin(\theta),\quad y = r \cos(\theta). \label{ccord}
\eeq
The boundary string energy in Cartesian coordinates is
\beq
E_{\rm b}(x,y) = \sqrt{A(r)}\left( \frac{J^2}{x} + x \mu \right) = \sqrt{A(r)} f(x), \label{EqEbXY}
\eeq
where $r = r(x,y) = \sqrt{x^2+y^2}$. The function $A(r)$ reflects the spacetime properties, while $f(x)$ those of the string loop. The behavior of the boundary energy function is given by interplay of the functions $A(r)$ and $f(x)$. Let us stress that we use in the following a simplification enabled by the assumption of purely axisymmetric oscillations of the string loop, namely $\rho = \sqrt{x^2+z^2} \lightarrow x$, assuming $x>0$.

The local extrema of the boundary energy function $E_{\rm b}$ are of crucial importance since they determine the regions of different character of the string loop motion. Stationary points of $E_{\rm b}(x,y)$ are determined by the conditions
\bea
  (E_{\rm b})_{x}' &=& 0  \,\, \ \Leftrightarrow \,\,  x A_{r}' f = - 2 r A f_{x}' \label{extr_a1} \\
  (E_{\rm b})_{y}' &=& 0  \,\, \ \Leftrightarrow \,\,   A_{r}' y  = 0, \label{extr_a2}
\eea
where we assume $f(x) > 0$ for $x>0$. The prime $()_{m}'$ denotes derivation with respect to the coordinate $m$.
In order to determine character of the stationary points at $(x_{\rm e},y_{\rm e})$ given by the stationarity conditions, i.e.,whether it is a maximum (``hill'') or minimum (``valley'') of the energy boundary function $E_{\rm b}(x,y)$, we have to consider the conditions
\bea
 && [(E_{\rm b})_{yy}''] (x_{\rm e},y_{\rm e}) \quad < 0 \,\, \mathrm{(max)} \quad > 0 \,\, \mathrm{(min)}  \label{extr_b1}\\
 && [(E_{\rm b})_{yy}'' (E_{\rm b})_{xx}'' - (E_{\rm b})_{yx}'' (E_{\rm b})_{xy}''](x_{\rm e},y_{\rm e}) > 0  \label{extr_b2}
\eea
Behavior of the energy boundary function $E_{\rm b}(x,y)$ in each of the directions $x$ and $y$ is relevant for the character of the string-motion boundary. The behavior in the $x$ direction in equatorial plane ($y=0$) gives the information if the boundary is open to the origin of coordinates (BH horizon), and the behavior in the $y$ direction gives the information if the boundary is open to the infinity.

It is obvious from equations (\ref{BRelation}),(\ref{EqEbXY}) that we can make the rescaling $E_{\rm b} \rightarrow E_{\rm b} / \mu $ and $J \rightarrow J / \sqrt{\mu} $ assuming $\mu > 0$. 
This choose of ``units'' will not affect string boundary equations, and this is equal to set the string tension $\mu=1$ in Eqs (\ref{BRelation}),(\ref{EqEbXY}).

%
%
%
%

\section{Analysis of the motion of string loops}
We shall successively study properties of the axisymmetric string loop motion for the special cases of spherically symmetric spacetimes: flat, \dS{}, \Schw{} and \Schw\nnd\dS{} in order to clearly demonstrate the role of the cosmological constant. 

In our numerical calculations of the string-loop motion we will set the string tension $\mu=1$. The starting point $P$ of the string motion will be properly chosen, usually little bit below the equatorial plane. If the string will be starting from the rest ($\dot{r}=0, \dot{\theta}=0$) with some given angular momentum $J$, the string energy can be calculated from the energy boundary function $E=E_{\rm b}(J)$.

First, we illustrate the role of the string parameters in the clearest form - for the motion in the flat spacetime.

%
%
%
%

\subsection{Flat spacetime}

For the flat spacetime the characteristic function of the line element (\ref{SfSymMetrika}) takes the form 
\beq
	A(r)=1
\eeq
and there is no characteristic length scale. The effective potential is given by the relation 
\beq
 V(r,\theta) = -E^2 + r^2 \sin^2 \theta (\SS^{\tau\tau})^2,
\eeq
while the boundary energy function takes the form
\beq
E_{\rm b}(x) = \frac{J^2}{x}  + x.
\eeq
It diverges $E_{\rm b} \lightarrow + \infty$ for both $x \lightarrow 0$ and $x \lightarrow \infty$.
Using the condition for the extrema of the boundary energy function (\ref{extr_a1}-\ref{extr_a2}), we find that 
the minimum of $E_{\rm b}(x)$ is located at 
\beq
x_{\rm min} = J, 
\eeq
for all $y$ and the  boundary energy 
\beq
	 E_{\rm min} = 2 J. \label{EnFlat} 
\eeq
Notice that in the $y$ - direction the energy boundary function is totally flat.

\begin{figure}[t!]
\subfigure[ ~Energy $ E_{\rm b}(x)$ \label{stringFIG_W00a}
]{\includegraphics[height=4.3cm]{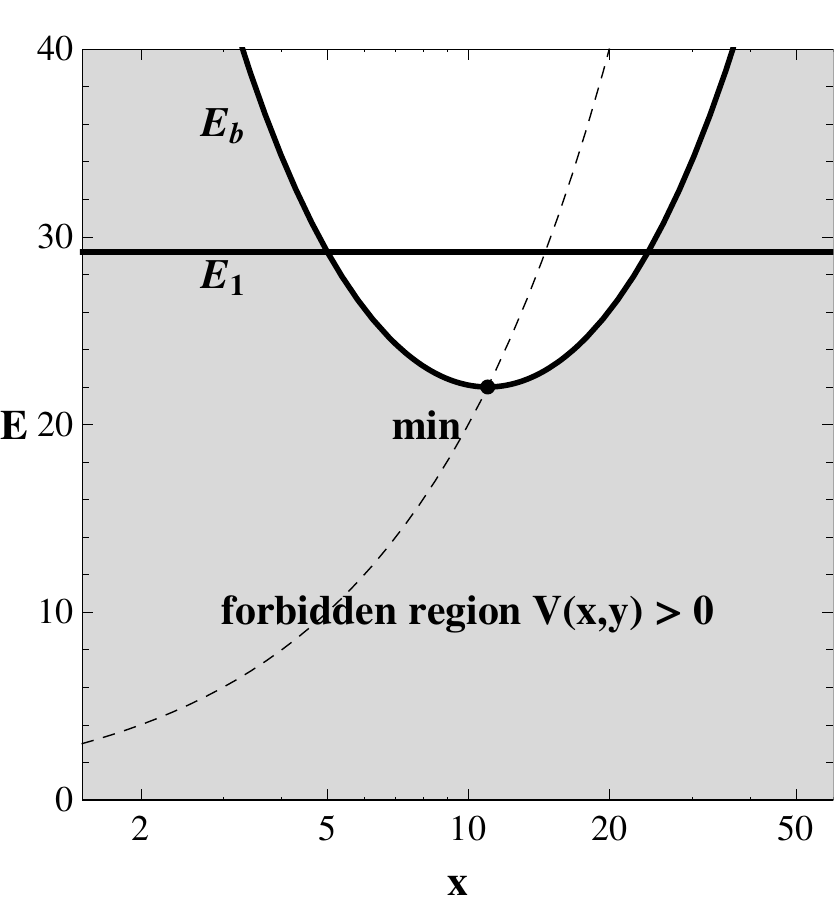}}
\subfigure[ ~ Energy $E_{\rm b}(J)$ \label{stringFIG_W00b}
]{\includegraphics[height=4.2cm]{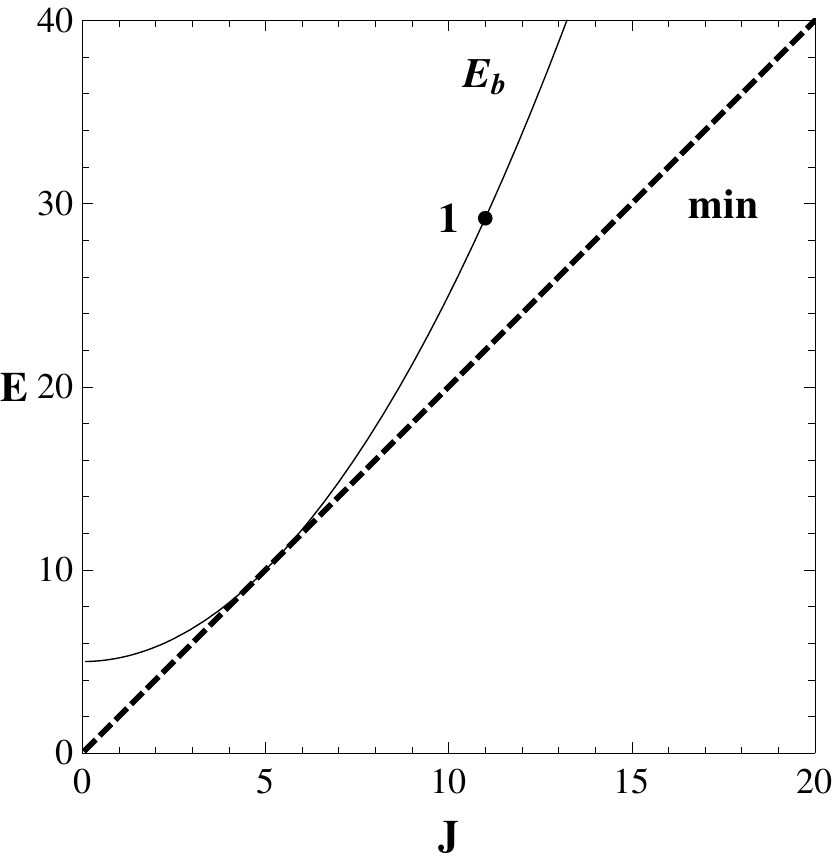}}
\vspace{-0.3cm}
\caption{
The boundary energy $E_{\rm b}$ as a function of Cartesian coordinate $x$ (\ref{ccord}) (case (a)) and the string parameter $J$ (case (b)) for the flat spacetime. String-loop initial conditions marked by the black point are given by coordinates $x_0=5, y_0=1$ and the string parameter $J=11$, corresponding to the energy $E\doteq 29$. The dashed curve (line) represents the minima of the  boundary energy function $E_{\rm b(min)}$. \label{stringFIG_W00}}
\vspace{0.2cm}
\begin{minipage}{4.2cm}
\vspace{-0.2cm}\includegraphics[width=4cm]{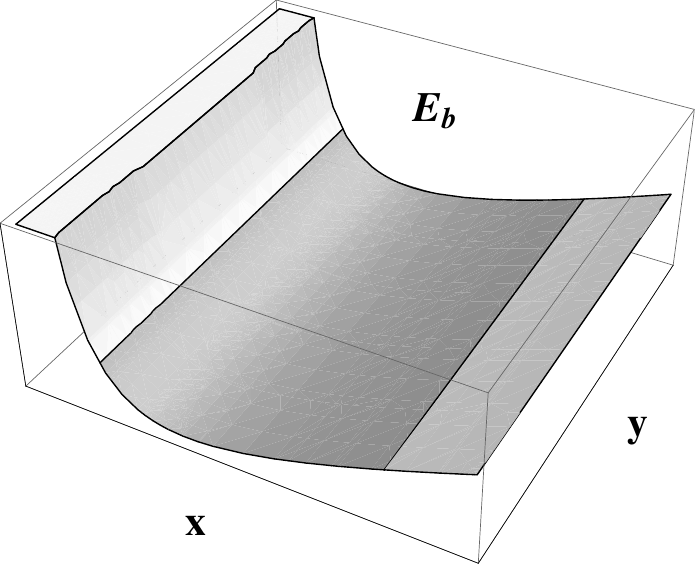}
\end{minipage}
\begin{minipage}{3.8cm}
\includegraphics[width=4cm]{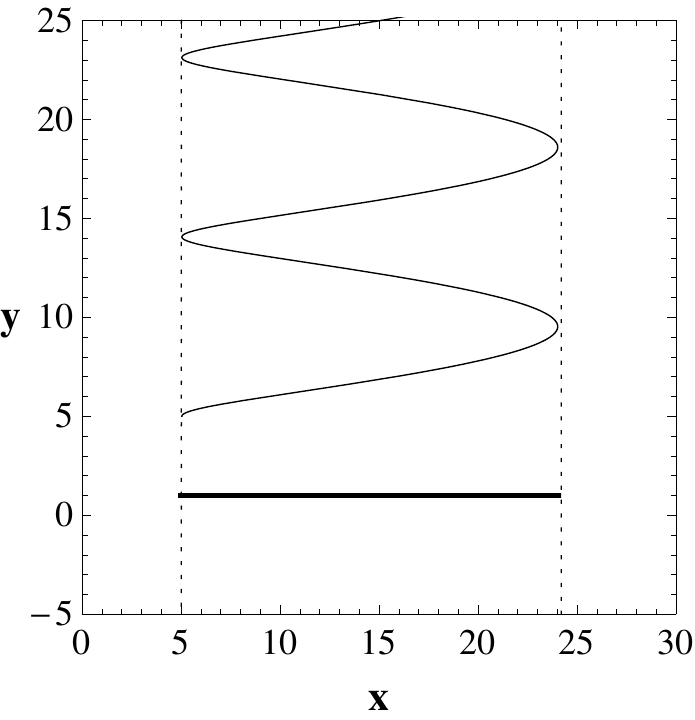}
\end{minipage}
\caption{\label{stringFIG_00} The boundary energy $E_{\rm b}(x,y)$ for given $J$ (case (a)) and a string-loop trajectory (case (b)) in the flat spacetime. The thick solid curve represents path of the string loop starting from rest ($\dot{x}=0$, $\dot{y}=0,$) at $x_0=5, y_0=1$, while the thin solid curve represents string with initial momentum in $y$-direction $\dot{y}=0.5$, $\dot{x}=0$ at $x_0=5, y_0=5.04$. We use string parameter $J=11$ and string energy $E=29.2$ in both cases. The dotted lines represent boundaries $E=E_{\rm b}$ for the string motion.}
\end{figure}

The behavior of the energy boundary $E_{\rm b}$ is illustrated in Figure \ref{stringFIG_W00}. Clearly, the convex shape of the function $E_{\rm b}(x)$ for all $y$ ensures that there will be always an inner and outer boundary for the motion in the$x$ direction. 
If the string is located at the radius $x_{\rm min}$ with energy $ E_{\rm min} = 2J $ it will not oscillate, keeping its position at this stable equilibrium point. The minimum of the  boundary energy $E_{\rm min}$ corresponds to the dashed line at Fig. \ref{stringFIG_W00b}. Motion of the string is allowed if its energy satisfy the condition $E > E_{\rm min}$.

Equation (\ref{MovingEq01}) for the motion of strings with non-zero tension and current (angular momentum) takes for the flat spacetime expressed in the spherical coordinates the form
\beq
 \ddot{r} = {\dot{\theta}}^2 r -\dot{r} \frac{\partial_{\tau}\Sigma^{\tau\tau}{}}{\Sigma^{\tau \tau}}  + r \sin^2\theta  \frac{\Sigma^{\sigma\sigma}}{\Sigma^{\tau \tau}}, \label{motion_flat}
\eeq
while (\ref{MovingEq02}) remains unchanged. These equations can be integrated numerically. The typical results are represented in Figure \ref{stringFIG_00}. 

It is intuitively clear that while moving in the $y$-direction, the amplitude of the string loop oscillations in the $x$-direction, representing oscillations of the loop as a whole, remains constant. Properties of the flat spacetime does not change while transporting the loop in any direction since the center of the spherical symmetry can be chosen at any point of the spacetime, keeping thus constant amplitude of oscillations. This fact can be easily demonstrated formally expressing the equations of the string motion in the $x$-$y$ coordinates. The equations take the form
\bea
     \ddot{x} \Sigma^{\tau \tau} + \dot{x} \partial_\tau \Sigma^{\tau \tau} - x \Sigma^{\sigma \sigma} &=& 0, \label{Neq01} \\
     \ddot{y} \Sigma^{\tau \tau} + \dot{y} \partial_\tau \Sigma^{\tau \tau} &=& 0, \label{Neq02}
\eea 
where
\beq
     \Sigma^{\tau\tau} = \frac{J^2}{x^2} + 1 , \,\, \Sigma^{\sigma\sigma} = \frac{J^2}{x^2} - 1 ,
     \partial_\tau \Sigma^{\tau\tau} = -\frac{2J^2}{x^3} \dot{x}. 
\eeq
Since the first equation governing the oscillations in the $x$-direction does not depend on $y$, it directly implies that magnitude of the oscillations is constant during the motion in the $y$-direction since we do not consider dissipative phenomena; for example, there is no gravitational radiation of the axially oscillating string. 

%
%
\begin{figure*}
\subfigure[~ Energy $E_{\rm b}(J)$]
{\label{stringFIG_13c} \includegraphics[width=5.5cm]{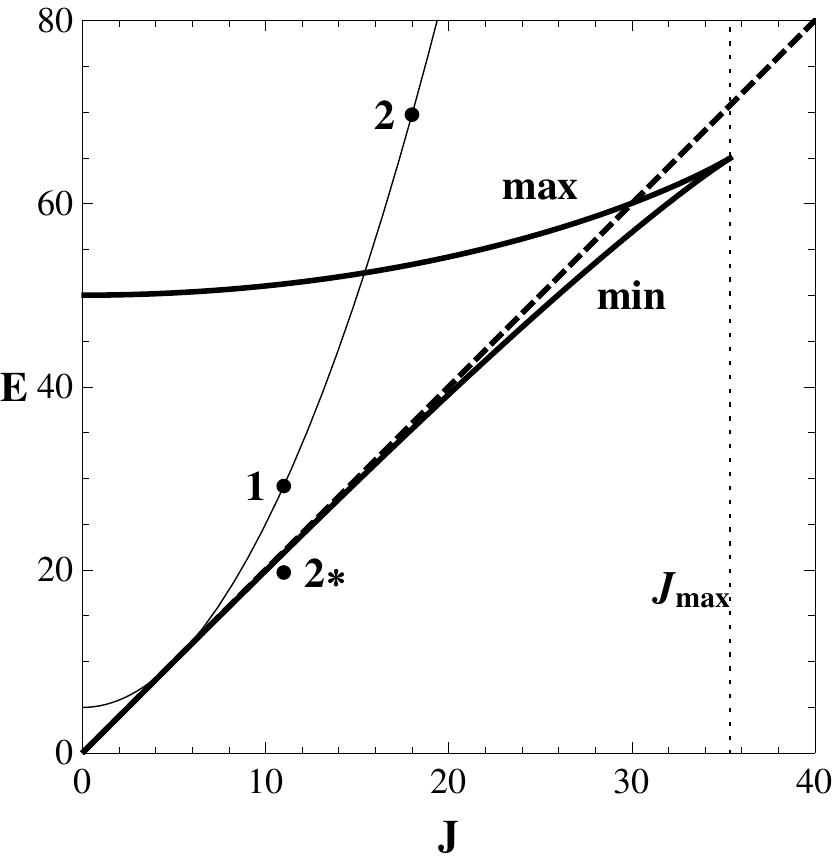}}
\subfigure[~Energy $E_{\rm b}(x)$ for different $J$]
{\label{stringFIG_13a} \includegraphics[width=5.6cm]{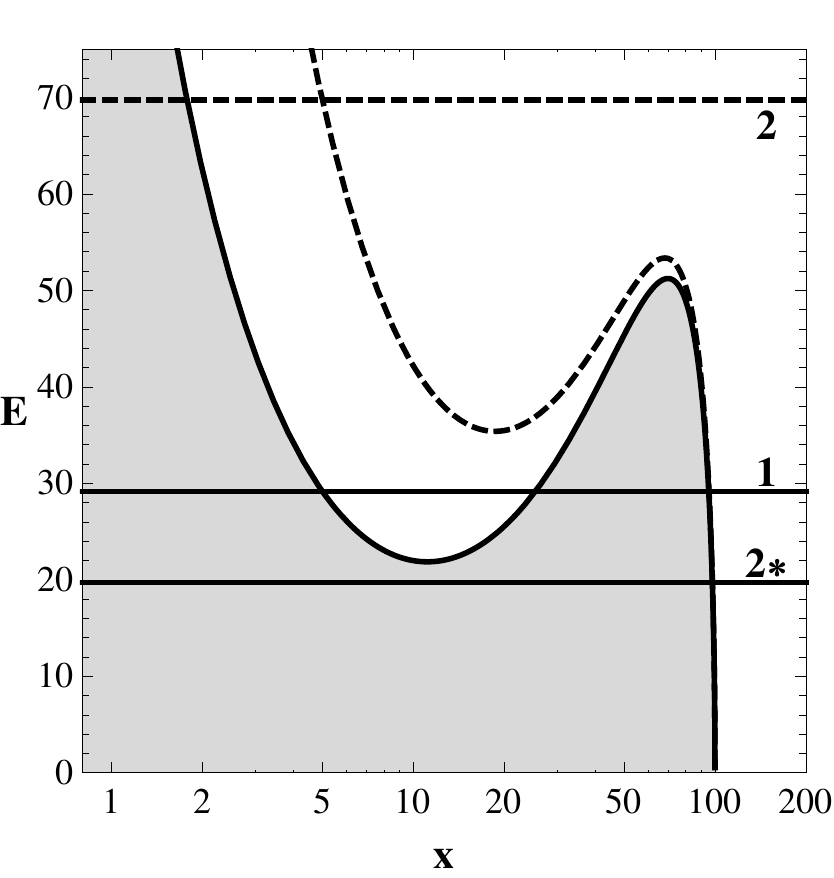}}
\subfigure[~ Energy $E_{\rm b}(y)$ for different $J$]
{\label{stringFIG_13b} \includegraphics[width=5.4cm]{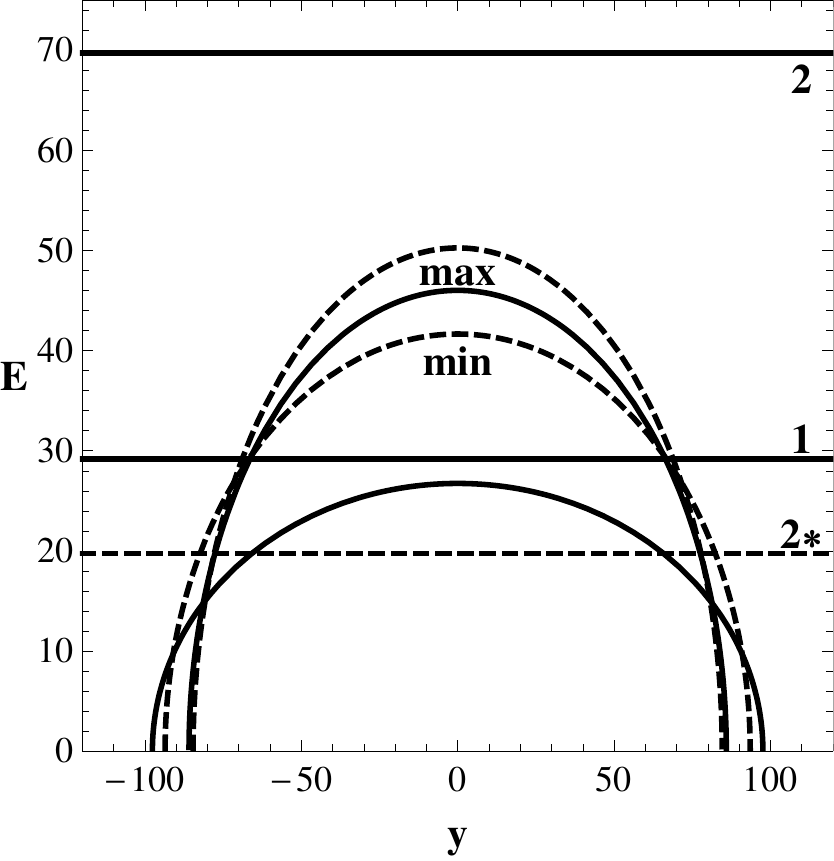}}
\vspace{-0.3cm}
\caption{ \label{stringFIG_13}
Boundary energy function $E_{\rm b}(x,y,J)$ for the {\bf \dS{} spacetime}. (a) Thick solid curves correspond to the maxima and minima of $E_{\rm b}$ in the $x$-direction that exist for $J<J_{\rm max}$. For comparison we also show minimum for the flat spacetime as thick dashed curve. Thin solid line represents the energy $E_{\rm b}(x_0,y_0,J)$ for $x_0 = 5, y_0 = -1$.
Figures (b) and (c) give representative $y=\rm{const}$ and $x=\rm{const}$ sections of the boundary energy function. We present $E_{\rm b}(x,0,J)$ and $E_{\rm b}(x_{\rm max/min},y,J)$, where the $x$-coordinate is taken in the minimum and maximum of the  boundary energy function $x$-profile. We choose representative values of string parameter $J$: $J_1 = 11$ (solid line) and $J_2 = 18$ (dashed line).
The string loop is assumed to be starting from the point $x_0 = 5, y_0 = -1$ for the cases 1,2 and $x_0* = 98, y_0* = -1$ for the case 2*. Calculated energies for the string-loop motion are $E_{\rm 1} \doteq 30, E_{\rm 2} \doteq 70, E_{\rm 2*} \doteq 20$.  
In all cases we assume the string loop starting from the rest ($\dot{x}=0,\dot{y}=0$). Restricted area for the string motion $V(x,y) > 0$ is shaded.
}
\end{figure*}

The motion can be expressed in a simple explicit integrated form. Defining a new time coordinate $\tit$ by
\beq
 \d \tit = \frac{1}{\Sigma^{\tau\tau}} \d \tau, \label{Neq00}
\eeq
equations of the string motion (\ref{Neq01}-\ref{Neq02}) take the form
\bea
     \ddot{\xx} - \frac{J^4}{x^3} + x &=& 0, \label{eq03} \\
     \ddot{\yy} &=& 0, \label{eq04}
\eea
where $\dot{\xx} = \d x / \d \tit $. Analytic solutions of the equations (\ref{eq03}-\ref{eq04}) are
\bea
 x^2(\tit) &=& \frac{1}{2} \Big[ x_0^2 + \frac{J^4}{x_0^2} + \dot{\xx}_0^2 + 2 x_0 \dot{\xx}_0 \sin(2\tit) \nonumber \\
 & & + \left( x_0^2 - \frac{J^4}{x_0^2} - \dot{\xx}_0^2 \right) \cos(2\tit) \Big], \\
 y(\tit) &=& \dot{\yy}_0 \tit + y_0,
\eea
where $x_0$ ($y_0$) is initial position and $\dot{\xx}_0$ ($\dot{\yy}_0$) is initial speed in $x$ or $y$ direction.

%
%
%
%

\subsection{\dS{} spacetime}

For the \dS{} spacetime the characteristic function of the line element (\ref{SfSymMetrika}) takes the form
\beq
A(r) = 1 - \frac{1}{3} \Lambda r^2  = 1 - \frac{r^2}{r_{c}^2} ,
\eeq
where a characteristic length scale determined by the cosmological constant is introduced by
\beq
        r_{c}  = \sqrt{3 / \Lambda}  
\eeq
and gives extension of the so called cosmic horizon of the \dS{} spacetime \cite{Car:1973:BlaHol:,Stu:1983:BULAI:}

In the case of the \dS{} spacetime the boundary energy function has different asymptotical behavior in comparison the case of the flat spacetime due to the cosmic repulsion - there is $E_{\rm b} \lightarrow +\infty$ for $r \lightarrow 0$, but $E_{\rm b} \lightarrow -\infty$ for $r \lightarrow \infty$.
The equation for extrema of the boundary energy function (\ref{extr_a2}) gives $y=0$ while (\ref{extr_a1}) implies a quadratic equation in $x^2$ 
\beq
 2 \Lambda x^4 - 3 x^2 + 3 J^2 = 0.
\eeq

The extrema condition can be expressed in the simple form
\beq
    J^2 = J^2_{\rm E} \equiv x^2 \left(1 - \frac{2}{3} \Lambda x^2 \right) .
\eeq
The loci of minimum $x_{\rm E(min)}$ and maximum $x_{\rm E(max)}$ of the boundary energy function are given by
\beq
	x_{\rm E(min)}^2 = \frac{r_{\rm c}^2}{4} \left(1 - \XXB \right),\quad
	x_{\rm E(max)}^2 = \frac{r_{\rm c}^2}{4} \left(1 + \XXB \right).
\eeq
where a new parameter
\beq
\XXB = \sqrt{1-\frac{8J^2}{r_{\rm c}^2}}
\eeq
is introduced. We can see that the maxima and minima exist if $J<J_{\rm max}$, where
\beq
 J_{\rm max} = r_{\rm c}/\sqrt{8} = 1/\sqrt{8\Lambda}.
\eeq

Assuming $J/r_c << 1$, we obtain asymptotic formulae
\beq
      x_{\rm E(min)}  \sim J, \qquad x_{\rm E(max)} = \frac{r_c}{\sqrt{2}} \left(1 - \frac{J^2}{r_{c}^2}\right),
\eeq
and we directly see that for very small values of the cosmological constant the internal solution corresponding to the minimum coincides with the solution for the flat spacetime, while the outer solution corresponding to the maximum is located near $r_c/\sqrt2$ with the string-parameter $J/r_c$ representing a small correction. 

The extremal values of the  boundary energy $E_{\rm b(min)}$ and $E_{\rm b(max)}$ are given by the relations 
\bea
         E_{\rm b(min)} &=& E_{\rm b}(x=x_{\rm E(min)},J,\Lambda), \label{Enmin}\\
         E_{\rm b(max)} &=& E_{\rm b}(x=x_{\rm E(max)},J,\Lambda). \label{Enmax}
\eea

In \dS{} spacetime these extrema can be expressed in an explicit form 
\beq
 E_{\rm b(ext)} = \sqrt{\frac{3}{4} \pm \frac{\XXB}{4}} \left( \frac{2J^2}{r_{\rm c}\sqrt{1 \pm \XXB }} +
 \frac{r_{\rm c}}{2} \sqrt{1 \pm \XXB} \right),
\eeq
where ``$+$'' stands for the maximum and ``$-$'' for the minimum.

In the limit of $J^2/r_{\rm c}^2 << 1$, we find
\bea
 E_{\rm b(min)} &\sim& 2 J \left( 1 - \frac{1}{2} \frac{J^2}{r_{\rm c}^2} \right),\label{ESdSmin}\\
 E_{\rm b(max)} &\sim& \frac{r_{\rm c}}{2} \left( 1 + \frac{J^4}{r_{\rm c}^4} \right),   \label{ESdS}
\eea
The values of the boundary energy at the local extrema $E_{\rm min}(J)$ and $E_{\rm max}(J)$ are illustrated in Fig. \ref{stringFIG_13c}.

The motion of the string loop in the de Sitter spacetimes is of the oscillatory character analogical to the motion in the flat spacetime for energy $E < E_{\rm max}$, while it is unlimited for $E > E_{\rm max}$ - see Fig. \ref{stringFIG_13}. We give the sections of constant $x$ and $y$, i.e., in the directions reflecting character of the oscillatory motion of the string loops.

The equation for the string motion in the spherical coordinates (\ref{MovingEq01}) can be written for \dS{} spacetime in the form
\bea
 \ddot{r} &=& {\dot{\theta}}^2 r -\dot{r} \frac{\partial_{\tau}\Sigma^{\tau\tau}{}}{\Sigma^{\tau \tau}}  + r \sin^2\theta  \frac{\Sigma^{\sigma\sigma}}{\Sigma^{\tau \tau}} \nonumber \\
  && + \frac{\Lambda}{3} \sin^2 (\theta) r^3 \left( \frac{2\mu}{\Sigma^{\tau \tau}} \right), \label{lambda_term}
\eea
while (\ref{MovingEq02}) remains unchanged. Comparing this equation for radial string motion with those for the motion in the flat spacetime (\ref{motion_flat}), we can see that the contribution of the cosmological constant is given by the last term in Eq. (\ref{lambda_term}). For $\Lambda > 0$ and $ \mu>0$ this term is positive, therefore, the repulsive cosmological constant has tendency to stretch the string.

\begin{figure}
\subfigure[~Case 1 on Fig. \ref{stringFIG_13} \label{stringFIG_01a}]{\includegraphics[width=4.2cm]{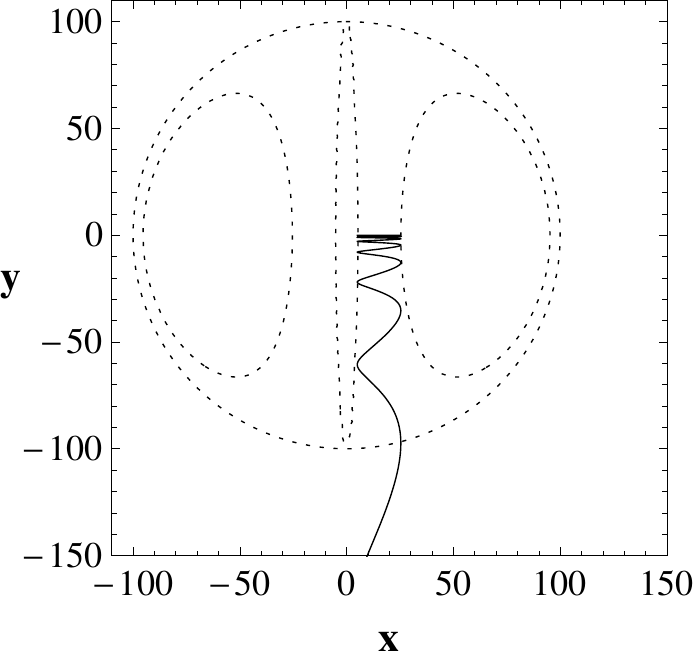}}
\subfigure[~Case 2 on Fig. \ref{stringFIG_13} \label{stringFIG_01b}]{\includegraphics[width=4cm]{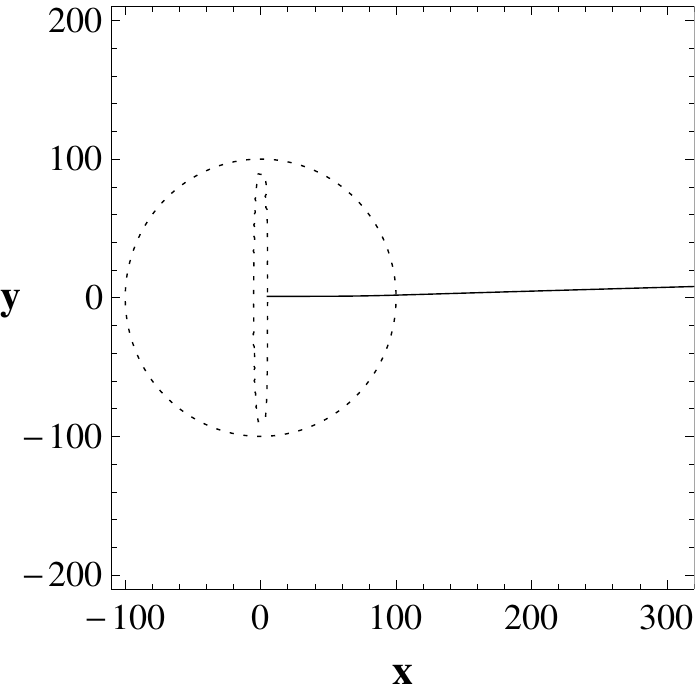}}
\vspace{-0.3cm}
\caption{\label{stringFIG_01} String-loop motion in the {\bf \dS{} spacetime} illustrated for initial conditions given above. The solid curves represent path of one point on the string, dotted curves represent boundaries determined by $E=E_{\rm b}$ established for the initial conditions of the string motion. Dotted circle represents the cosmological horizon located at $r_c = 100$.}
\vspace{0.2cm}
\begin{minipage}{4.2cm}
\vspace{-0.2cm}\includegraphics[width=4cm]{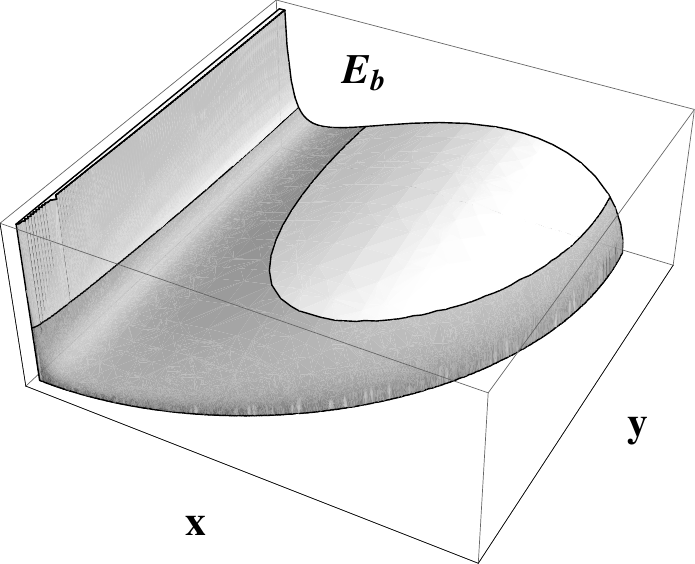}
\end{minipage}
\begin{minipage}{4cm}
\includegraphics[width=4cm]{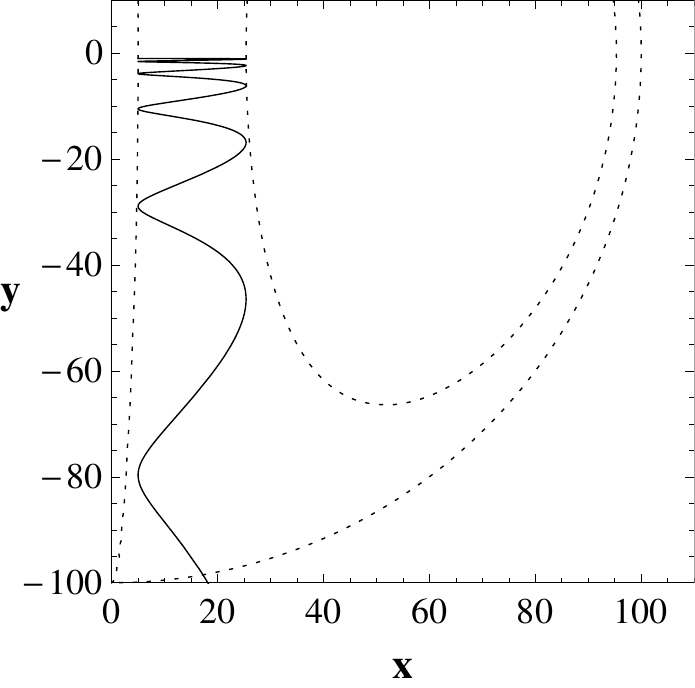}
\end{minipage}
\caption{Boundary energy function $E_{\rm b}(x,y)$ (a) and the string-loop path (b) for the case 1 given in Fig. \ref{stringFIG_13}. Amplitude of the string motion in the $x$-direction remains constant, while string is being accelerated in the $y$-direction.}
\end{figure}

The only plane where the string loop can stay at equilibrium position in \dS{} spacetime is the equatorial plane ($y = 0$), but this position is unstable. Any deviation from the equatorial plane leads to the motion in the $y$-direction. The influence of the repulsive cosmological constant on the string motion is illustrated in Fig. \ref{stringFIG_01}. As follows from the role of the term (\ref{lambda_term}), the string loop oscillates in the $x$ direction while moving and speeding up in $y$ direction, due to the influence of the cosmological constant. 

Since in the de Sitter spacetime the center of the spherical symmetry can be chosen at any point of the spacetime, similarly to the case of the flat spacetime, the amplitude of the oscillatory motion again remains constant. We can  demonstrate this fact formally by expressing the equations of the motion in the $x$-$y$ plane:
\bea
  \ddot{x} \Sigma^{\tau \tau} + \dot{x} \partial_\tau \Sigma^{\tau \tau} - x \Sigma^{\sigma \sigma} &=& \frac{2 \Lambda \mu}{3} x^3 ,\\
  \ddot{y} \Sigma^{\tau \tau} + \dot{y} \partial_\tau \Sigma^{\tau \tau} &=& \frac{2 \Lambda \mu}{3} x^{2} y. 
\eea 
The equation for the oscillatory motion in the $x$-direction is independent of $y$ and its amplitude will remain constant during the motion in the $y$-direction. Of course, the amplitude is now influenced by the cosmological constant term as clear from the equation of the energy boundary (see Fig.6).

Using the relation (\ref{Neq00}) we arrive to
\bea
     \ddot{\xx} - \frac{J^4}{x^3} + x &=& \frac{2 \Lambda}{3} x (J^2+x^2),  \label{eq05} \\
     \ddot{\yy} &=& \frac{2 \Lambda}{3} y (J^2+x^2). \label{eq06}
\eea
In order to obtain the solution for the $x$ motion we have to find the roots of the cubic equation in $x$
\beq
 \Lambda x^3 + (2J^2 \Lambda - 3) x^2 + 6 C_1 x - 3 J^4 = 0.
\eeq
Denoting the three roots $a,b,c$, we find the equation of $x$-motion in terms of the Jacobi elliptic function ${\rm JacobiSN}(u,m)$
\beq
 x^2 = (b-c)\,{\rm JacobiSN}\left[ \sqrt{\frac{a-c}{3}} \, \tau + C_2, \frac{b-c}{a-c} \right]^2 + c,
\eeq
where $C_1, C_2$ are constants of integration. The $y$-motion can be integrated numerically only.

%
%
%
%

%
%
\begin{figure*}
\subfigure[~ Energy $E_{\rm b}(\JJ)$]
{\label{stringFIG_15a} \includegraphics[width=5.5cm]{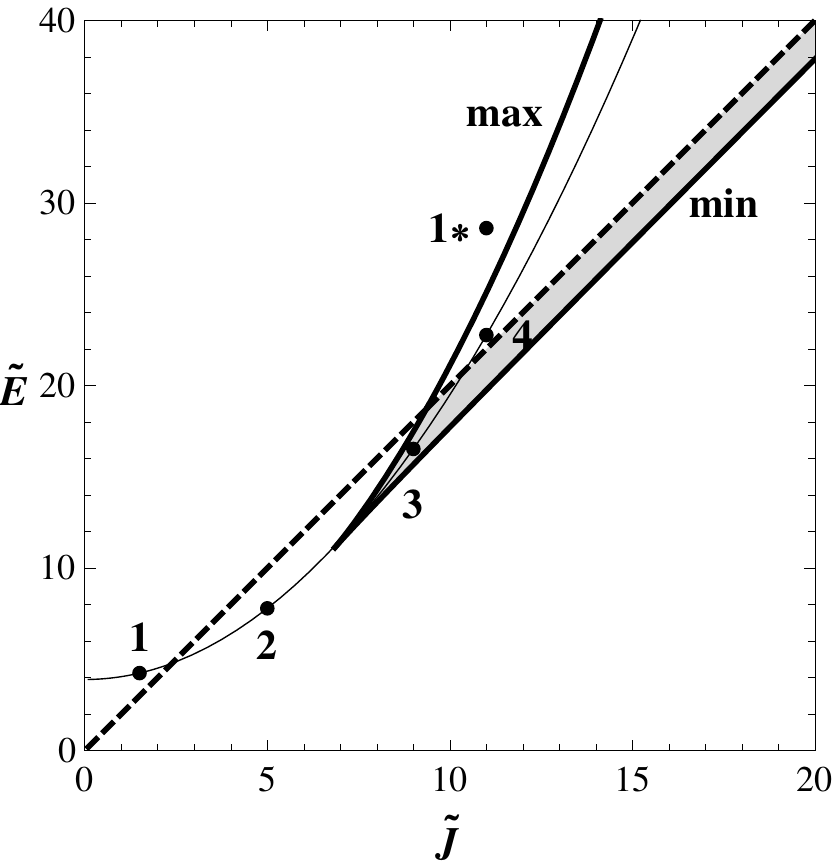}}
\subfigure[~ Energy $E_{\rm b}(\xx)$ for different $\JJ$.]
{\label{stringFIG_15b} \includegraphics[width=5.5cm]{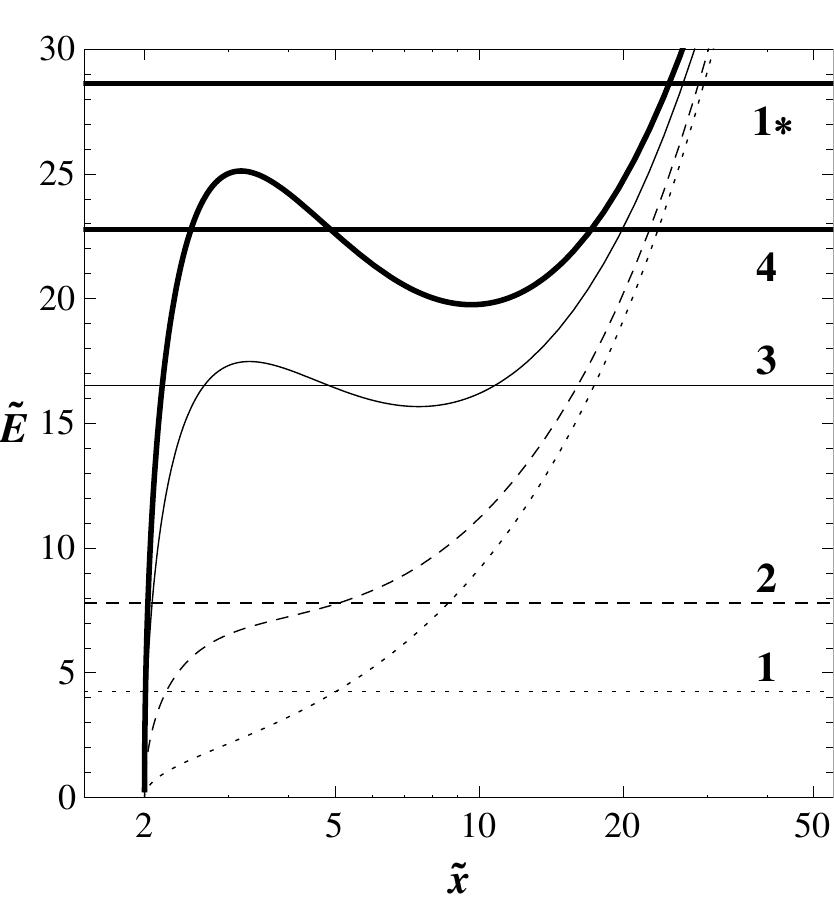}}
\subfigure[~ Energy $E_{\rm b}(\tilde{y})$ for different $\JJ$.]
{\label{stringFIG_15c}\includegraphics[width=5.5cm]{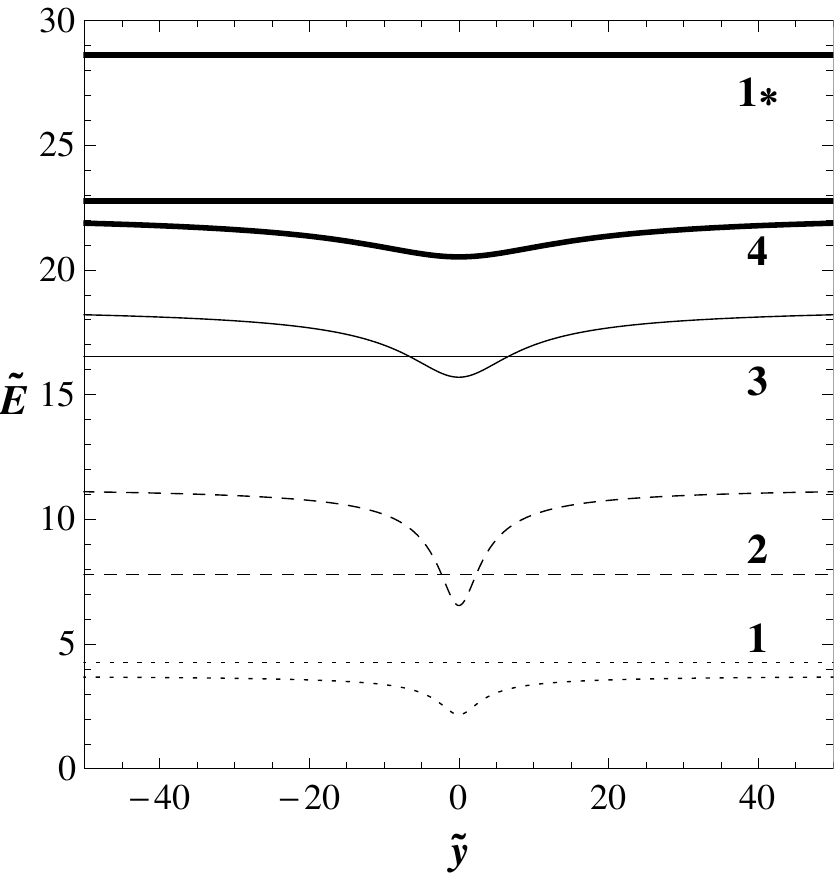}}
\vspace{-0.3cm}
\caption{\label{stringFIG_15}
Boundary energy function $\EE_{\rm b}(\xx,\yy,\JJ)$ of the {\bf \Schw{} spacetime}. (a) Thick solid curves correspond to the maxima and minima of $\EE_{\rm b}$ in the $\xx$-direction. Minimum of the boundary energy function for the flat spacetime $\EE = 2 \JJ$ represents the dashed curve, while the thin solid line represents the energy $\EE_{\rm b}(\xx_0,\yy_0,\JJ)$ taken at the point with coordinates $\xx_0 = 5, \yy_0 = 1$. We choose points with representative values of the string parameter: $\JJ_1 = 1.5$, $\JJ_2 = 5$, $\JJ_3 = 9$ and $\JJ_4 = 11$. Region where ``lakes'' corresponding to the trapped states can exist is shaded.
The figures (b) and (c) show $\xx$- and $\yy$- profiles of the boundary energy function; we use $\EE_{\rm b}(\xx,0,\JJ)$ (b) and $\EE_{\rm b}(\xx_{\rm min},\yy,\JJ)$ (c). (For cases 1,2 there is no minimum in $\xx$-direction and we use $\EE_{\rm b}(3,\yy,\JJ)$ instead.) 
The profiles are presented for the chosen representative values of string parameter: $\JJ_1 = 1.5$ (dotted line), $\JJ_2 = 5 $ (dashed line), $\JJ_3 = 9 $ (thin line) and $\JJ_4 = 11 $ (thick line).
We assume a string loop starting from the rest ($\dot{\xx}=0,\dot{\yy}=0$) at the point $\xx_0 = 5, \yy_0 = 1$ for cases 1-4 and at the point $\xx_0 = 25, \yy_0 = 1$ for the case 1* (trajectory of the type 1). The energies calculated for the string motion are given as follows: $\EE_{\rm 1} \doteq 4, \EE_{\rm 2} \doteq 8, \EE_{\rm 3} \doteq 17, \EE_{\rm 4} \doteq 23, \EE_{\rm 1*} \doteq 29$.  
}
\subfigure[~ $\yy=0$]
{\label{stringFIG_15Ba} \includegraphics[width=5.5cm]{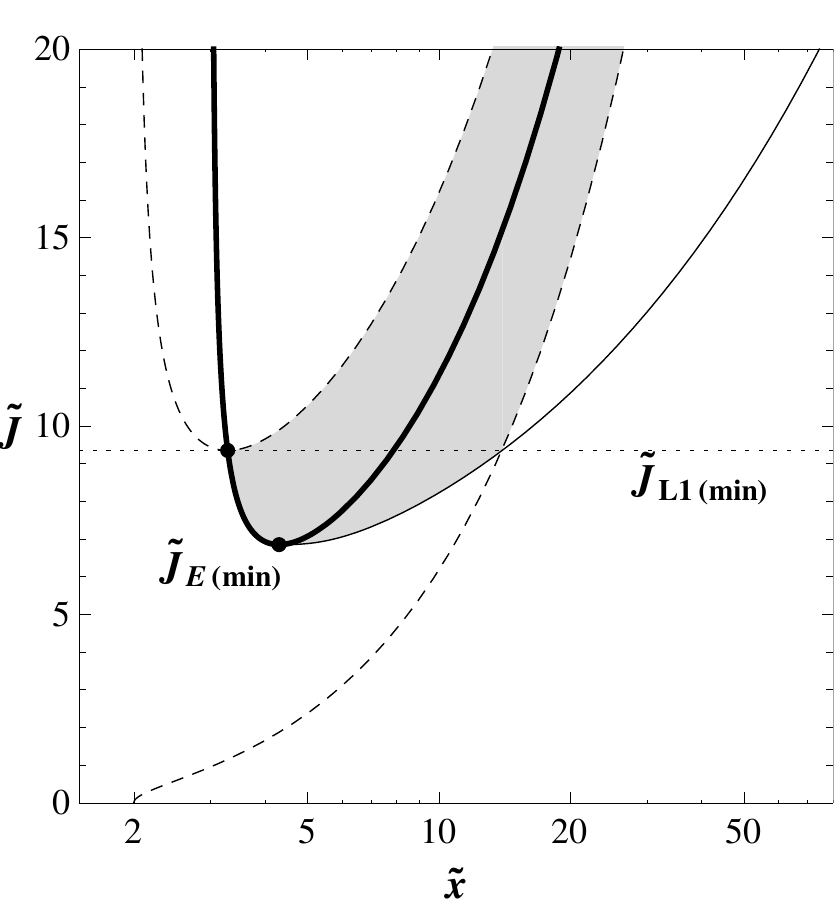}}
\subfigure[~ $\yy= 1$]
{\label{stringFIG_15Bb}\includegraphics[width=5.5cm]{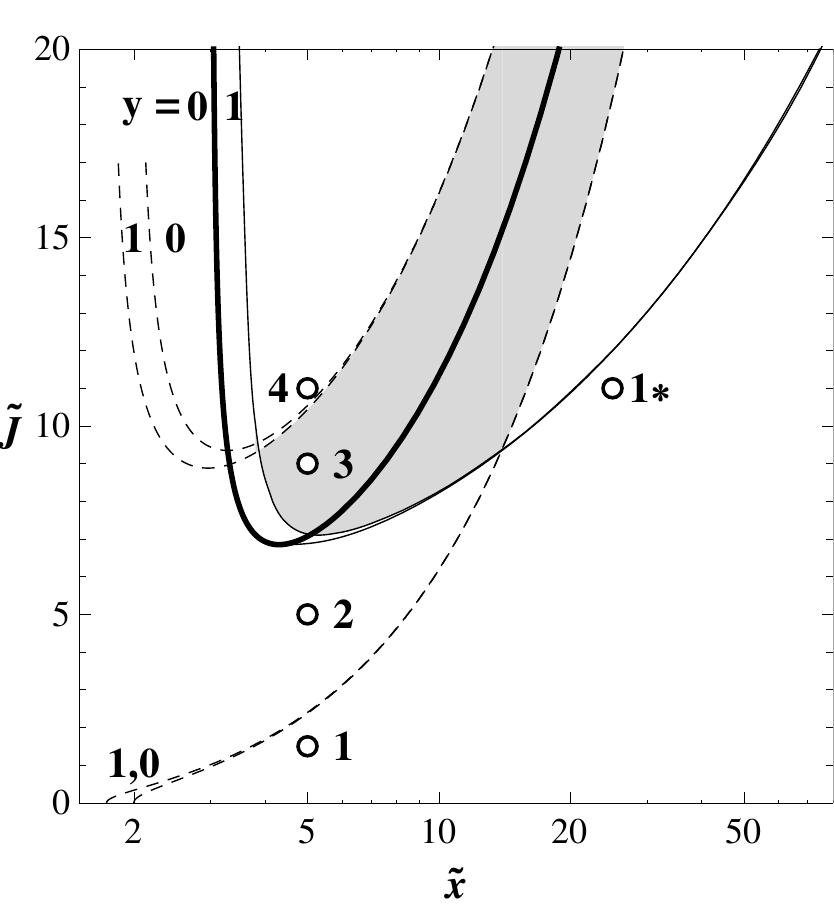}}
\subfigure[~ $\yy=\infty$]
{\label{stringFIG_15Bc} \includegraphics[width=5.5cm]{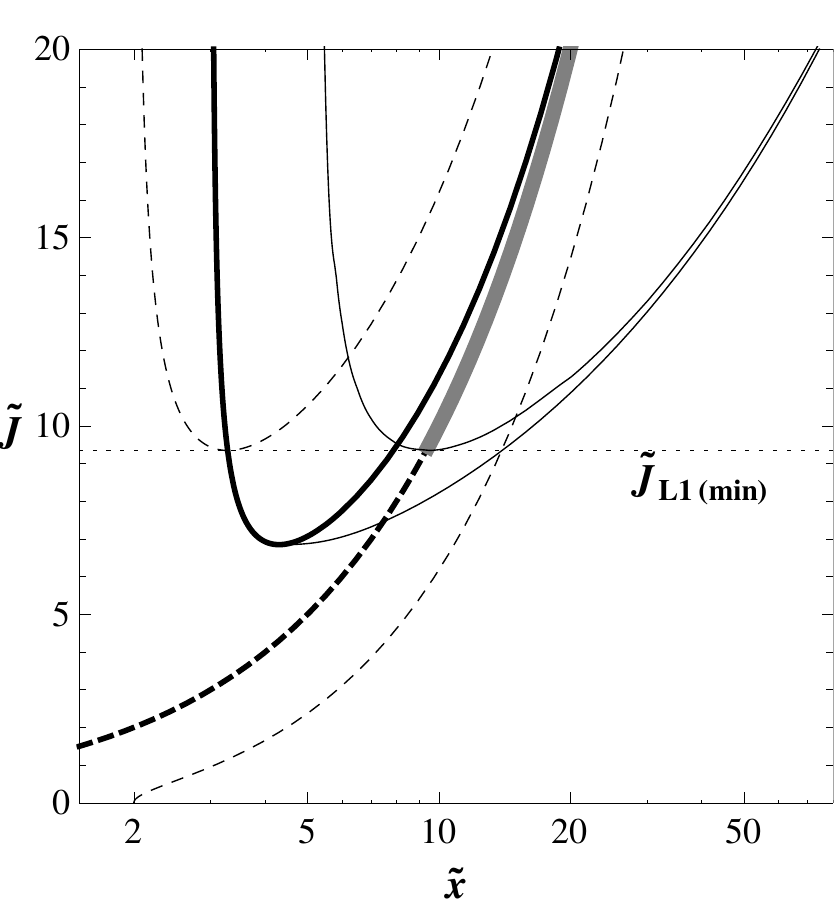}}
\vspace{-0.3cm}
\caption{\label{stringFIG_15B}
String parameter functions $\JJ(\xx)$ determining character of the boundary energy function $\EE_{\rm b}$, given for three representative values of $\yy$. The function $\JJ_{\rm E}(\xx)$ (thick solid curve) determining extrema of $E_{\rm b}$ is independent of $\yy$. Dashed curves represent the functions $\JJ_{\rm L1}(\xx,\yy), \JJ_{\rm L1}(\xx,\yy)$. For $\yy \lightarrow \infty$ these curves coalescence to the function $ \JJ = \xx $ (thick dashed curve on the third Fig.).
The thin solid curves represent numerically calculated ``projected'' angular momentum function $\JJ_{\rm p}(\xx,\yy)$ that is determined by the projection of the extremal (maximal) energy level $\EE_{\rm b( max)}(\xx_{\rm max},\yy,\JJ)$ onto the energy boundary function itself. Regions where ``lakes'' can exist are shaded. The lowest ``big lake'' exist for $\JJ=\JJ_{\rm L1(min)}$, extending to $\yy=\infty$ in the $\yy$-direction and down to $\xx_{\rm mt}$ in the $\xx$- direction. For $\JJ>\JJ_{\rm L1(min)}$ the lake can extend asymptotically to $\yy=\infty$ at $\xx=\JJ$. For $\JJ_{\rm E(min)}<\JJ<\JJ_{\rm L1(min)}$ the ``lake'' does not reach infinity.
}
\end{figure*}

\subsection{\Schw{} spacetime}

For the \Schw{} spacetime the characteristic function of the line element (\ref{SfSymMetrika}) takes the form
\beq
            A(r) = 1 - \frac{2M}{r}.
\eeq
This geometry introduces a characteristic length scale corresponding to the radius of the black hole horizon that is given by the condition $A(r)=0$ and is determined by $r_{\rm h} = 2M$. Recall that the innermost stable circular orbit of the free particle motion is located at $r_{\rm ISCO} = 6M$, the radius of marginally bound orbit is located at $r_{\rm mb}=4M$,  while the photon circular orbit is located at $r_{\rm ph} = 3M$ \cite{Mis-Tho-Whe:1973:Gra:}.

In the \Schw{} spacetime, there is $E_{\rm b} \lightarrow +\infty$ for $r \lightarrow \infty$, but a new kind of asymptotic behavior appears near the black hole horizon - $E_{\rm b} \lightarrow 0$ for $r \lightarrow 2M$. Typical behavior of the boundary energy function is demonstrated by its profiles in the $x$- and $y$-directions - see Fig. \ref{stringFIG_15}.

The extrema equation (\ref{extr_a1}) leads to the cubic equation in $x$ coordinate
\beq
 	 x^3 -   Mx^2 - J^2 x + 3M J^2 = 0, 
\eeq
while (\ref{extr_a2}) gives $y=0$. Using the dimensionless coordinate $\xx = x/M$, ($\yy = y/M$, $\rr = r/M$) and dimensionless string parameter $\JJ = J/M$ (dimensionless energy $\EE = E/M$), the extrema equation can be expressed in the form
\beq
      \JJ^2 = \JJ_{\rm E}^2 \equiv \frac{\xx^2(\xx-1)}{\xx-3}. 
\eeq
Clearly, we have to restrict our considerations to the region above the black hole horizon $r > r_{\rm h}$. The extrema function $\JJ_{\rm E}^2$ diverges at $\xx = 3$ and at infinity - see Fig. \ref{stringFIG_15B}. The local extrema of the function $\JJ_{\rm E}^2$ are located at radii given by the condition
\beq
           \xx^2 - 5 \xx + 3 = 0.
\eeq
For our discussion, the minimum of $\JJ_{\rm E}^2$ located at
\beq
           \xx_{\rm min} = \frac{5+\sqrt{13}}{2} \sim 4.303 ,
\eeq
is only relevant. The corresponding minimal value of $\JJ_{\rm E}^2$ reads
\beq
 \JJ_{\rm E(min)}^2 = \frac{47+13\sqrt{13}}{2} \sim 46.936 .
\eeq
The boundary energy function has two extrema, maximum and minimum, located above the black-hole horizon (at $\xx > 2$), when (in the following, we can assume $\JJ>0$ due to the spherical symmetry of the spacetime)
\beq
        \JJ > \JJ_{\rm E(min)}.
\eeq
For $\JJ = \JJ_{\rm E(min)}$, the energy boundary function $\EE_{\rm b}(\xx,\JJ)$ has an inflex point. For $\JJ < \JJ_{\rm E(min)}$ there are no extrema of the energy boundary function above the horizon. Using the general formulae (\ref{Enmin}), (\ref{Enmax}), we give the extremal values of the boundary energy function in dependence on the string parameter $\JJ_{\rm E}$ in Fig.\ref{stringFIG_15a}.
The $x$-profiles of the  boundary energy function are for selected typical values of the string parameter illustrated in Fig. \ref{stringFIG_15b}. The oscillatory motion in the $x$-direction is allowed for string loops with $\JJ > \JJ_{\rm E(min)}$ and energy satisfying the condition $\EE_{\rm b(min)}(J) < \EE < \EE_{\rm b(max)}(J)$. String loops with $\JJ < \JJ_{\rm E(min)}$, or with $\EE < \EE_{\rm b(max)(J)}$ and $\JJ > \JJ_{\rm E(min)}$, can be captured by the black hole.

The $y$-profiles of the boundary energy function (illustrated in Fig. \ref{stringFIG_15c} for the same selected values of the string parameters as in Fig. \ref{stringFIG_15b}) determine whether the oscillating strings are trapped in the black-hole field or escape to infinity. The string loop can escape to infinity if its energy is bigger then the ``rest'' energy in infinity, i.e., for  $\EE > \EE_{\rm min(flat)}$, given by (\ref{EnFlat}). The escape is thus determined by the limiting energy in the flat spacetime. (The captured and trapped states of the oscillating strings are shaded in the Fig. \ref{stringFIG_15a}).

It is quite interesting to determine the conditions for existence and extension of regions of trapped states of the oscillating string loops in dependence on the motion constants $\JJ$ and $\EE$. Such states, in which the string loop may not escape to infinity neither may not be captured by the black hole, correspond to ``lakes'' determined for appropriately chosen energy levels by the energy boundary function $\EE_{\rm b}(x,y;\JJ)$. To do so, restrictions from the $x$-profiles and $y$-profiles of the boundary energy function has to be properly combined, in dependence on the string (angular momentum) parameter $\JJ$.

First, the restrictions for the trapped oscillations in the $x$-direction has to be represented by an $\it{projected}$ angular momentum function $\JJ_{\rm p}(\xx,\yy)$ determined by the projection of the extremal (maximal) energy level $\EE_{\rm b( max)}(\xx_{\rm max},\yy,\JJ)$ onto the energy boundary function itself. It is numerically constructed in a simple way: we choose for a fixed $\yy$ a value of $\JJ_{\rm p}$, find the corresponding value of the energy boundary function maximum $\EE_{b(\rm max)}(\xx_{\rm max}, \yy, \JJ_{\rm p})$ and the related coordinate $\xx_{\rm p}$ where $\EE_{\rm b(max)}(\xx_{\rm max}, \yy, \JJ_{\rm p}) = \EE_{\rm b}(\xx_{\rm p}, \yy, \JJ_{\rm p})$.

Second, we can show by solving equation
\beq
       \EE_{\rm b}(\xx, \yy, \JJ) = \EE_{\rm min(flat)}(\JJ) = 2 \JJ  \label{eq_Econ}
\eeq
that the trapped states of the oscillating string loops could exist if 
\beq
       \JJ_{\rm L1}(\xx, \yy) < \JJ < \JJ_{\rm L2}(\xx, \yy)
\eeq
where the ``lake'' angular momentum functions $\JJ_{\rm L1}(\xx, \yy)$ and $\JJ_{\rm L2}(\xx, \yy)$ are solutions of (\ref{eq_Econ}). In the equatorial plane ($\yy=0$), the trapped regions are most extended and the ``lake'' angular momentum functions take the simple form 
\beq
 \JJ_{\rm L1} = \frac{\xx(\sqrt{\xx}+\sqrt{2})}{\sqrt{\xx-2}}, \quad 
 \JJ_{\rm L2} = \frac{\xx(\sqrt{\xx}-\sqrt{2})}{\sqrt{\xx-2}}.
\eeq
The functions $\JJ_{\rm L1}(\xx)$ and $\JJ_{\rm L1}(\xx)$ are illustrated in Fig. \ref{stringFIG_15Ba}. 

The minimum of the function $\JJ_{\rm L1}(\xx)$ coincides with its intersection with the function $\JJ_{\rm E}(\xx)$. It is located at the radius
\beq
       \xx_{\rm mt} = \xx_{\rm L1(min)} = \frac{9 + \sqrt{17}}{4} \sim 3.3
\eeq
that determines the so called marginally trapped radius giving the smallest value of the $x$-coordinate allowed for the oscillating string loops. The related value of the string parameter $\JJ_{\rm L1(min)}$ is given by
\beq
        \JJ_{\rm L1(min)}^2 = \frac{349 + 85\sqrt{17}}{8} . 
\eeq
Therefore, the region of trapped motion is restricted to radii $\xx > \xx_{\rm mt} \sim 3.3$; the critical value of the string parameter reads $\JJ_{\rm L1(min)} = \JJ_{\rm mt} \sim 9.35$.

\begin{figure}
\begin{minipage}{4.2cm}
\vspace{-0.2cm}
\subfigure[ ~Case 1 on Fig. \ref{stringFIG_15} \label{stringFIG_04a}]{\includegraphics[height=3.4cm]{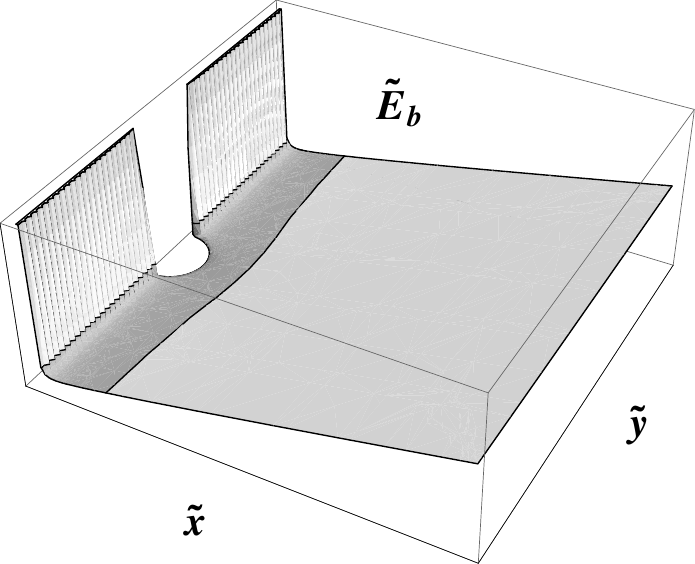}}
\vspace{0.1cm}
\subfigure[ ~Case 2 on Fig. \ref{stringFIG_15} \label{stringFIG_04b}]{\includegraphics[height=3.4cm]{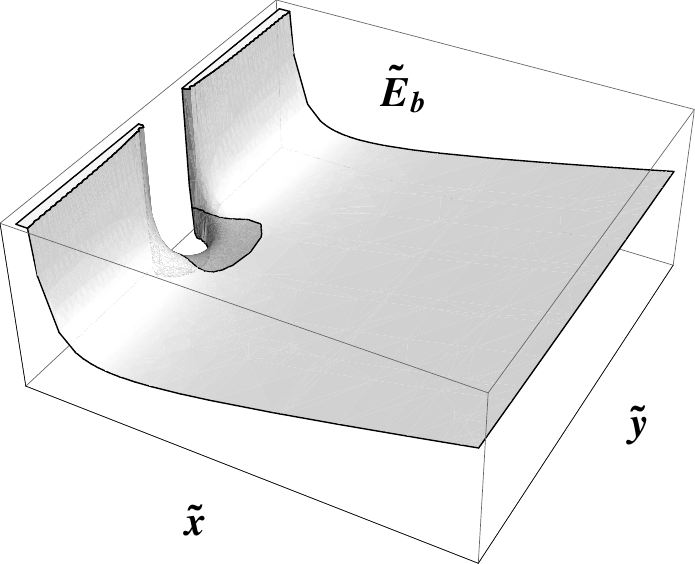}}
\vspace{0.1cm}
\subfigure[ ~Case 3 on Fig. \ref{stringFIG_15} \label{stringFIG_04c}]{\includegraphics[height=3.4cm]{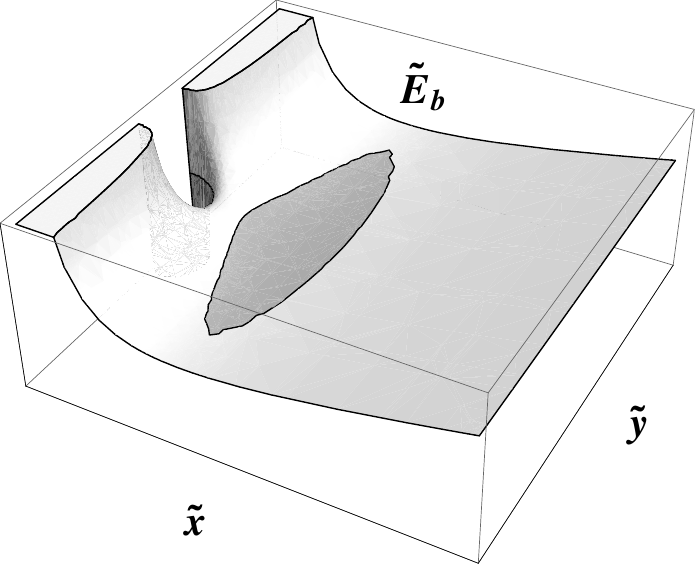}}
\vspace{0.1cm}
\subfigure[ ~Case 4 on Fig. \ref{stringFIG_15} \label{stringFIG_04d}]{\includegraphics[height=3.4cm]{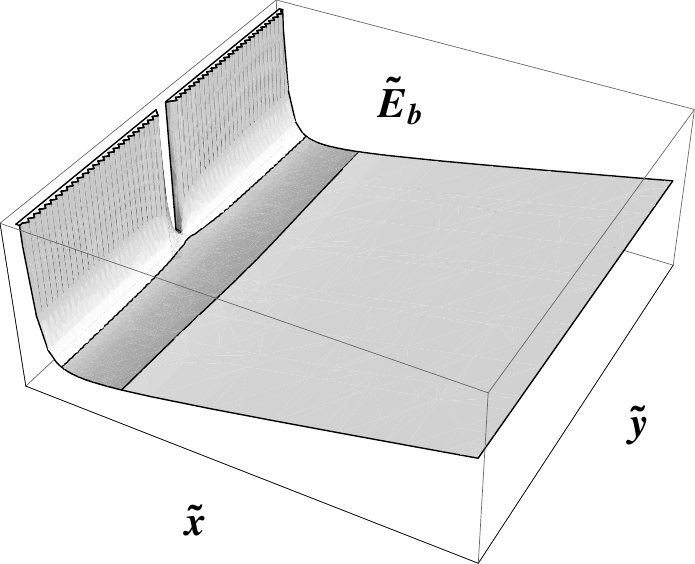}}
\end{minipage}
\begin{minipage}{4cm}
\includegraphics[width=4cm]{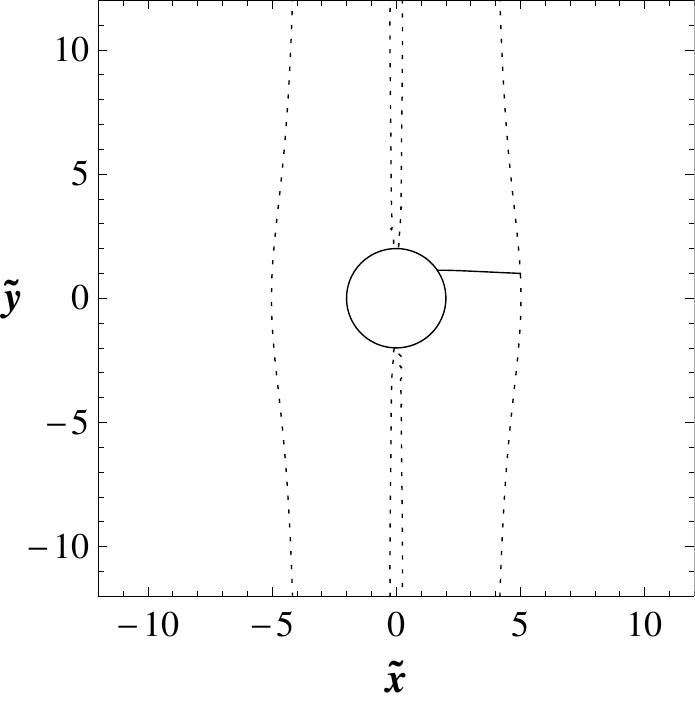}
\vspace{0.2cm}
\includegraphics[width=4cm]{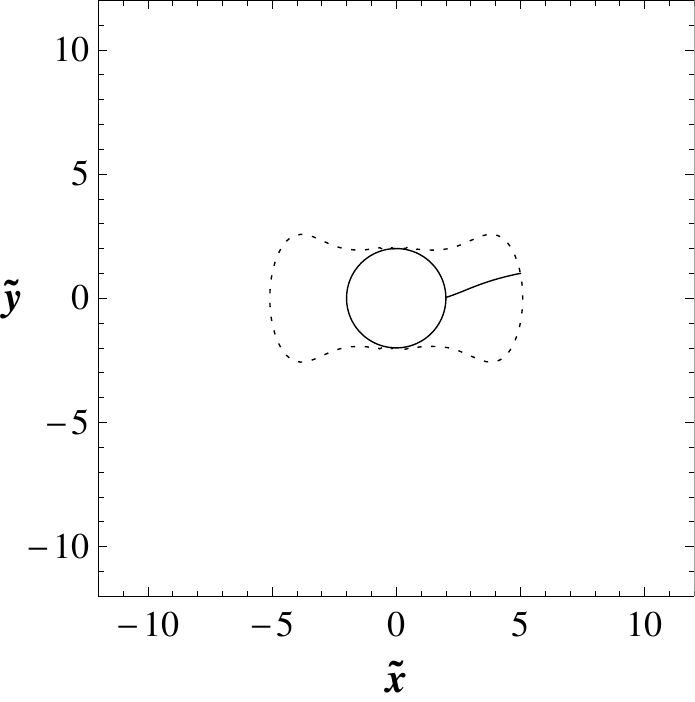}
\vspace{0.2cm}
\includegraphics[width=4cm]{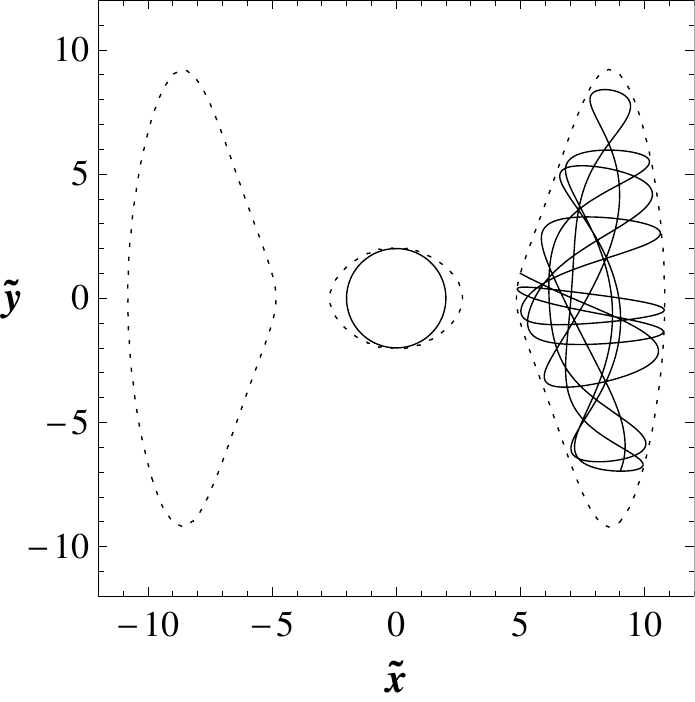}
\vspace{0.2cm}
\includegraphics[width=4cm]{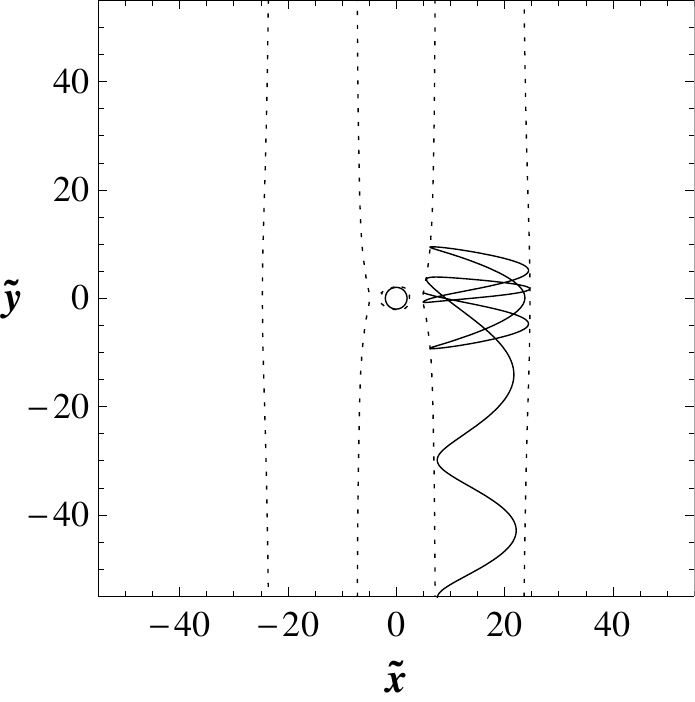}
\end{minipage}
\caption{String-loop motion in the {\bf  \Schw{}{} spacetime}. Four types of the motion as given in Fig. \ref{stringFIG_15} are presented. The boundary energy function is given for the selected values of the string parameter $\JJ$ and the trajectories are constructed for the related energies calculated under the assumption of the rest initial state of the string. \label{stringFIG_04}}
\end{figure}

The results relevant for the trapped states are summarized in Fig.\ref{stringFIG_15B}. The regions of trapped oscillations are most extended at the equatorial plane ($\yy=0$) and shrink with $|\yy|$ increasing. For $\JJ_{\rm E(min)}(\xx, \yy) < \JJ < \JJ_{\rm L1}(\xx, \yy)$ the trapped region is given by the projected angular momentum function $\JJ_{\rm p}(\xx, \yy)$, for $\JJ > \JJ_{\rm L1(min)}$ it is given by $\JJ_{\rm L1}(\xx, \yy)$ and $\JJ_{\rm L2}(\xx, \yy)$. For $|\yy| \lightarrow \infty$ the region reduces to $\xx = \JJ$. The most extended ``lake'' of trapped oscillations exists for $\JJ = \JJ_{\rm mt}$ that extends up to $|\yy| \lightarrow \infty$ and down to $\xx = \xx_{\rm mt}$ (for $\yy=0$).

In the \Schw{} spacetimes, we can distinguish four different types of the behavior of the boundary energy function and the character of the string loop motion; in Fig. \ref{stringFIG_15} we denote them by points numbered by 1 to 4. 
The first case $J~<~J_{\rm L2}$ corresponds to no inner and outer boundary and the string can be captured by the black hole or escape to infinity, in the second case, $\JJ_{\rm L2}~<~\JJ~<~\JJ_{\rm E}$, there is an outer boundary and the string loop cannot escape to infinity, and it must be captured by the black hole. The third case, $\JJ_{\rm E}~<~J~<~\JJ_{\rm L1}$, corresponds to the situation when both inner and outer boundary exist and the string is trapped in some region forming a potential ``lake'' around the black hole. In the fourth case $\JJ_{\rm L1}~<~\JJ$ the  string cannot fall into the black hole but it can escape to infinity (or be trapped). 
The situation is illustrated in Fig. \ref{stringFIG_04}. The states of the oscillating string loop are allowed for the intervals of coordinates $\xx$, $\yy$ (and string parameter $\JJ$) in the shaded region given by a corresponding energy level. 

The trajectories of the string loop are obtained by integrating the equation for the string motion in the spherical coordinates (\ref{MovingEq01}) that can be written for the \Schw{} spacetime in the form
\bea
 \ddot{\rr} &=& {\dot{\theta}}^2 (\rr-3) -\dot{\rr} \frac{\partial_{\tau}\Sigma^{\tau\tau}{}}{\Sigma^{\tau \tau}} \nonumber\\
 && + \sin^2\theta \left( \frac{\Sigma^{\sigma\sigma}}{\Sigma^{\tau \tau}} (\rr-2) - 1  \right),
\eea
while the equation (\ref{MovingEq02}) remains unchanged. Now, the amplitude of the string-loop motion is changed during the motion off the equatorial plane due to the gravitational effects of the mass located at the center of symmetry - in the \Schw{} (and SdS) spacetime there is the only center of symmetry, contrary to the cases of flat and dS spacetimes.

The string path for all four types of motion can be found in Fig. \ref{stringFIG_04}. The string loop is starting from the rest $\dot{\xx}=0, \dot{\yy}=0$, and the starting point is chosen at $\xx_0 = 5, \yy_0 = 1$, i.e., above the equatorial plane. The string parameter $\JJ$ is chosen, while the energy $\EE$ is calculated from the equation $\EE=\EE_{\rm b}(\JJ)$. For energy $\EE_1 \doteq 4, \EE_2 \doteq 8$ (cases 1,2), the string falls into black hole, Fig. \ref{stringFIG_04a}, Fig. \ref{stringFIG_04b}, for energy $\EE_3 \doteq 17$, the string is trapped in some region above the black hole horizon, Fig. \ref{stringFIG_04c}, and for energy $\EE_4 \doteq 23$, the string escapes to infinity, Fig. \ref{stringFIG_04d}. 

\begin{figure}%
\includegraphics[width=4cm]{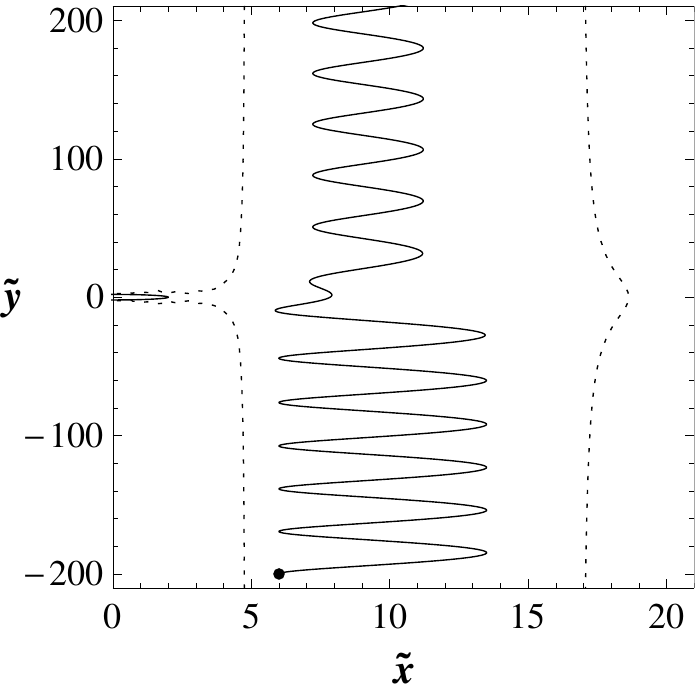}\includegraphics[width=4cm]{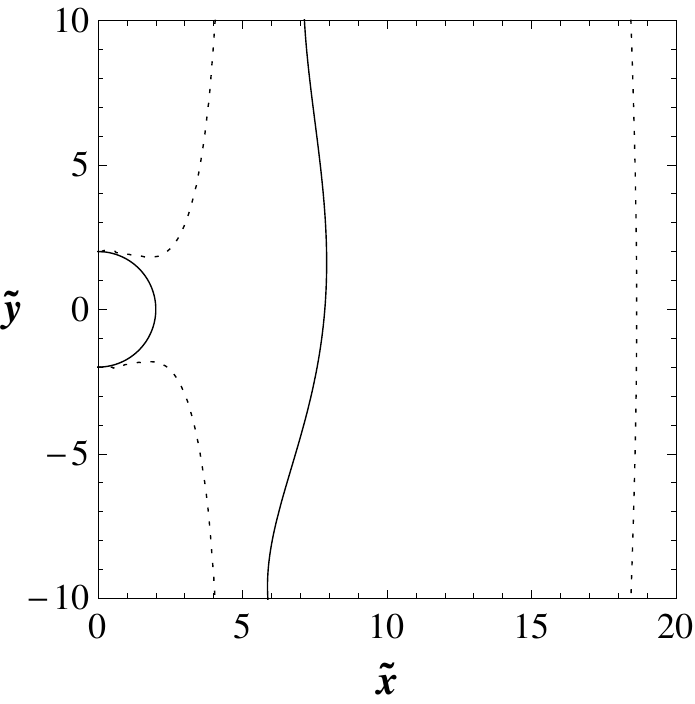}
\includegraphics[width=4cm]{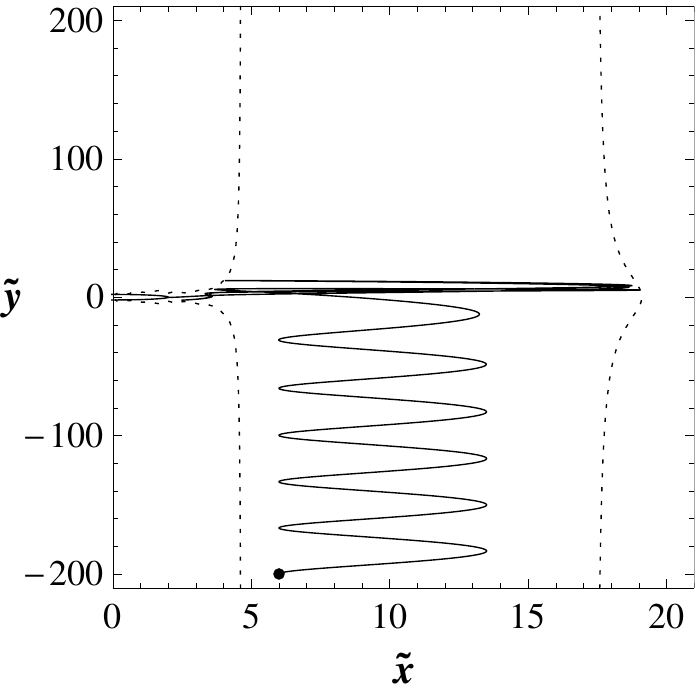}\includegraphics[width=4cm]{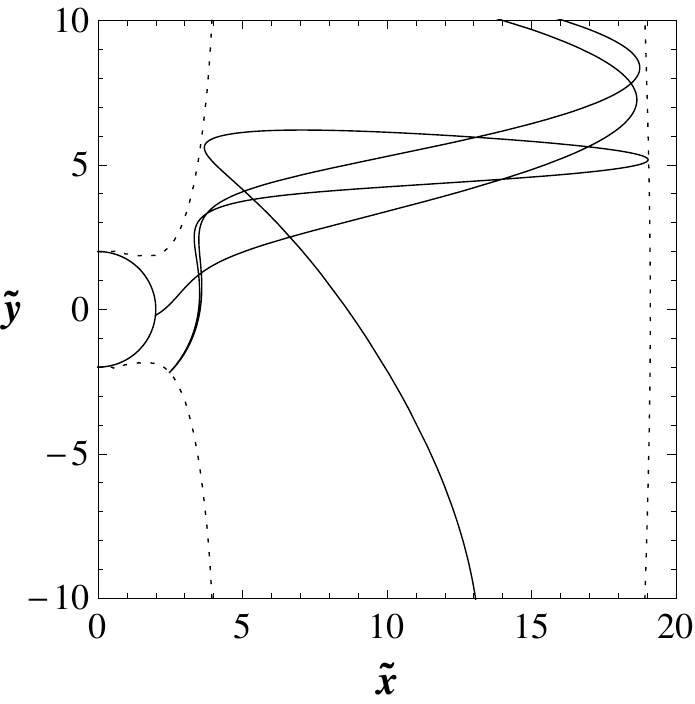}
\includegraphics[width=4cm]{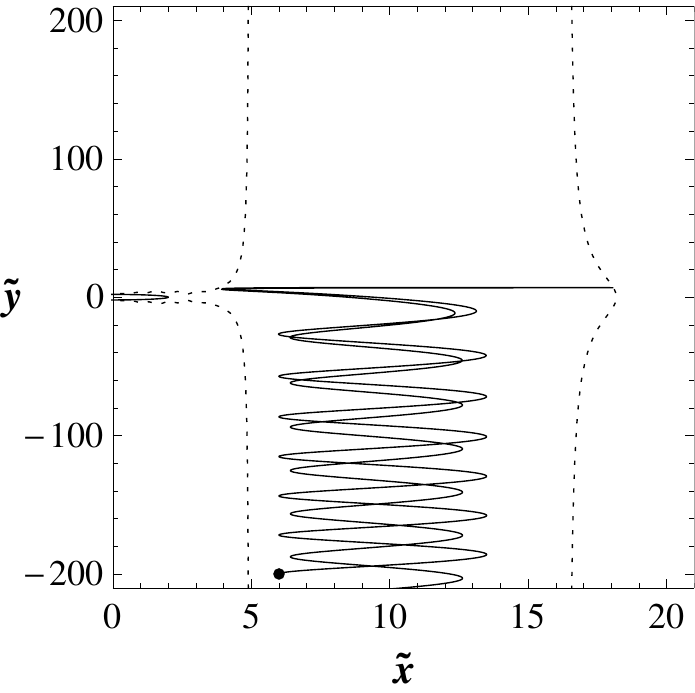}\includegraphics[width=4cm]{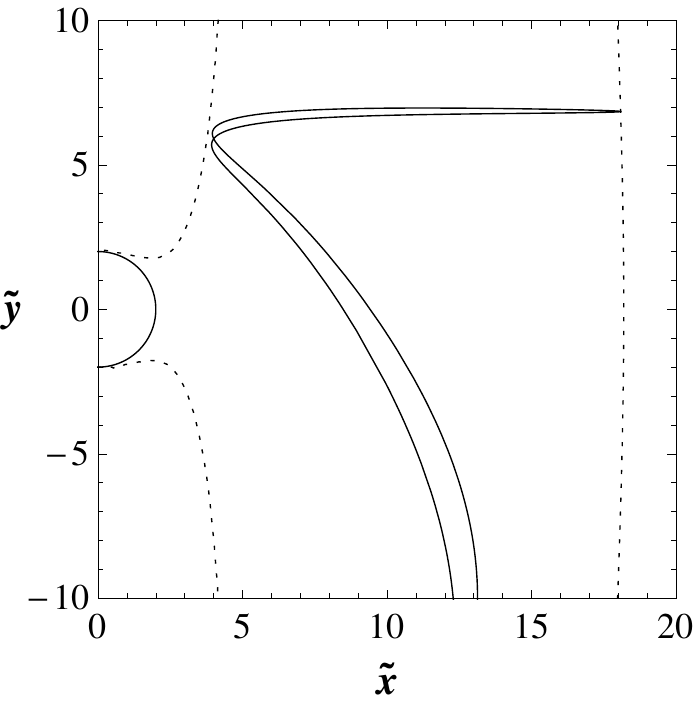}
\vspace{-0.2cm}
\caption{Scattering, capturing and backscattering of a string loop in the \Schw{} spacetime. The string is starting from the point with coordinates $\tilde{x}_0 = 6, \tilde{y}_0 = -200$, having the same string (angular momentum) parameter $\JJ = 9$ and various momentum in the $\yy$-direction and related energy: $\dot{\tilde{y}} = 3$ and $\EE \doteq 21.7$ for scattering , $\dot{\tilde{y}} = 3.25$ and $\EE \doteq 22.1$ for capturing, $\dot{\tilde{y}} = 2.75$ and $\EE \doteq 21.4$ for backscattering.
After the scatter, the amplitude of the string loop oscillations in the $x$-direction is reduced; there is conversion of oscillatory energy to the kinetic energy in the $y$-direction. 
This effect was described as ``string transmutation'' in \citep{Lar:1994:CLAQG:,Jac-Sot:2009:PHYSR4:}. 
\label{stringFIG_22}
}
\end{figure}

In the cases 1 and 4, when the energy boundary function reaches infinity, the oscillating string loop approaching the black hole from large distances (infinity) can be scattered, rescattered, or captured by the black hole field, as demonstrated in Fig. \ref{stringFIG_22}.

Motion of current-carrying string loop in \Schw{} spacetime has been also studied in \citep{Lar:1994:CLAQG:} using a different approach of \citep{Lar:1993:CLAQG:}. The results obtained in those work agree with those presented here.

%
%
%
%

\subsection{\Schw\nnd\dS{} spacetime and the effect of the cosmological constant}

%
%
\begin{figure*}[p]
\subfigure[~ Energy $E_{\rm b}(\JJ)$ ]
{\label{stringFIG_17c} \includegraphics[width=5.3cm]{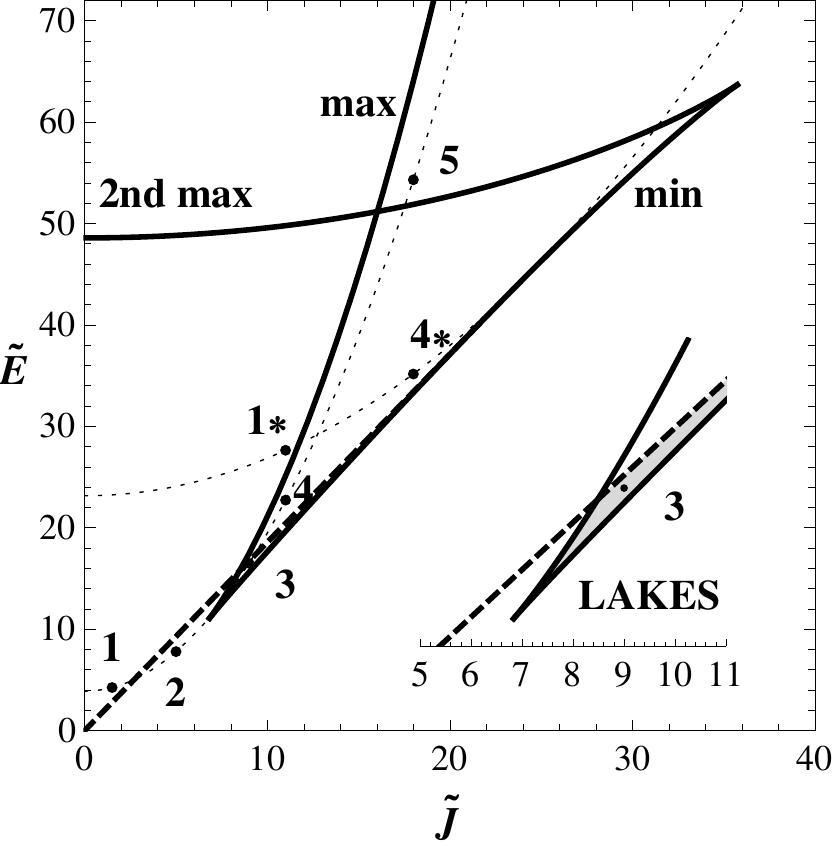}}
\subfigure[~ Energy $E_{\rm b}(\xx)$ for different $\JJ$.]
{\label{stringFIG_17a} \includegraphics[width=5.3cm]{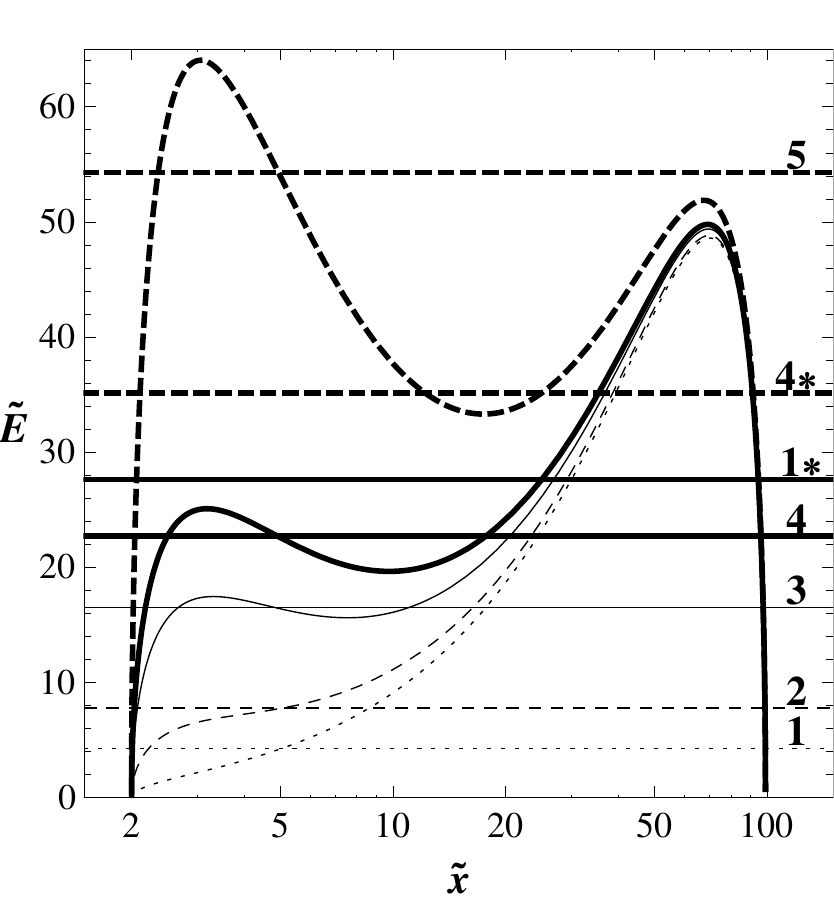}}
\subfigure[~ Energy $E_{\rm b}(\tilde{y})$ for different $\JJ$]
{\label{stringFIG_17b} \includegraphics[width=5.3cm]{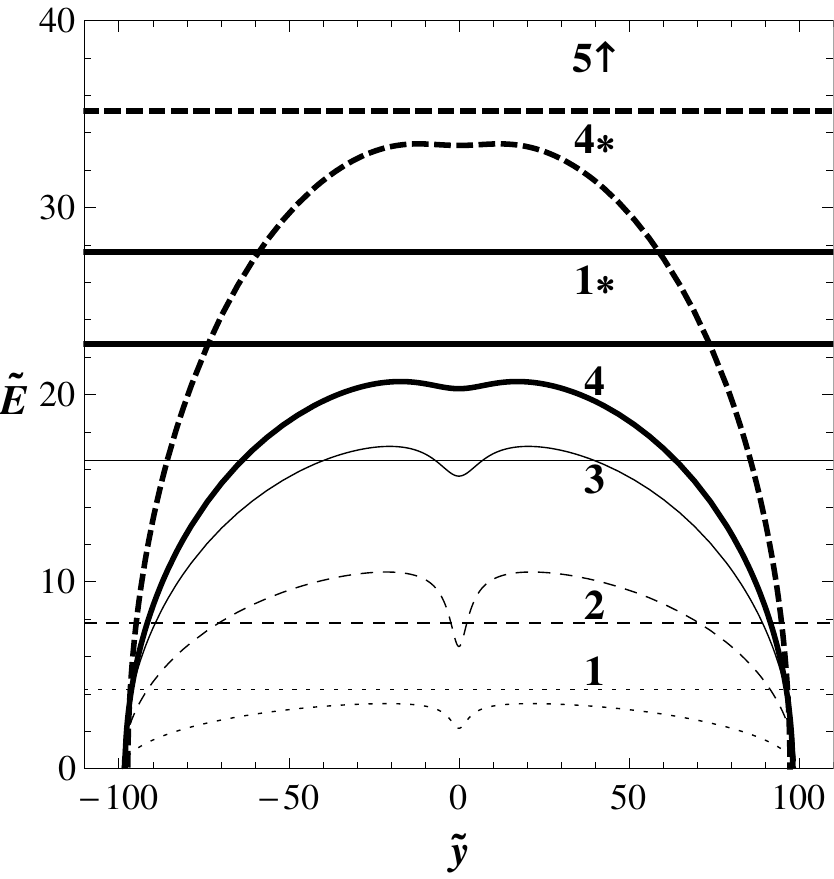}}
\vspace{-0.35cm}
\caption{ \label{stringFIG_17A}
Boundary energy function $\EE_{\rm b}(\xx,\yy,\JJ,\cosm)$ of the SdS spacetime with the cosmological parameter $\cosm=10^{-4}$. (a) Thick solid curves correspond to the maxima, minima and second maxima of $\EE_{\rm b}$ in the $\xx$-direction. Maximal energy for trapped states, $\EE = 2 \JJ \sqrt{A(\rr_{\rm s})}$ represents the dashed curve, while the thin dotted lines represent the energy profiles $\EE_{\rm b}(\xx_0,\yy_0,\JJ)$ taken at the spacetime points with coordinates $\xx_0 = 5, \yy_0 = 1$ (for the states represented by points 1-5) and $\xx_0 = 25, \yy_0 = 1$ (for $1_{*},4_{*}$). The states are chosen with representative values of the string 
 parameter: $\JJ_1 = 1.5$, $\JJ_2 = 5$, $\JJ_3 = 9$, $\JJ_4 = 11$, and $\JJ_5 = 18$. (There is $\JJ_{1_*} = \JJ_4$ and $\JJ_{4_*} = \JJ_5$.) Region where ``lakes'' corresponding to the trapped states can exist is shaded. 
The figures (b) and (c) show $\xx-$ and $\yy-$ profiles of the boundary energy function; we use $\EE_{\rm b}(\xx,0,\JJ)$ (b) and $\EE_{\rm b}(\xx_{\rm min},\yy,\JJ)$ (c). (For cases 1,2 there is no minimum in the $\xx$-direction and we use $\EE_{\rm b}(3,\yy,\JJ)$ instead.) The profiles are presented for the chosen representative values of string parameter:  $\JJ_1 = 1.5$ (dotted curves), $\JJ_2 = 5 $ (dashed curves), $\JJ_3 = 9 $ (thin curves), $J_4 = 11 $ (thick curves) and $\JJ_5 = 18 $ (thick dotted curves).
We assume a string loop starting from the rest 
at the point $\xx_0 = 5, \yy_0 = 1$ for cases 1-5 and at the point $\xx_0 = 25, \yy_0 = 1$ for the cases $1_*,4_*$ (trajectories of the type 1 and 4). The energies calculated for the string motion are given as follows: $\EE_{\rm 1} \doteq 4, \EE_{\rm 2} \doteq 8, \EE_{\rm 3} \doteq 16, \EE_{\rm 4} \doteq 23, \EE_{\rm 1*} \doteq 28, \EE_{\rm 4*} \doteq 35 $ and $\EE_{\rm 5} \doteq 54 $.
Trajectories of the type 1-4 correspond to the cases 1-4 in the \Schw{} spacetime (see Figs. \ref{stringFIG_15}, \ref{stringFIG_04}), while trajectories of the type 5 correspond to the type-2 trajectories in the de Sitter spacetime (see Figs. \ref{stringFIG_13}, \ref{stringFIG_01b}).
}

\subfigure[~ $ \yy = 0$]
{\label{stringFIG_17Ba} \includegraphics[width=5.3cm]{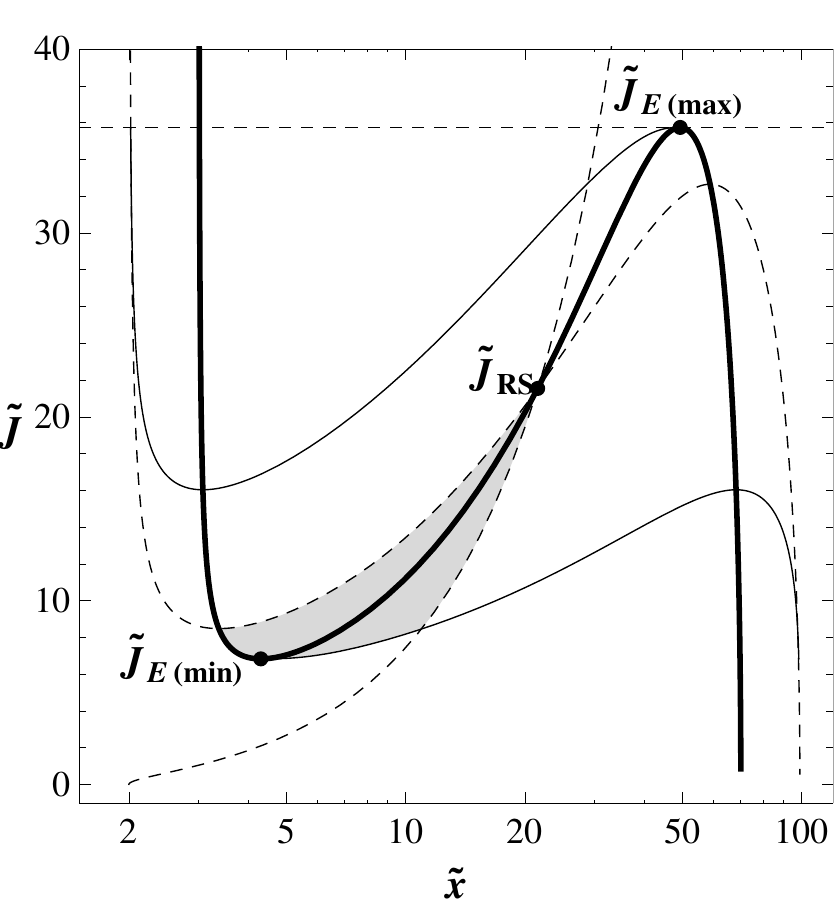}}
\subfigure[~ $ \yy = 1$]
{\label{stringFIG_17Bb}\includegraphics[width=5.3cm]{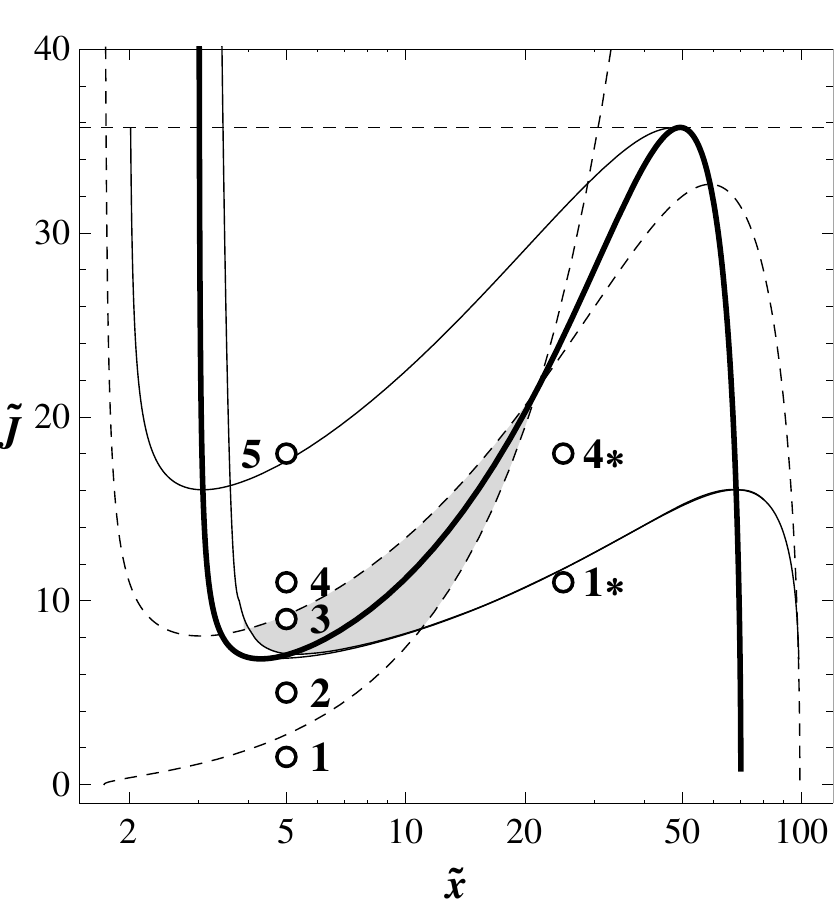}}
\subfigure[~ $ \yy = \yy_{\rm max}$]
{\label{stringFIG_17Bc} \includegraphics[width=5.3cm]{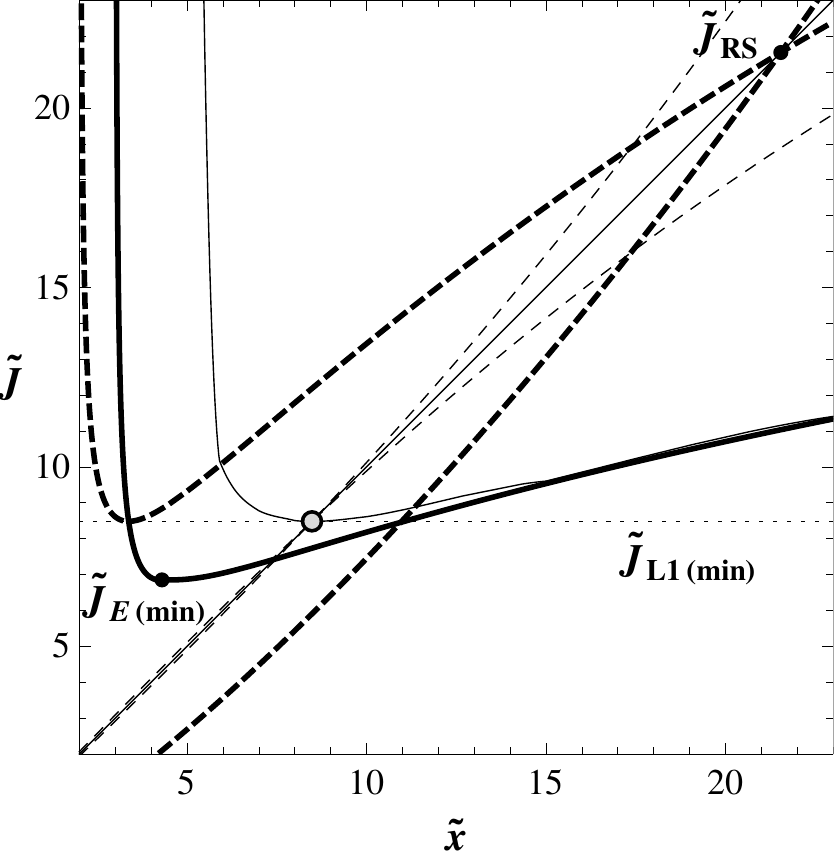}}
\vspace{-0.35cm}
\caption{\label{stringFIG_17B}
String parameter functions $\JJ(\xx,\cosm)$ determining character of the boundary energy function $\EE_{\rm b}(\xx,\yy,\JJ,\cosm)$. The functions are given for three representative values of $\yy$ in the SdS spacetime with $\cosm = 10^{-4}$. (In the case of $\yy=1$, we present position of all the states chosen to describe the character of the boundary energy function.) The function $\JJ_{\rm E}(\xx)$ (thick solid curve) determining extrema of $E_{\rm b}$ is independent of $\yy$. Dashed curves represent the functions $\JJ_{\rm L1}(\xx,\yy), \JJ_{\rm L2}(\xx,\yy)$. These functions have a common point at $\xx=\JJ=\JJ_{\rm RS}$, but this intersection point exists only above $\JJ_{\rm L1(min)}$, i.e., for $\yy^2 < \yy_{\rm max}^2 = \rr_{\rm s}^2 - x^2_{\rm min}$, where $ x_{\rm min} = \JJ_{\rm L1(min)}$. The value of $\yy_{\rm max}$ gives maximal extension of the ``lake'' corresponding to the trapped states in the $\yy$-direction - for $\cosm=10^{-4}$ it takes the value $\yy_{\rm max} \doteq 19.8$.
The thin solid curves represent numerically calculated ``projected'' angular momentum functions $\JJ_{\rm p}(\xx,\yy)$ that are determined by the projection of the extremal (maximal) energy level $\EE_{\rm b( max)}(\xx_{\rm max},\yy,\JJ)$ onto the energy boundary function itself. (Note that in the SdS spacetimes there are two branches, and the upper branch of these functions is irrelevant for the trapped states.) 
In the case of $y=\yy_{\rm max}$, there is an extra thin line $\xx = \JJ$ where the common point of $\JJ_{\rm L1}, \JJ_{\rm L2}$ occurs. Regions where the ``lakes'' can exist are shaded. We see that the ``lake'' area is shrinking with increasing $\yy$, and for $y=y_{\rm max}$ it is reduced to the one point corresponding to $\xx = \JJ = \JJ_{\rm L1(min)}$. 
}
\end{figure*}

The most general spacetime we consider here is the SdS one. The characteristic function of the line element (\ref{SfSymMetrika}) then takes the form
\beq
		A(r) = 1-\frac{2M}{r} - \frac{1}{3} \Lambda r^2 = 1 - \frac{2}{\rr} - \cosm \rr^2.
\eeq
In this case two characteristic length scales given by the mass parameter $M$ and the cosmological constant $\Lambda$ are introduced. It is therefore convenient to use the dimensionless coordinate $\rr = r/M$ ($\xx = x/M, \yy = y/M$) and a dimensionless cosmological parameter 
\beq 
		\cosm = \frac{1}{3} \Lambda M^2 .
\eeq 
In order to clearly demonstrate the role of the cosmological constant, we will use in our figures $\cosm= 10^{-4}$ and $\cosm= 10^{-6}$. Of course, these values are very large in comparison with values expected for realistic supermassive black holes in active galactic nuclei, when even for the most extreme case of quasar $TON 618$ with mass of the central black hole estimated to be $M \sim 6.6 \times 10^{10} M_{\odot}$, there is $\cosm \sim 10^{-24}$ (\cite{Ziol:2008:CJA:,Stu-Sla-Kov:2009:CLAQG:}). However, for astrophysically realistic values of $\cosm$, the string-loop motion is of the same character as presented in our discussion, only the scales of the static radius (and the cosmological horizon) are shifted to values much larger than those considered here. Different situations (with much larger values of $\cosm$) are possible in the early universe \cite{Stu-Sla-Hle:2000:ASTRA:}. 

The horizons of the SdS spacetime are again given by $A(r) = 0$. For $\cosm < 1/27$, there are the cosmological $\rr_{\rm c}$ and black hole $\rr_{\rm h}$ horizons that are given by the relations \cite{Stu-Hle:1999:PHYSR4:,Stu-Sla-Hle:2000:ASTRA:}
\beq
 \rr_{\rm h} = \frac{2}{\sqrt{3 \cosm}} \cos\frac{\pi+\xi}{3}, \quad \rr_{\rm c} = \frac{2}{\sqrt{3 \cosm}} \cos\frac{\pi-\xi}{3}
\eeq
where
\beq
  \xi = \cos^{-1} \left( 3 \sqrt{3\cosm} \right).
\eeq
For $\cosm = 1/27$ the horizons coalesce at $\rr = 3$, while for $\cosm > 1/27$ the SdS spacetime describes a naked singularity.
The photon circular orbit is located at $\rr_{\rm ph} = 3$, independently of the cosmological parameter (see \cite{Stu:1990:BULAI:}). 
The crucial role plays the static radius
\beq
     \rr_{\rm s} = \cosm^{-1/3}
\eeq
where the gravitational attraction of the black hole acting on a test particle is just balanced by the cosmic repulsion \cite{Stu-Hle:1999:PHYSR4:}. Thus, no circular orbits of test particles are possible at $\rr > \rr_{\rm s}$ and $\rr < \rr_{\rm ph}$.
The ISCO location is implicitly determined by the condition
\beq
						\cosm = \cosm_{\rm ms} \equiv \frac{\rr-6}{\rr^3(4\rr-15)}
\eeq
The stable orbits exist at spacetimes with 
\beq
            \cosm < \cosm_{\rm ms} = \frac{12}{15^4} \sim 0.000237
\eeq 
when two ISCO orbits exist - the inner and outer ones \cite{Stu:1983:BULAI:}.

In the SdS spacetime, the asymptotic behavior of the boundary energy function is determined by the presence of the black-hole horizon and the cosmological horizon - there is $E_{\rm b} \lightarrow 0$ for both $r \lightarrow r_{\rm h}, r_{\rm c}$. Typical behavior of the boundary energy function (its $x$- and $y$-profiles) is represented in Fig. \ref{stringFIG_17A}.

Introducing the dimensionless string (angular momentum) parameter $\JJ = J/M$ and energy $\EE = E/M$, the extrema equation (\ref{extr_a1}) leads to the quintic equation in the $\xx$ coordinate
\beq
 - \frac{2}{3} \Lambda \xx^5 + \xx^3  -   \xx^2 - \JJ^2 \xx + 3  \JJ^2  = 0 , \label{eq_blabla}
\eeq
while (\ref{extr_a2}) gives again $\yy=0$, but also an additional condition - $A_{\rr}'=0$ determining a maximum of the energy boundary function in the $y$-direction that is located at $\rr_{\rm max}=\rr_{\rm s}$. Clearly, the static radius plays a fundamental role in the motion of string loops, similarly to the case of the motion of test particles. 

\begin{figure}
\includegraphics[width=4cm]{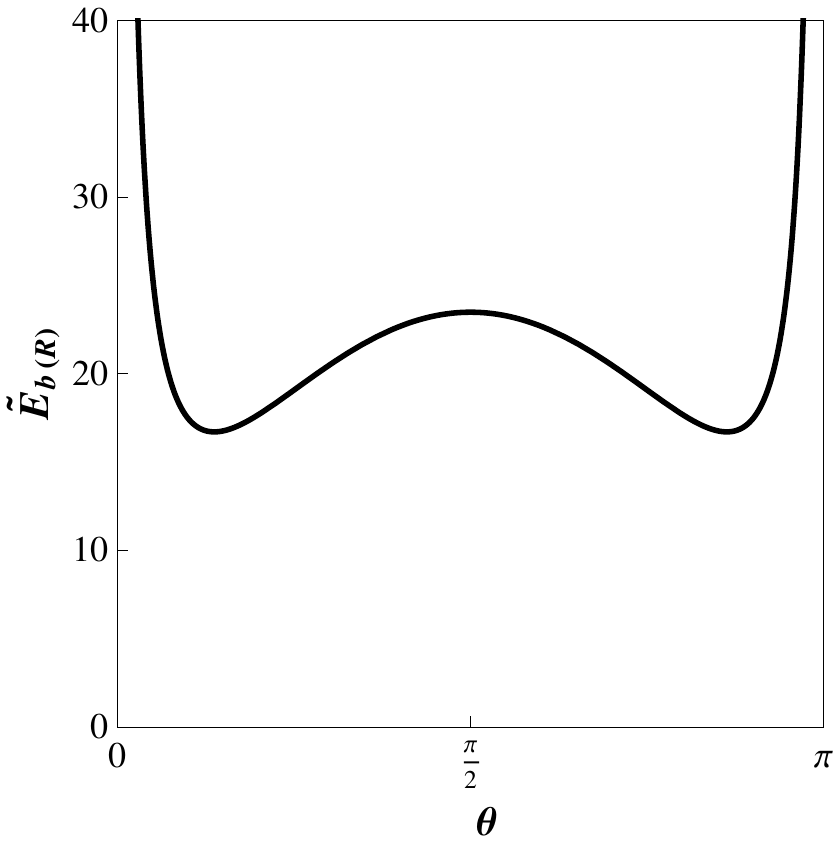}
\includegraphics[width=4cm]{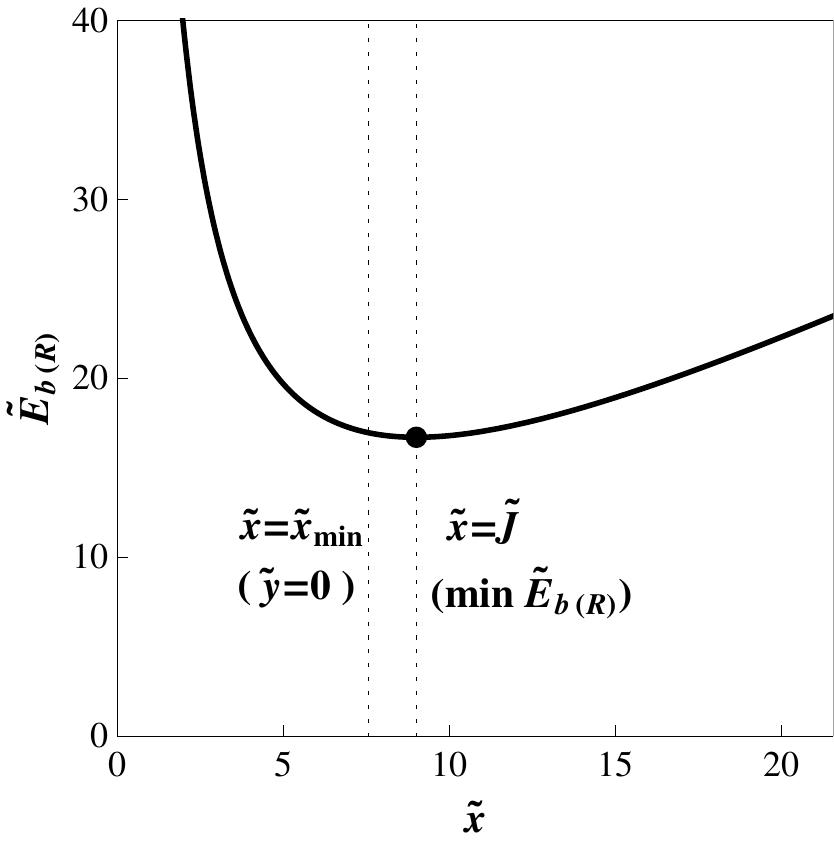}
\caption{Boundary energy $\EE_{\rm b(R)}$ at the top of the ``ridge'' at the static radius as a function of coordinates $\theta$ and $\xx$. The cosmological parameter is chosen to be $\yy = 10^{-4}$ and the string parameter $\JJ = 9$. We can see that the minimum of $\EE_{\rm b(R)}$ located at $\xx = \JJ$ is above $\xx_{\rm min}$ giving position of the minimum of the potential valey for trapped string loop oscillations; cf. (\ref{extr_c}).}
\vspace{0.2cm}
\includegraphics[height=4.1cm]{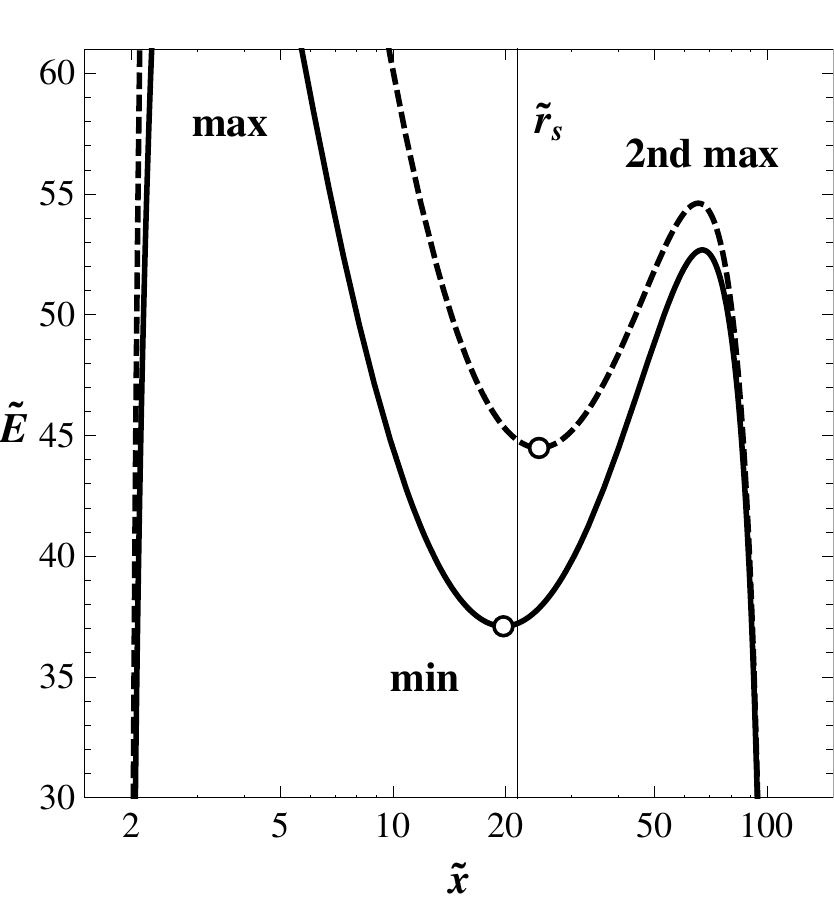}
\includegraphics[height=4cm]{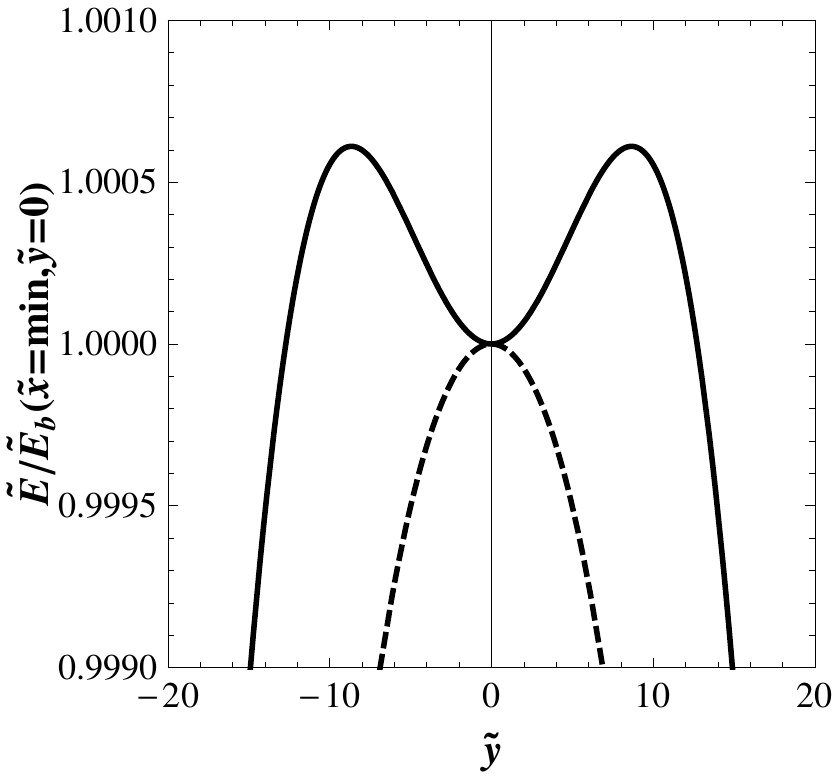}
\caption{\label{stringFIG_09}
Profiles of the boundary energy function of the string motion in the SdS spacetime with $\cosm = 10^{-4}$. The $\xx$-profiles (left) and the related $\yy$-profiles (right) demonstrate the differences in vicinity of the ``ridge'' at the static radius.  We show the case 1 when the ``lake'' of the trapped states can exist for $\JJ_1 = 20 < \rr_{\rm s}$ (solid lines) and the case 2 when the ``lake'' cannot exist for $\JJ_2 = 24 > \rr_{\rm s}$ (dashed lines) - there is minimum of the boundary energy function at $\xx=\xx_{\rm 1(min)}, y=0$ while saddle point at $\xx=\xx_{\rm 2(min)}, y=0$.
}
\end{figure}

First, we examine restrictions for motion in the $x$-direction. Equation (\ref{eq_blabla}) can expressed in the form
\beq
    \JJ^2 = \JJ_{\rm E}^2 \equiv \frac{\xx^2(\xx-1-2\cosm \xx^3)}{\xx-3}.
\eeq
The function $\JJ_{\rm E}$ is illustrated in Fig. \ref{stringFIG_17B} (we assume $\JJ>0$ again).  

The zero points of the function $\JJ^2_{\rm E}(\xx,\cosm)$ are given by the condition
\beq
    \cosm = \cosm_{\rm z} \equiv \frac{\xx-1}{2\xx^3} \label{zeropoints}
\eeq
while its divergent points are located at the radius of the photon circular orbit $\xx_{\rm ph} = 3$, independently of the value of the cosmological constant.

\begin{figure}%
\includegraphics[width=6.5cm]{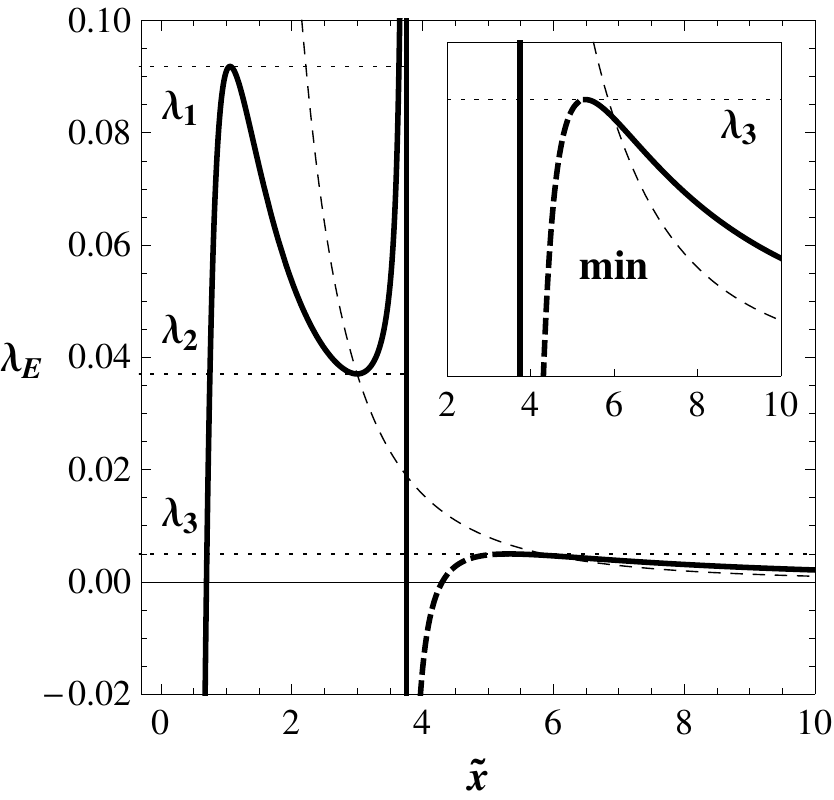}
\caption{\label{stringFIG_16} 
Existence of boundary-energy-function minima giving states of trapped oscillations. Critical dimensionless cosmological parameter function $\cosm_{\rm E}(\xx)$ (thick line) is compared to $\cosm_{\rm s} = \xx^{-3}$ (thin-dashed line). The part of $\cosm_{\rm E}(\xx)$ where the minima occur is dashed.}
\vspace{0.3cm}
\subfigure[ ~$\cosm=0.04$]{\includegraphics[width=4cm]{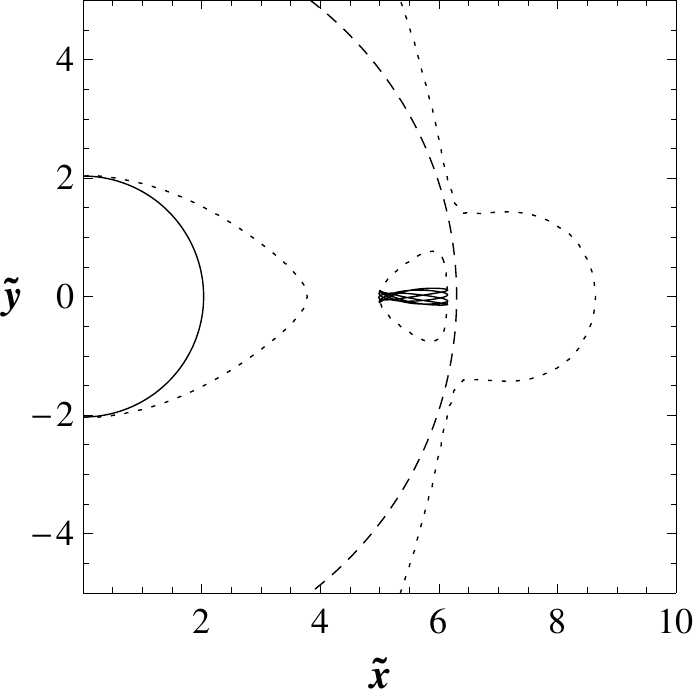}}
\subfigure[ ~$\cosm=0.05$]{\includegraphics[width=4cm]{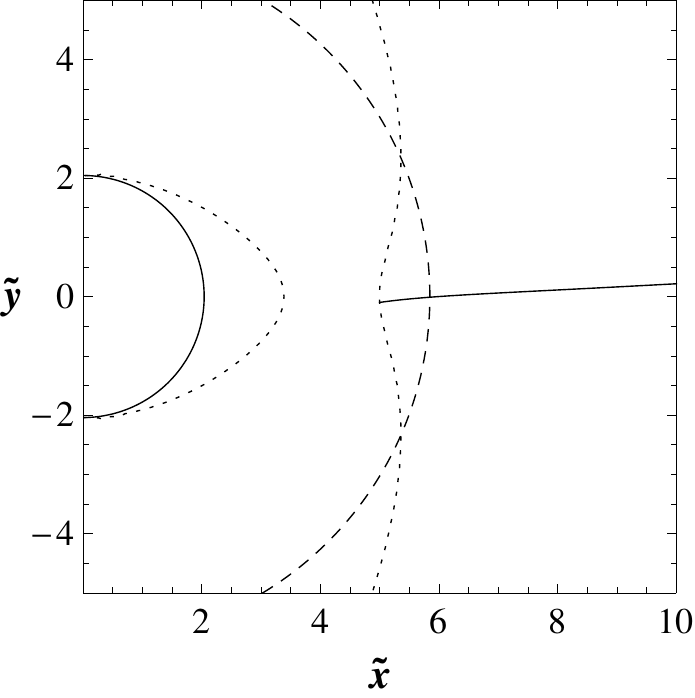}}
\vspace{-0.3cm}
\caption{\label{FIG_fik}
No trapped oscillations can exist (no ``lakes'') for $\cosm > \cosm_{\rm trap} \sim 0.00497$. Boundaries of the motion and trajectories are given for values of $\cosm$ close to the limiting values. Dotted curves represent the boundaries for the string motion, string trajectories are represented by the solid thin curves. Black-hole horizont, static radius and cosmological horizont are represented by the full, dashed and dotted circles.}
\end{figure}

The local extrema of the function $\JJ_{\rm E}^2 (\xx,\cosm)$ are located at radii given by the condition
\beq
         \cosm = \cosm_{\rm E}(\xx) \equiv \frac{\xx^2 - 5\xx + 3}{\xx^3 (4\xx-15)} .
\eeq
The zero point of the function $\cosm_{\rm E}(\xx)$ is located at $\xx_{\rm E(z)}=\xx_{\rm min} \sim 4.303$, given by (\ref{zeropoints}) relevant in the \Schw{} spacetimes. Its divergence point is located at $\xx_{\rm E(d)} = 15/4$ - notice that it coincides with the divergence point of the function $\cosm_{\rm ms}(\xx)$ governing the ISCO orbits of the SdS spacetimes.  
There are three extrema points of the function $\cosm_{\rm E}$ taking values (see Fig. \ref{stringFIG_16})
\beq
       \cosm_{\rm E(e)1,3} = \frac{1633 \pm 129\sqrt{129}}{33750},\quad \cosm_{\rm E(e)2} = \frac{1}{27}. 
\eeq
The only physically relevant extremal point, located at $\xx>15/4$, is given by the third (lowest) solution. It is located at 
\beq
        \xx=\frac{3(17+\sqrt{129})}{16} \sim 5.3 , 
\eeq
see Fig \ref{stringFIG_16}. 

The physically relevant extremal points of the function $\JJ_{\rm E}^2(\xx)$ thus exist only for 
\beq
         \cosm < \cosm_{\rm trap} \equiv \frac{1633 - 129\sqrt{129}}{33750}  \sim 0.00497 . 
\eeq
For such values of $\cosm$, there are two maxima of the boundary energy function enabling the oscillations in the $x$-direction. The role of the cosmological parameter in the behavior of the boundary energy function is illustrated in Figs\ref{stringFIG_09} and \ref{stringFIG_16}.

Now, assuming $\cosm < \cosm_{\rm trap}$, we find that the function $\JJ_{\rm E}(\xx, \cosm)$ has a minimum $\JJ_{\rm E(min)}(\xx_{\rm min}, \cosm)$ and a maximum $\JJ_{\rm E(max)}(\xx_{\rm max}, \cosm)$ - see Fig. \ref{stringFIG_17B}. Oscillatory motion in the $x$-direction is then possible only for $\JJ_{\rm E(min)} < \JJ < \JJ_{\rm E(max)}$. In SdS spacetimes with $\cosm > \cosm_{\rm trap}$, the oscillations in the $x$-direction does mot appear and the string loop is directly captured by the black hole or escapes to infinity (see Fig. \ref{FIG_fik}. for an illustration).

Restrictions for the motion in the $\yy$-direction are given by the condition $A_{\rr}'=0$ - see (\ref{extr_a2}) - that has to be combined with the condition giving extremal points of the boundary energy function at $\yy = 0$ for all values of $\JJ$ and independently on $\xx$. For $0 < \cosm < 1/27$, there exists a maximum (``ridge'') of $\EE_{\rm b}(\xx,\yy)$ that is located at $\rr_{\rm s}$. At $\yy=0$, there is a minimum for $\JJ < \rr_s$ and a maximum for $\JJ > \rr_s$. (For $\cosm > 1/27$, there is a maximum at $\yy=0$.) The ``ridge'' represents a barrier for the motion in both directions. The ``ridge'' can represent a boundary of the motion in the $\yy$-direction only for string loops starting at $\rr<\rr_{\rm s}$; there is no boundary in the $\yy$-direction for strings starting their motion in the outer direction at $\rr > \rr_{\rm s}$, see Fig. \ref{stringFIG_09}. 

In order to determine the character of the string-loop motion, especially the possibility to be in the trapped states, 
we have to find the lowest value of boundary energy at the static radius $E_{\rm b(R)} = E_{\rm b}(\rr_{\rm s},\theta)$. From eq. (\ref{BRelation}) we obtain
\beq
  (\EE_{\rm b})_{\theta}' = 0 \,\, \ \Leftrightarrow \,\, 
  \cos(\theta) \left( \rr_{\rm s} - \frac{\JJ^2}{\rr_{\rm s}\sin^2(\theta)} \right) = 0 \label{extr_c}.
\eeq
Being interested in $\cos(\theta) \neq 0$ points, we see that the minimal energy on the ``ridge'' is located at $\theta_{\rm R}$ given by the relations
\beq
 \rr^{2}_{\rm s} \sin^{2} \theta_{\rm R} =  \xx^{2}_{\rm R} = \JJ^2
\eeq
and its value is given by
\beq
 \EE_{\rm b(R)(min)} = 2 \JJ \sqrt{A(\rr_{\rm s})} = 2 \JJ \sqrt{1 - 3\cosm^{1/3}}.
\eeq
For $\cos\theta = 0$ (at $\theta = \pi/2$, i.e., $\yy = 0$), there is another extremal point at the ``ridge'', with energy being given by
\bea
 \EE_{\rm b(R)(max)} &=&  \left( \frac{\JJ^2}{\rr_s} + \rr_s \right) \sqrt{A(\rr_{\rm s})} \\
 &=& (\JJ^2 \cosm^{1/3} + \cosm^{-1/3}) \sqrt{1 - 3\cosm^{1/3}}. \nonumber
\eea
It is a maximum for $\JJ < \rr_s$ and a minimum for $\JJ > \rr_s$. The situation is illustrated in Figs 13 and 14 - for the boundary energy profiles in the $\yy$-direction, the character of the local extrema at $\yy=0$ is determined by the magnitude of the string (angular momentum) parameter $\JJ$. Clearly, no trapped states are possible outside the static radius, since for $\JJ > \rr_s$, there is a saddle point of the boundary energy function at $\yy=0$ where minimum in the $\xx$-direction is located. The string loop can escape to infinity if its energy overcomes the minimum energy on the ``ridge'', i.e., when $E > E_{\rm b(R)(min)}$.

Following the procedure introduced for the \Schw{} spacetime, we again determine conditions for existence and extension of regions of the string-loop trapped states, i.e., for the ``lakes'' given by appropriately chosen energy levels restricted by the boundary energy function $\EE_{\rm b}(\xx,\yy,\JJ,\cosm)$. We have to combine the restriction put on the motion in the $x$- and $y$- directions. 
In the $x$-direction, the restrictions are represented by the ``projected'' angular momentum functions $\JJ_{\rm p \pm}(\xx_{p \pm},\yy,\cosm)$. In the SdS spacetimes (with $\cosm < \cosm_{\rm trap}$), the projected angular momentum function is constructed by the same procedure as in the \Schw{} spacetime, but it has two branches due to the special character of the SdS spacetimes. Near the static radius of such SdS spacetimes, there is an outer maximum of the boundary energy function $\EE_{\rm b(max+)}(\xx_{+},\yy,\JJ,\cosm)$ for all values of $\JJ < \JJ_{\rm E(max)}$, while for the range $\JJ_{\rm E(min)} < \JJ < \JJ_{\rm E(max)}$ there is an additional maximum $\EE_{\rm b(max-)}(\xx_{-},\yy,\JJ,\cosm)$ close to the black hole horizon, and there is an energy minimum located between the maxima, where oscillations of the string loop in the $x$-direction are allowed. The first branch (standard one) is constructed by projecting the inner maximum $\EE_{\rm b(max-)}(\xx_{-},\yy,\JJ_{\rm p},\cosm)$,  located near the black hole horizon, onto the boundary energy function $\EE_{\rm b}(\xx,\yy,\JJ_{\rm p},\cosm)$ when condition $\EE_{\rm b(max-)}(\xx_{-},\yy,\JJ_{\rm p},\cosm) < \EE_{\rm b(max+)}(\xx_{+},\yy,\JJ_{\rm p},\cosm)$ is satisfied. We have to construct the second branch when condition $\EE_{\rm b(max-)}(\xx_{-},\yy,\JJ_{\rm p},\cosm) > \EE_{\rm b(max+)}(\xx_{+},\yy,\JJ_{\rm p},\cosm)$ holds, by projecting the outer maximum onto the energy function $\EE_{\rm b}(\xx_{-},\yy,\JJ_{\rm p},\cosm)$. (Note, however, that the additional branch of the projected angular momentum function is shown to be irrelevant for the trapped states.) 

\begin{figure}
\includegraphics[width=4cm]{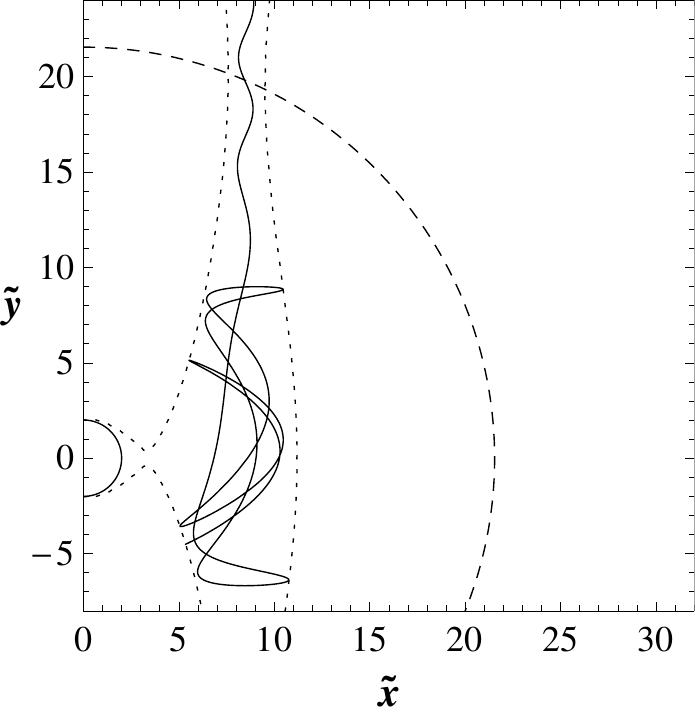}\includegraphics[width=4.07cm]{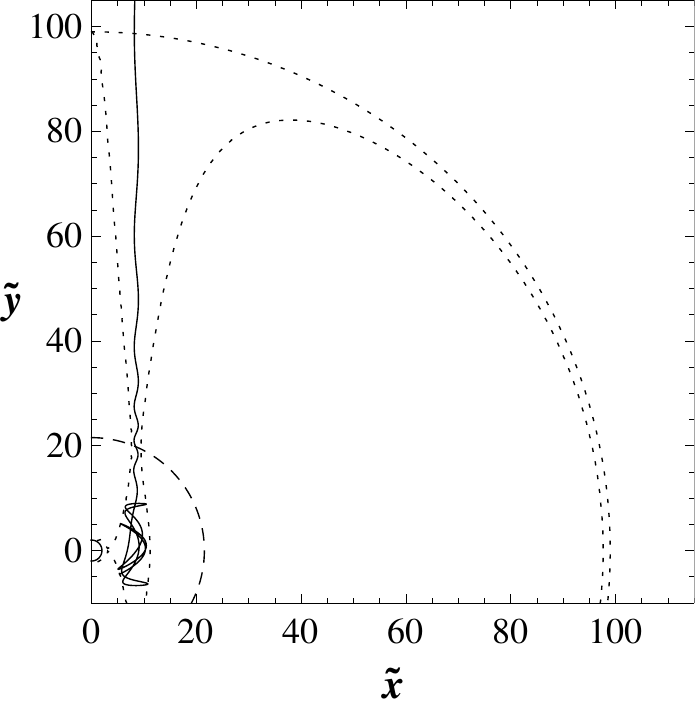}
\vspace{-0.3cm}
\caption{Escaping string loop in the SdS spacetime with $\lambda = 10^{-4}$. The boundary of the motion and the trajectory is given for $\JJ= 8.5$, $\EE \simeq 15.9$ and $\rr = 7, \tt = \simeq 2.3$. Notice the throat of the $E_{\rm b}$ boundary located at $\rr = \rr_{\rm s}$, $\xx = \JJ$. \label{stringFIG_17}}
\vspace{0.2cm}
\includegraphics[width=4cm]{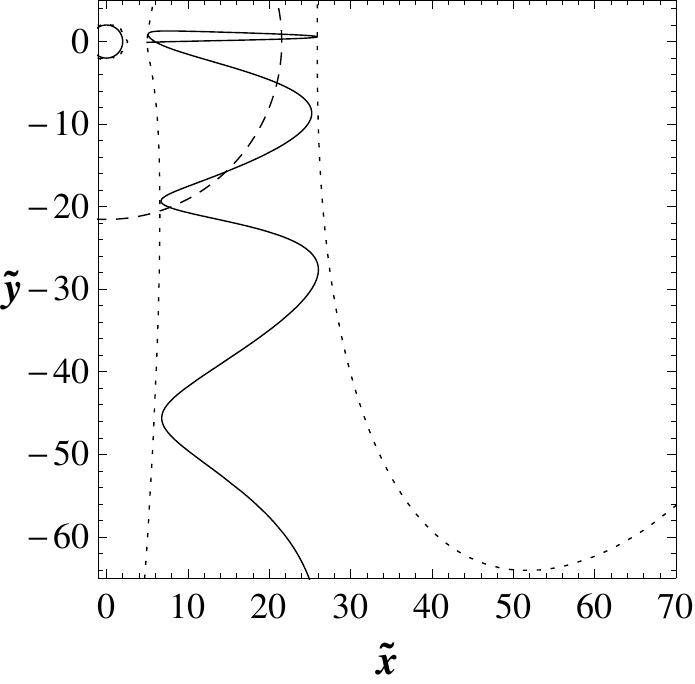}\includegraphics[width=4cm]{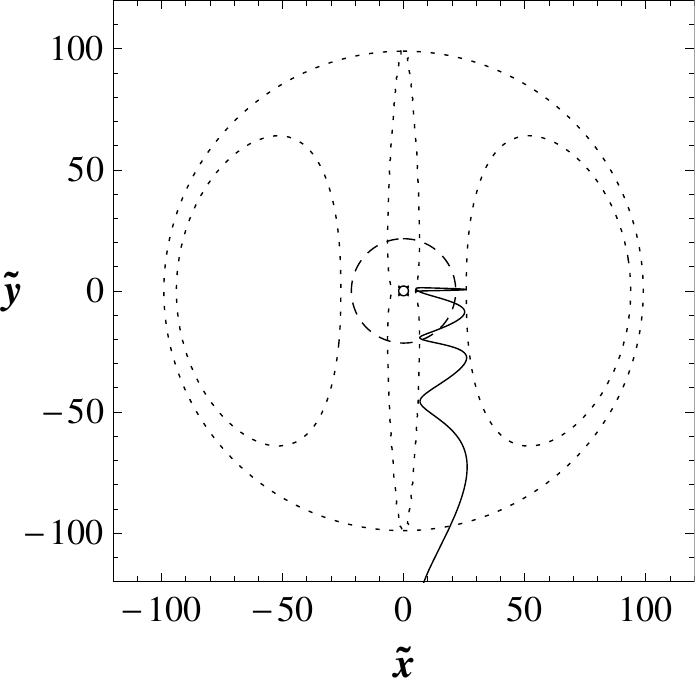}
\vspace{-0.3cm}
\caption{String-loop motion in the SdS spacetime with $\cosm = 10^{-4}$, demonstrating the case 4 in Fig. \ref{stringFIG_17A}. We can see that the string motion is accelerated in the $\yy$-direction due to the cosmic repulsion. \label{stringFIG_06}}
\vspace{0.2cm}
\subfigure[ ~Region near $\rr_{\rm h}, \rr_{\rm s}$ \label{stringFIG_08a}]{\includegraphics[width=4cm]{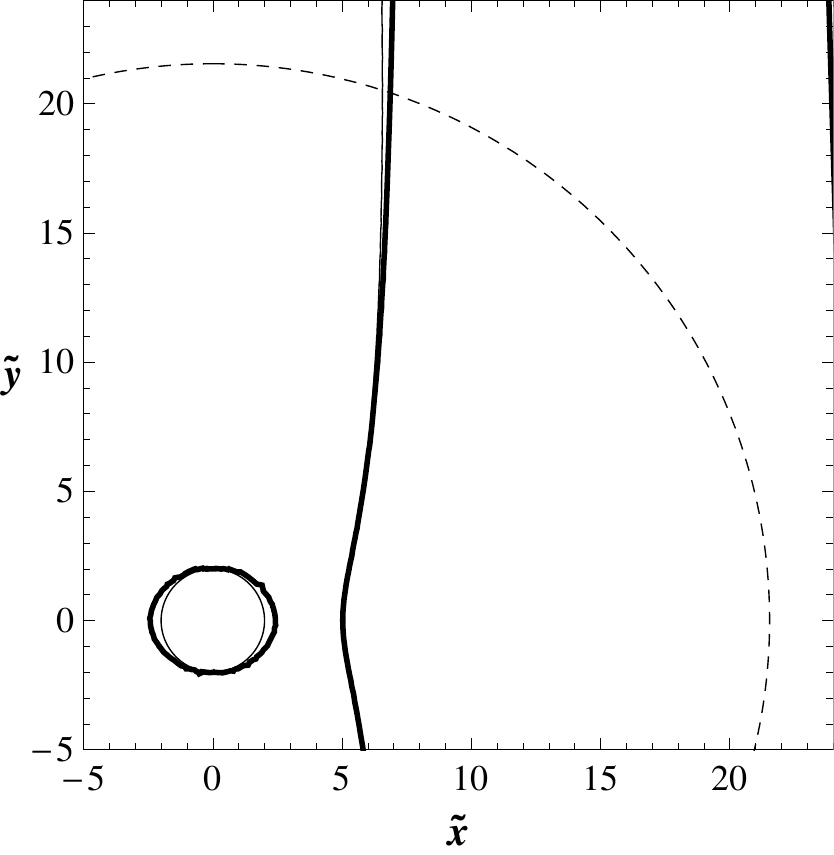}}
\subfigure[ ~Region near $\rr_{\rm c}$ \label{stringFIG_08b}]{\includegraphics[width=4.07cm]{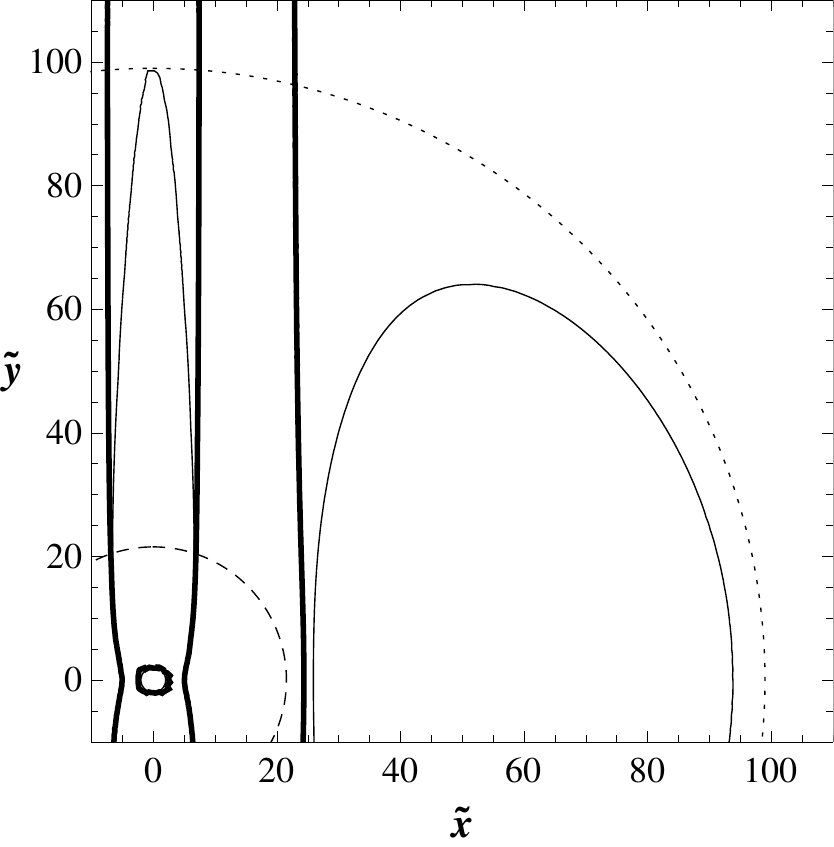}}
\vspace{-0.3cm}
\caption{\label{stringFIG_08} Influence of the repulsive cosmological constant demonstrated in behavior of the boundary energy function. Boundary for the motion $E=E_{\rm b}$ in the \Schw{} spacetime (thick curves) is compared to the related boundary in the \Schw\nnd\dS{} spacetime (thin curves) with $\cosm=10^{-4}$. There are no differences near the black hole horizon $\rr_{\rm h}$ (solid circle), small above the static radius (dashed circle), and large near the cosmological horizon $\rr_{\rm c}$ (dotted circle).}
\end{figure}

In the $y$-direction the restrictions are given by the two ``lake'' angular momentum functions $\JJ_{\rm L1}(\xx,\yy,\cosm)$ and $\JJ_{\rm L2}(\xx,\yy,\cosm)$ that are given by the solutions of the equation 
\beq
             E_{\rm b}(\xx,\yy,\JJ,\cosm) = E_{\rm b(R)(min)}.
\eeq
In the equatorial plane ($y=0$), the solutions take the form
\bea
 \JJ_{\rm L1} = \frac{\tilde{x}(\sqrt{\tilde{x}(1-3 \lambda^{1/3})} - \sqrt{2 -3\tilde{x} \lambda^{1/3} + \lambda 
\tilde{x}^3})}{\sqrt{\tilde{x}-2- \lambda \tilde{x}^3}}, \\
 \JJ_{\rm L1} = \frac{\tilde{x}(\sqrt{\tilde{x}(1-3 \lambda^{1/3})} + \sqrt{2 -3\tilde{x} \lambda^{1/3} + \lambda 
\tilde{x}^3})}{\sqrt{\tilde{x}-2- \lambda \tilde{x}^3}}.
\eea
The functions $\JJ_{\rm L1}$ and $\JJ_{\rm L2}$ are illustrated in Fig. \ref{stringFIG_17Ba} where typical sections of $\yy={\rm const}$ are given. These two functions have a common point at $\xx_{R}=\JJ$ ($\rr=\rr_{\rm s}$), where they have a common point also with the function $\JJ_{\rm E}(\xx,\cosm)$; we denote the point $\JJ_{\rm RS}$. 
The minimum of the function $\JJ_{\rm L1}$ also coincides with the intersection of this function with $\JJ_{\rm E}$. It is located at $\xx_{\rm L1(min)}$ given by the equation
\bea
\tilde{x}^6 \lambda^2 + (3-\tilde{x}) \sqrt{\tilde{x} - 3 \tilde{x} \lambda^{1/3}} \sqrt{2-3\tilde{x} \lambda^{1/3} + \tilde{x}^3 \lambda} && \nonumber \\
  - 2 \tilde{x}^4 \lambda + 4 \tilde{x}^3 \lambda + 3 \tilde{x}^2 \lambda^{1/3} - \tilde{x}( 1+9 \lambda^{1/3}) + 4 &=& 0. \nonumber \\ \label{xL1pos} 
\eea
The related value of the string parameter can be calculated from the condition
$\JJ_{\rm L1(min)} = \JJ_{\rm L1}( \xx_{\rm L1(min)})$.

For trapped states the energy level of the oscillating string loop must be chosen in such a way that the motion is limited by the energy boundary function in both $x$- and $y$- directions.
The trapped states are thus limited by the projected angular momentum function for $\JJ_{\rm E(min)} < \JJ < \JJ_{\rm L1(min)}$ and by the ``lake'' angular momentum functions for $\JJ_{\rm L1(min)} < \JJ < \JJ_{\rm RS}$. The results for the trapped states are summarized in Fig. \ref{stringFIG_17B}. It is clear that the trapped states are always limited by radius located under or approaching the static radius. Their extension is largest in the equatorial plane $y=0$. There are no trapped states for string starting above the static radius, i.e., at $\rr>\rr_{\rm s}$.

In order to ensure existence of the trapped states of the string-loop motion, the existence of some minimum (valey) of the boundary energy function in the $\xx$-direction and fulfilling of the condition $E_{\rm b(R)(min)} > E_{\rm b(min)}$ have to be guaranteed simultaneously. Therefore, the condition $\JJ_{\rm E(min)} < \rr_{\rm S}$ has to be satisfied that can be put into the form
\beq
 \cosm_{\rm E(min)}(\xx) < \cosm_s \equiv \frac{1}{\xx^{3}}, 
\eeq
where $\cosm_{\rm E(min)}(\xx)$ denotes part of the function $\cosm_{\rm E}(\xx)$ corresponding to the minima.
Since the maximum of the function $\cosm_{\rm E}(\xx)$ is located at $\xx=3(17+\sqrt{129})/16 \sim 5.3$, while the equation $\cosm_{\rm E}(\xx) = \xx^{-3}$ is satisfied at $\xx=6$, it is clear that the condition relevant for existence of trapped states is satisfied - see Fig \ref{stringFIG_16}.

The trapped states can appear only in the SdS spacetimes with the cosmological parameter smaller than the critical value of $\cosm_{\rm trap} \sim 0.00497$ that is the other important limit on the cosmological parameter relevant for the motion in the field of SdS black holes. Notice that this limiting value of the cosmological parameter is by more than one order larger in comparison with the restricting value of $\cosm$ relevant for the existence of stable circular geodesics $\cosm_{\rm ms} \sim 0.000237$ that could be considered as a test-particle analogy for trapped string states.

Behavior of the energy boundary function $\EE_{\rm b}(\xx, \yy, \JJ, \cosm)$ in terms of the string parameter $\JJ$ is represented in Fig. \ref{stringFIG_17c} together with the representative sections of the energy functions with $\xx={\rm const}$ and $\yy={\rm const}$. 

In the SdS spacetimes, the classification of the behavior of the boundary energy function is similar to the classification relevant in the \Schw{} spacetimes, giving captured, trapped and escaped (including scattered or rescattered) motion  (see Fig.\ref{stringFIG_04}).

\begin{figure}
\includegraphics[width=4cm]{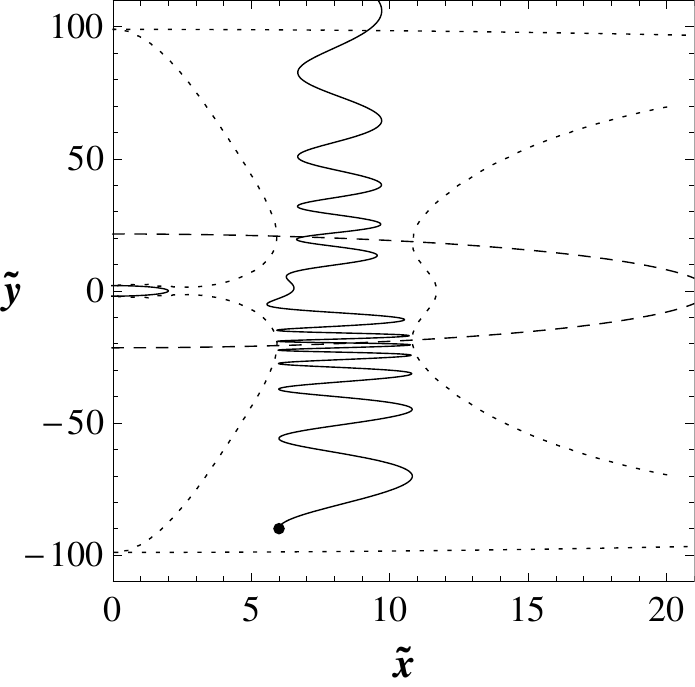}\includegraphics[width=4cm]{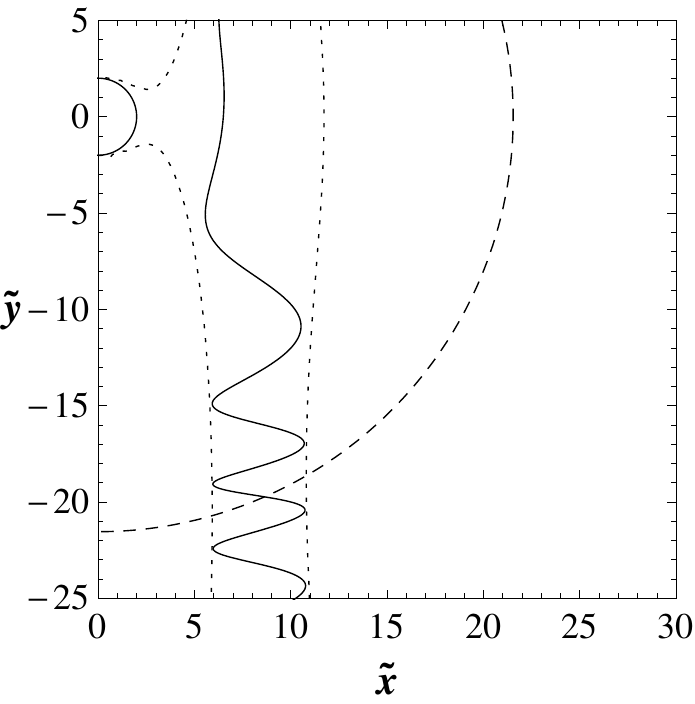}
\includegraphics[width=4cm]{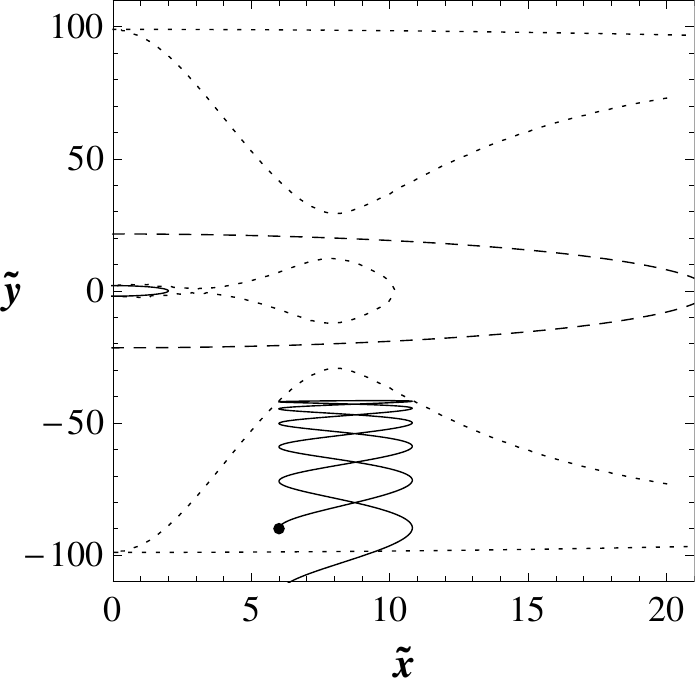}\includegraphics[width=4cm]{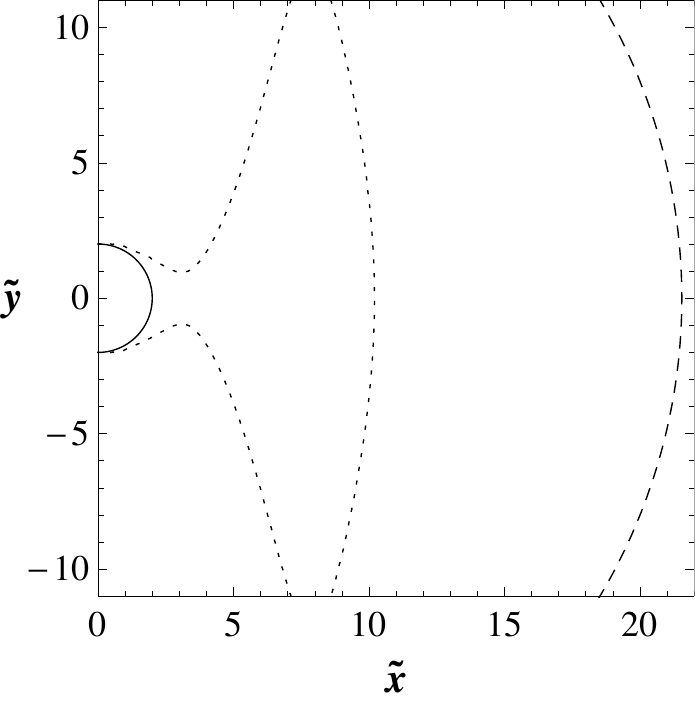}
\includegraphics[width=4cm]{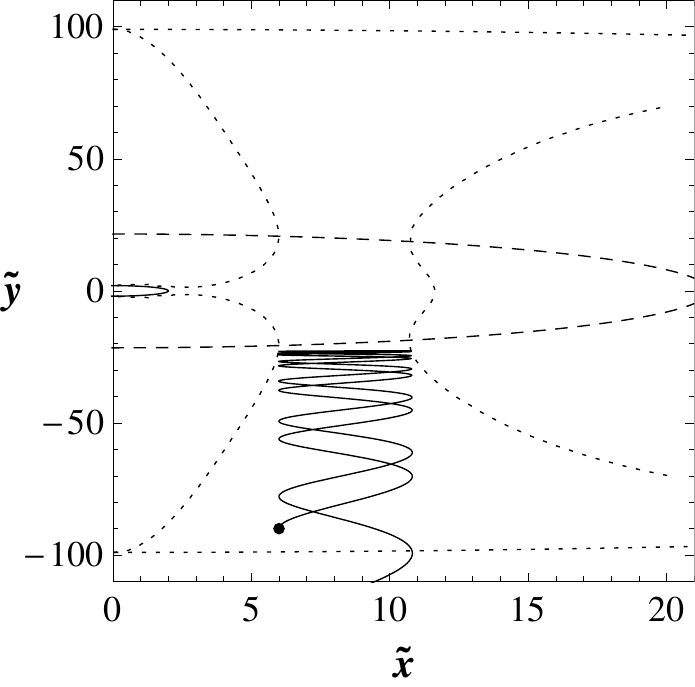}\includegraphics[width=4cm]{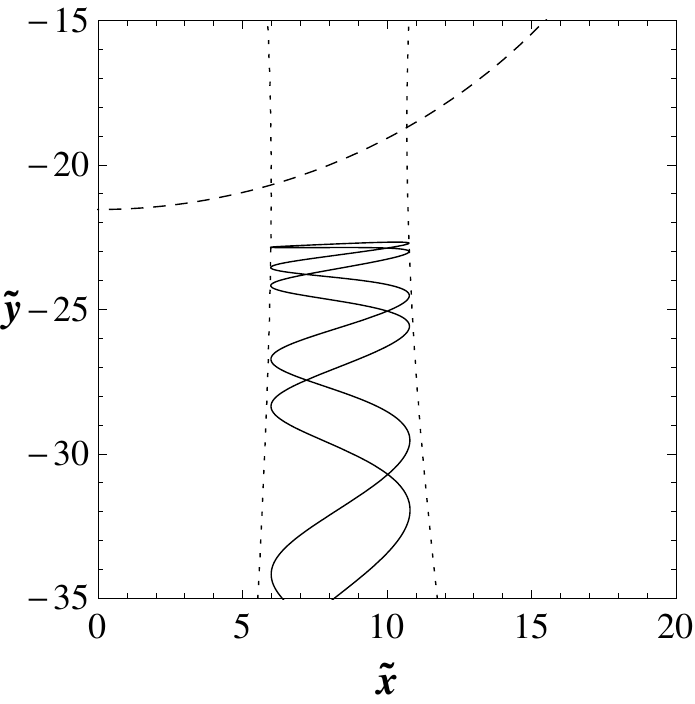}
\includegraphics[width=4cm]{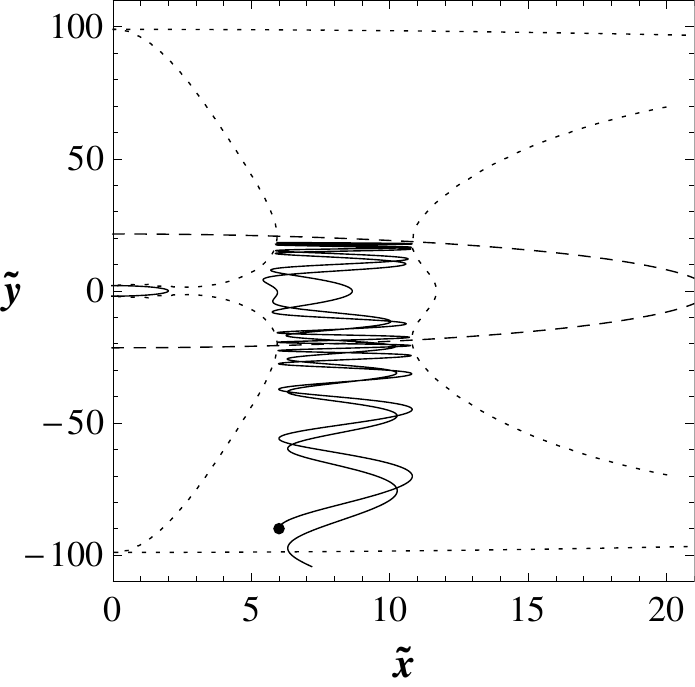}\includegraphics[width=4cm]{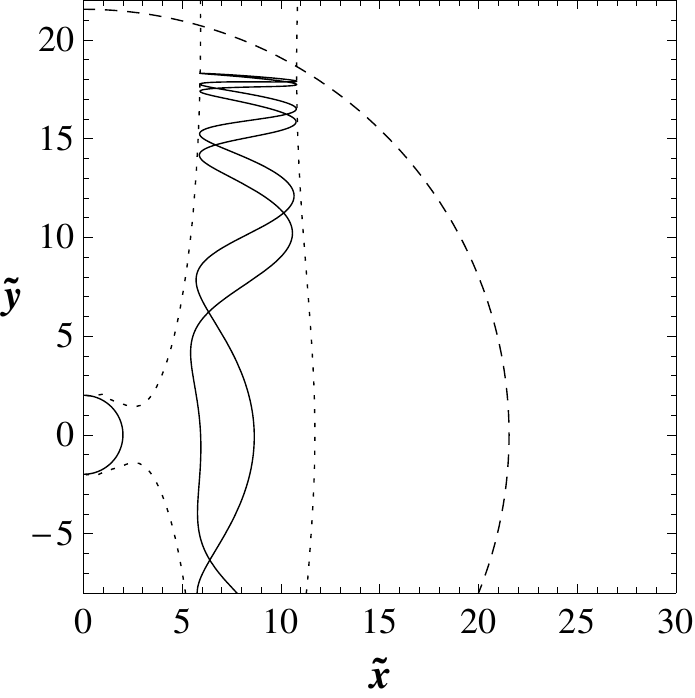}
\caption{Scattering and rescattering of string loops in the SdS spacetime with $\cosm=10^{-4}$. The strings are starting from $\tilde{x}_0 = 6, \tilde{y}_0 = -90$, with $\JJ=8$ and various momenta in the $\yy$-direction. 
On the first pair of figures (scattering), we can see that for the string it is difficult to overcome the static radius $r_{\rm s}$. Further, after crossing the vicinity of the black hole horizon, the oscillatory amplitude is diminished and the string is accelerated in the $\yy$-direction.
In the following pairs of figures, various kinds of rescattering are demonstrated showing new types of trajectories typical for the SdS spacetimes - first, the string cannot cross the static radius to enter the black-hole region because of a repulsive barriere; second, the string has not energy enough to enter the black-hole region and is turned back slightly above  $\rr_{\rm s}$; third, the string crosses the black-hole region and is backscattered slightly under $\rr_{\rm s}$.
\label{stringFIG_20}
}
\end{figure}

Equation for the string motion in the spherical coordinates (\ref{MovingEq01}) can be written for \Schw\nnd\dS{} spacetime in the form
\bea
 \ddot{\rr} &=& {\dot{\theta}}^2 (\rr-3)   + \sin^2\theta \left( \frac{\Sigma^{\sigma\sigma}}{\Sigma^{\tau \tau}} (\rr-2) -1  \right)\nonumber \\
 && -\dot{\rr} \frac{\partial_{\tau}\Sigma^{\tau\tau}{}}{\Sigma^{\tau \tau}} + \frac{\cosm}{3} \sin^2 (\theta) \rr^3 \left( \frac{2}{\Sigma^{\tau \tau}} \right) \label{EqMotion_SdS},
\eea
while (\ref{MovingEq02}) remains unchanged. The first line of Eq.(\ref{EqMotion_SdS}) corresponds formally to the relation relevant in the  \Schw{} spacetime, while the last term represents the string stretching originating in the repulsive cosmological constant (\ref{lambda_term}).

We consider the test string starting from the rest and from the same starting point $\xx_0 = 5, \yy_0 = 1$ as in the \Schw{} spacetime. This choice of the starting point gives five different types of the string motion - see Fig. \ref{stringFIG_15} for classification. First four of them correspond to the same cases as in the \Schw{} spacetime, but the fifth one is an extra case typical for the SdS (\dS{}) spacetimes. For energy $\EE \doteq 54$ (case 5) there will be no outer boundary for the string motion in $\xx$-direction, so the string radius can exponentially grow.

In the SdS spacetime we can also directly demonstrate how the repulsive cosmological constant is speeding up the string loop in $y$-direction, see Figs. \ref{stringFIG_06} and \ref{stringFIG_17}. Comparing the boundaries (i.e. starting points) for the string motion (see Fig. \ref{stringFIG_08}), we observe no differences with respect to the case of \Schw{} black holes in vicinity of the black hole horizon Fig. \ref{stringFIG_08a}, but significant differences occur in vicinity of the cosmological horizon Fig. \ref{stringFIG_08b}.

Notice that the results of the numerical integration demonstrate quite interesting phenomenon of the static radius serving as a repulsive barriere for the backscattered motion (in the $\yy$-direction) of the oscillating string (Fig.\ref{stringFIG_20}). There is another important effect observed as a result of the numerical integration of the string-loop motion - the possibility to lower amplitude of the string oscillations (in the $\xx$-direction) while moving in vicinity of the black hole horizon and to accelerate the string motion in the $\yy$-direction that occurs both for the \Schw{} and SdS spacetimes. In this case the internal energy of the string loop is transformed to the translation kinetic energy of the string. (However, an opposite effect of amplitude amplification of the oscillations in the $\xx$-direction occuring in the vicinity of the horizon is also possible. Then the translation kinetic energy is converted to the internal energy.) Of course, the cosmic repulsion itself is a source of string acceleration. Surely, all these phenomena are worth of further studies.

\begin{figure*}[p]
\subfigure[~\Schw{} $\JJ = 0$  \label{stringFIG_10a}]{\includegraphics[width=5.5cm]{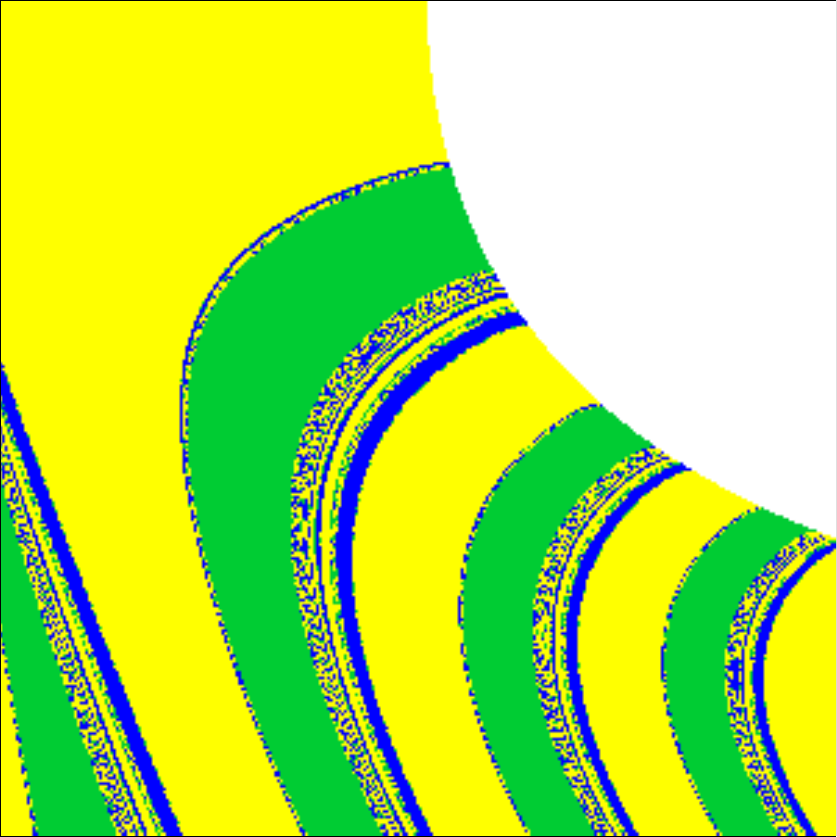}}
\subfigure[~\Schw{} $\JJ = 6$ \label{stringFIG_10b}]{\includegraphics[width=5.5cm]{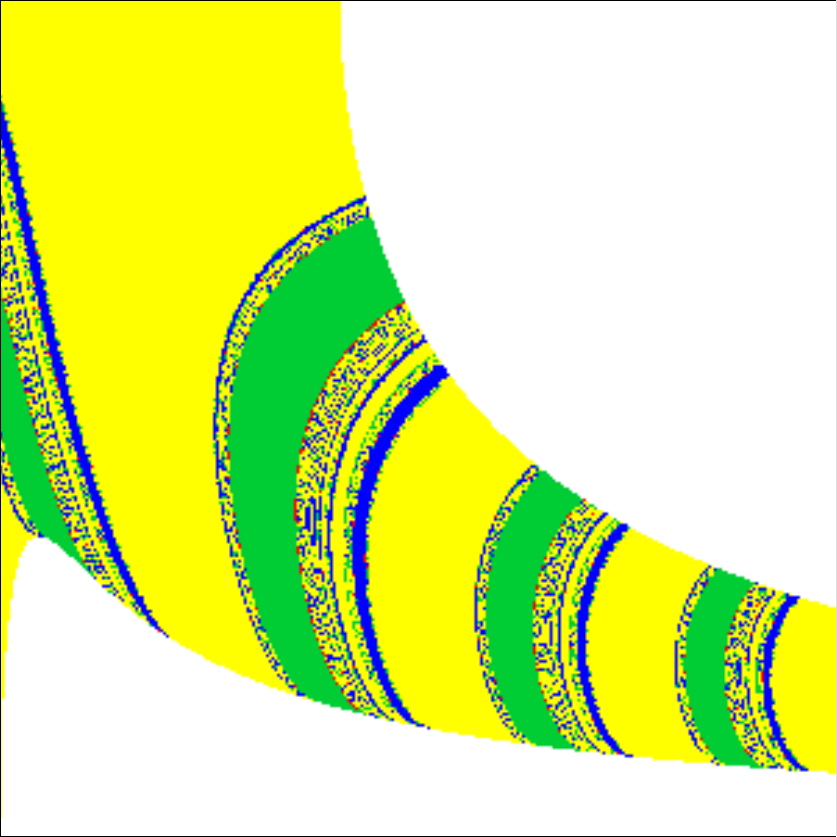}}
\subfigure[~SdS $\JJ = 6, \cosm = 10^{-4}$ \label{stringFIG_10c}]{\includegraphics[width=5.5cm]{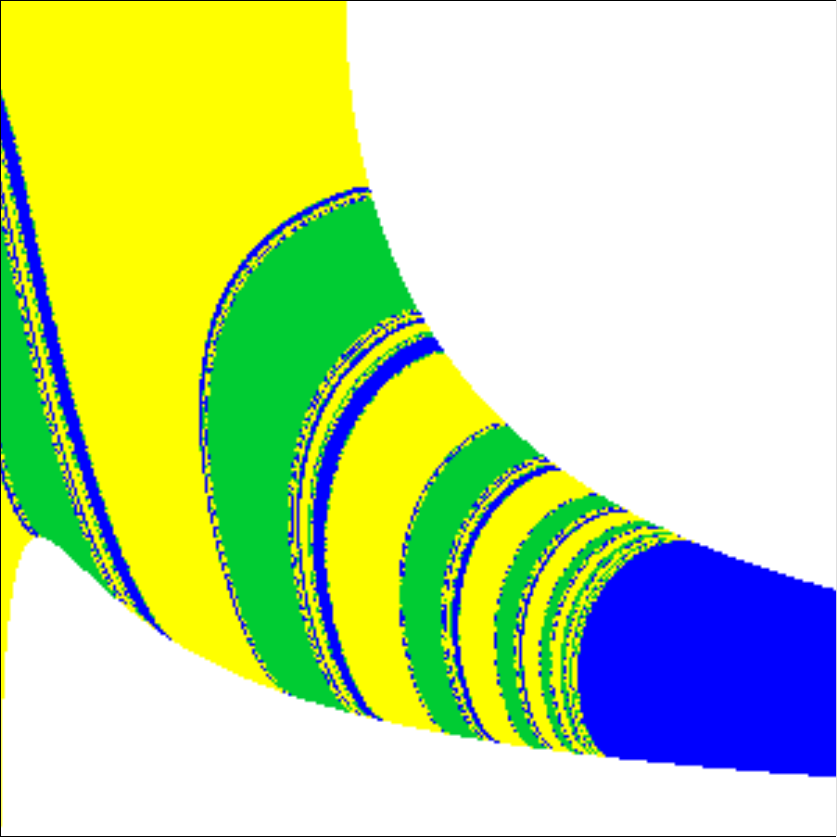}}\\
\vspace{-0.4cm}
\caption{\label{stringFIG_10} 
Basin boundary of the phase space for the string motion in the \Schw{}, cases a) with $\JJ=0$ and b) with $\JJ=6$, and SdS spacetimes, case c) with $\JJ=6$. The slice is constructed for $\EE = 14, \frac{\d}{\d \tau} (\rr \cos\tt) = \dot{\yy} = 0$, with $\rr \in (2, 27.5)$ on the horizontal axis, and $\tt \in (0, \pi / 2)$ on the vertical axis, see Fig. 4(b) from \citep{Fro-Lar:1999:CLAQG:}. In SdS case, Fig \ref{stringFIG_10c}., we use cosmological parameter $\cosm = 10^{-4}$.
\textcolor{yellow}{\bf{Yellow}} color represents trajectories captured by the black hole, \textcolor{green}{\bf{green}} represents escaped trajectories $\theta>\pi/2$, while \textcolor{blue}{\bf{blue}} represents escaped $\tt<\pi/2$ (backscattered) trajectories and \textcolor{red}{\bf{red}} represents trapped trajectories when the string is neither captured neither escaped.
Void white space represents the region $V(\rr,\tt) > 0 $ where the motion of the string is not allowed. 
}
\subfigure[~\Schw{} $\JJ = 0$ \label{stringFIG_12a}]{\includegraphics[width=5.5cm]{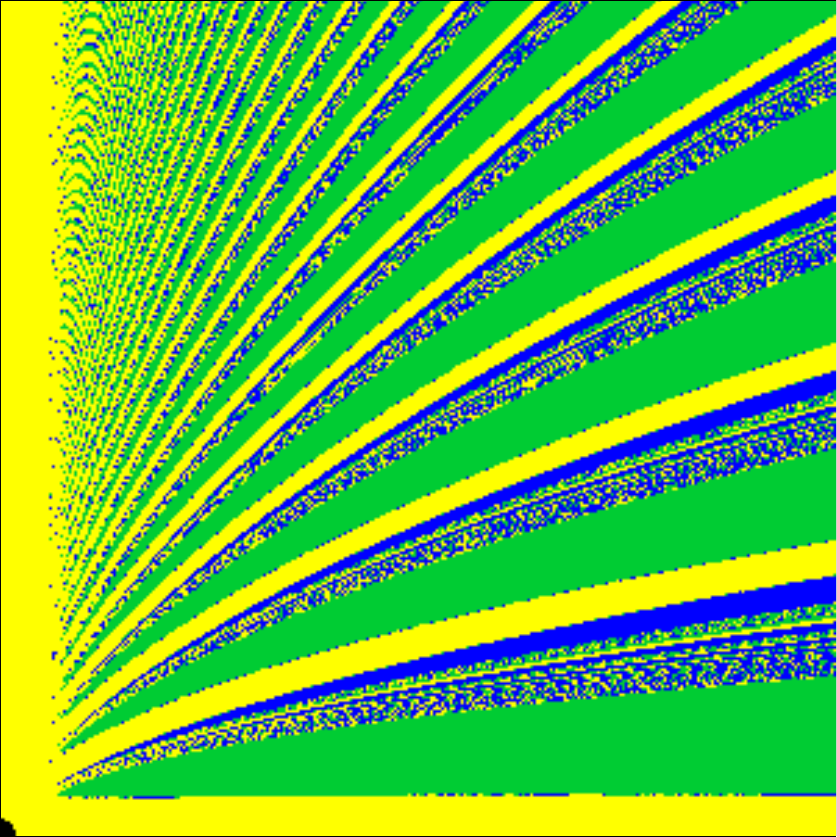}}
\subfigure[~\Schw{} $\JJ = 13$\label{stringFIG_12b}]{\includegraphics[width=5.5cm]{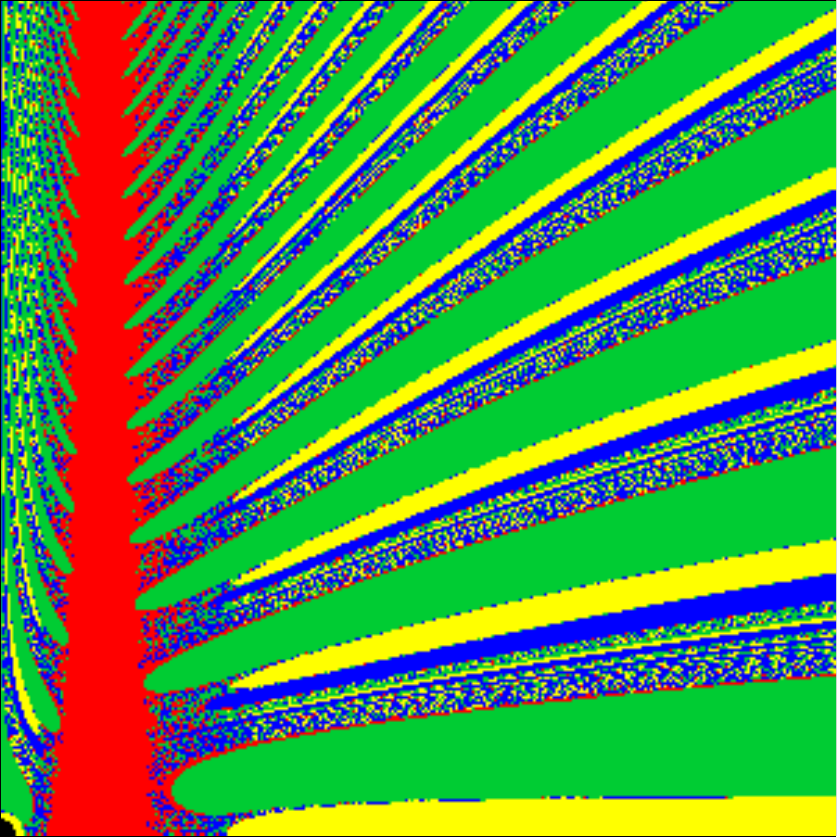}}
\subfigure[~SdS $\JJ = 13, \cosm = 10^{-6}$ \label{stringFIG_12c}]{\includegraphics[width=5.5cm]{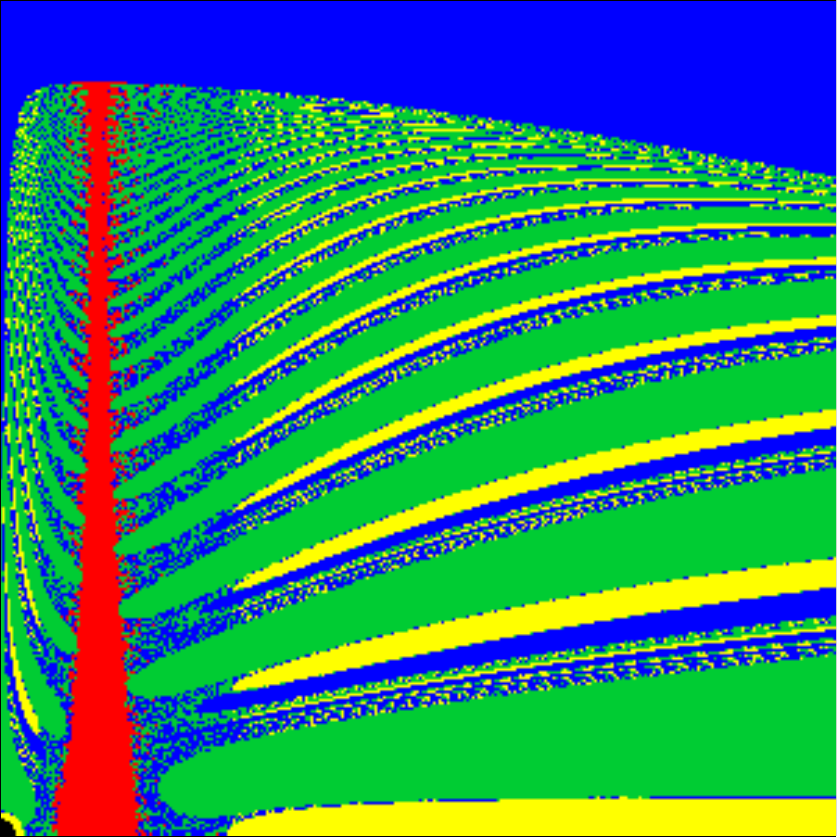}}
\subfigure[~\Schw{} $\JJ = 0$ \label{stringFIG_11a}]{\includegraphics[width=5.5cm]{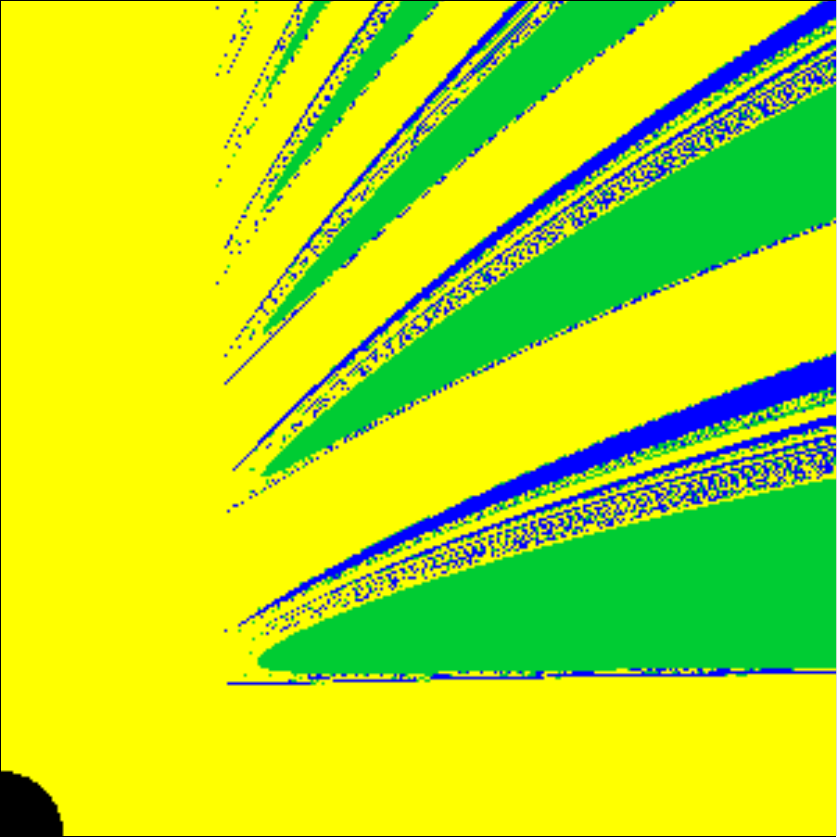}}
\subfigure[~\Schw{} $\JJ = 13$\label{stringFIG_11b}]{\includegraphics[width=5.5cm]{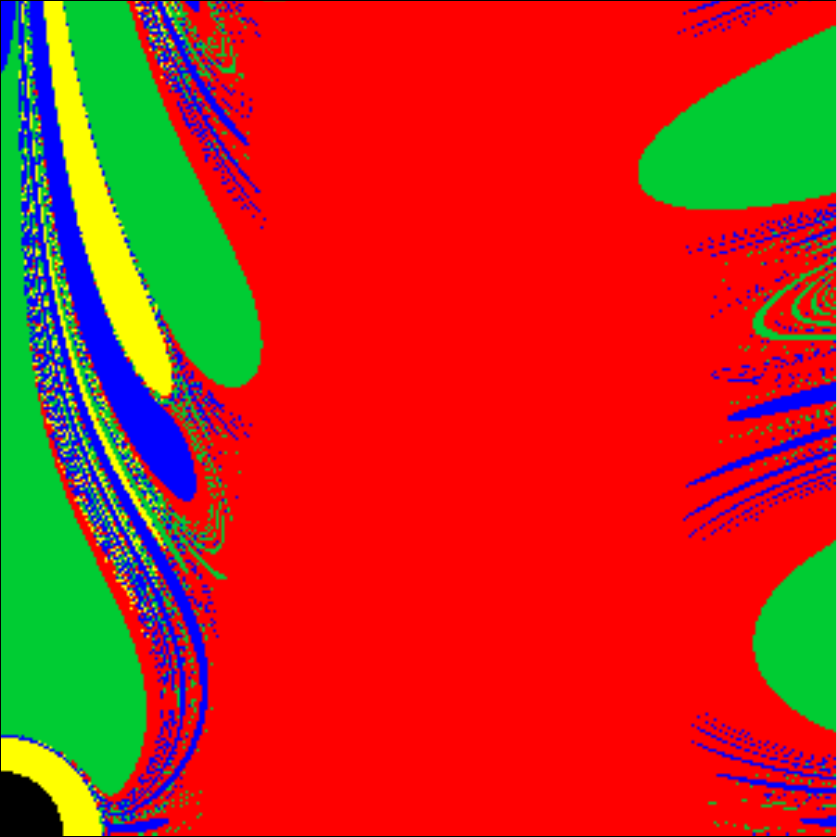}}
\subfigure[~SdS $\JJ = 13, \cosm = 10^{-4}$ \label{stringFIG_11c}]{\includegraphics[width=5.5cm]{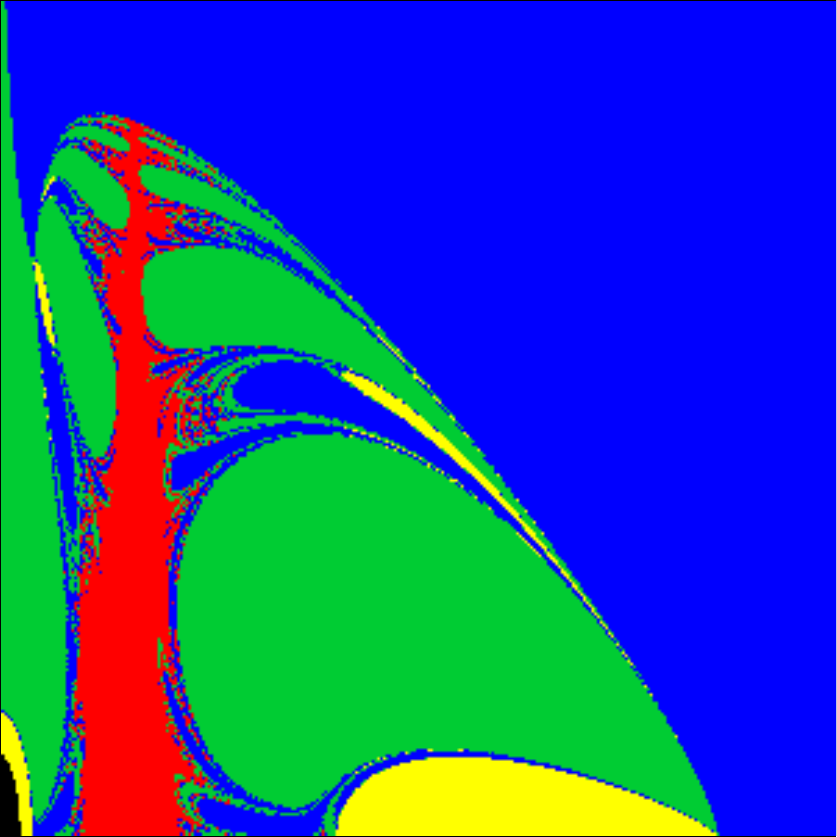}}
\vspace{-0.4cm}
\caption{\label{stringFIG_11-12} 
Basin boundaries of the phase space for the string motion. \textcolor{yellow}{\bf{Yellow}} color represents captured  trajectories, \textcolor{green}{\bf{green}} escaped trajectories $\yy\rightarrow-\infty$, while \textcolor{blue}{\bf{blue}} escaped trajectories $\yy\rightarrow\infty$ scattered or backscattered and \textcolor{red}{\bf{red}} represents trapped  trajectories. Black points in the low-left corner represent area below the horizon of the black hole.
For Figs. \ref{stringFIG_12a},\ref{stringFIG_12b}, \ref{stringFIG_12c} we use $\xx \in (0, 110)$ on horizontal axis, and $\yy \in (0, 110)$ on vertical axis. In SdS case, Fig \ref{stringFIG_12c}, we use cosmological parameter $\cosm = 10^{-6}$. There is blue region above $\yy > ~ \rr_{\rm s} = 100$ where the string directly escapes to infinity ($y\rightarrow\infty$). The red slope on the Figs. \ref{stringFIG_12b}, \ref{stringFIG_12c} corresponds to the trapped trajectories, see Fig. \ref{stringFIG_04c}.
For Figs. \ref{stringFIG_11a}, \ref{stringFIG_11b} we use on axis $\xx \in (0, 25), \yy \in (0, 25)$ - zoom- of the two Figs. above. In SdS case with $\cosm = 10^{-4}$, Fig. \ref{stringFIG_11c} we distorted scaling of the axis $\xx \in (0, 80), \yy \in (0, 20)$, to demonstrate clearly the behavior of the ``blue'' escaped region for $\yy > ~ \rr_{\rm s} \doteq 21.5, \xx > ~ \rr_{\rm c} = 100$. 
}
\end{figure*}

String loops initially located above the static radius and approaching the black hole could be captured by the black hole, or scattered (rescattered) by its gravitational field. Typical scattered or rescattered trajectories are given in Fig.\ref{stringFIG_20}. The string-loop motion is of chaotic character generally and we can reflect this fact by the basin boundary method representing different final outcomes of given initial states as shown in the next section.
The role of the static radius $\rr_{\rm s}$ is demonstrated in the Fig. \ref{stringFIG_11c}. There is only blue color above static radii in $\yy$-direction - string starting from the rest above the static radius in $\yy$-direction has to go directly to infinity.

%
%
%
%

\section{Chaotic motion of the string}

It is well known that motion of strings has to be of chaotic character due to the presence of the string tension (and angular momentum) terms in the equations of motion \cite{Fro-Lar:1999:CLAQG:} - this can be directly demonstrated by properties of the string trajectories in the phase space. We shall use here methods of chaos theory (Poincare surface sections) in order to characterize possible states of the string-loop motion in terms of appropriatelly taken sections of the phase space of initial conditions of the motion.

The string-loop dynamics can be represented in a standard way by appropriate 2D slices of the 4D phase space of initial conditions. We focus attention to construct distribution of the four types of the string loop motion (captured, trapped, escaped, scattered or backscattered) in the phase space of the initial conditions of the motion using the basin boundary method. We compare the results obtained for the \Schw{} and SdS spacetimes in order to clearly demonstrated the role of the cosmic repulsion.

Usually, it is convenient for a string loop, with a given string parameter $\JJ$, to fix the constant energy parameter $\EE$ and impose a relation between values of ($\rr, \tt, \dot{\rr}, \dot{\tt}$) at the initial time $\tau=0$. Using the basin boundary method for chaotic scattering \citep{Ott:1993:book:,Fro-Lar:1999:CLAQG:}, the 2D slice of initial conditions is separated to the four regions corresponding to the four types of the string loop motion. Due to numerical reasons, the determination of two escape asymptotic outcomes is not realized at infinity, but at some large but finite $\rr$. In the Schwarzschild spacetimes the string that passes the cutoff radius could, in principle, go back, not
escaping definitely, but such procedure causes only a limitednumber of wrongly determined asymptotic outcomes of the string motion \citep{Fro-Lar:1999:CLAQG:}. In the \Schw\nnd\dS{} spacetimes, there is a radius large enough to enable definite decision on the character of the string-motion asymptotic outcome. Such a radius has to be larger than the static radius due to the prevailing influence of the cosmic repulsion above the static radius.

The quantities $\rr,\tt,\dot{\rr},\dot{\tt}$ (or $\xx,\yy,\dot{\xx},\dot{\yy}$) form four-dimensional phase space of initial conditions of the  string-loop motion. Chaotic behavior of the string-motion was demonstrated in \cite{Fro-Lar:1999:CLAQG:}, here we discuss the differences and similarities occurring in the \Schw{} and \Schw\nnd\dS{} spacetimes comparing also the cases with string parameter $\JJ=0$ (strings with no currect and angular momentum) given in \cite{Fro-Lar:1999:CLAQG:} with $\JJ\neq 0$ considered in this paper. 

We chose two different kinds of the initial-state sections. First, following \cite{Fro-Lar:1999:CLAQG:} we use the standard approach taking slices of $\EE = 14$ and $\frac{\d}{\d \tau} (\rr \cos\tt) = \dot{\yy} = 0$, i.e., we have a fixed energy and zero initial momentum in the $\yy$-direction. The results are presented in Fig. \ref{stringFIG_10}. We can directly see a signature of the cosmic repulsion given by the ``backscattered'' (blue) region at the radii exceeding the static radius of the SdS spacetime. 
The presence of the non-zero angular momentum of the string loop is demonstrated in both the \Schw{} and SdS spacetimes by the presence of the lower limit in the sections.

Second, we use an alternative approach in mapping the distribution of the types of the string-loop motion in the phase space of the initial states that could give a more complete and complex view of the situation. We consider sections given simply by assumption of string loops (with the same angular momentum as in the previous case) starting from the rest, i.e., the slice of the phase space is chosen to be $\dot{\rr} = \dot{\tt} = 0$ ($\dot{\xx} = \dot{\yy} = 0$); of course, now different initial states (with different $\rr, \tt$) have different energy. Such a non-standard approach gives a good illustration of the distribution of the trapped states. Notice that the trapped states can be determined quite regularly and exactly by the energy conditions; of course, the motion in the trapped states is chaotic. On the other hand, the chaotic character is directly expressed in the distribution of the escaped and captured trajectories in the phase space of the initial conditions, since different types of the motion can have the same energy. The results expressing the differences between the \Schw{} and SdS spacetimes and different string currents are illustrated in Fig. \ref{stringFIG_11-12}. (For Figs. \ref{stringFIG_10}, \ref{stringFIG_11-12} we use resolution grid of $300\times300$ points.) 
We first consider SdS spacetime with $\cosm = 10^{-6}$ (and $\JJ=13$) and related regions of the \Schw{} spacetime for $\JJ=0$ and $\JJ=13$ and then SdS spacetime with $\cosm = 10^{-4}$ ($\JJ = 13$) and related enlargements of the phase space regions of the \Schw{} spacetime. Both the role of the cosmological constant and the chaotic character of the motion reflected in the distribution of the asymptotic outcomes of the string-loop motion are clearly demonstrated.

Recall that we use astrophysically implausibly large values of the cosmological parameter in order to clearly illustrate the role of the cosmic repulsion. For astrophysically relevant values the figures are qualitatively of the same character, but the regions of the cosmic-repulsion relevance are shifted to much larger dimensionless radii $\rr$. 

\section{Conclusions}

We have demonstrated how the tension and angular momentum (current) of a string loop, and the repulsive cosmological constant, affect the string loop motion in the field of SdS black holes. In order to obtain a deeper understanding of the related phenomena, we studied in detail also the motion of current (angular momentum)-carrying string loops also in the \Schw{} spacetime, and in the flat and de Sitter spacetimes giving asymptotic structure of the black-hole spacetimes. Due to the spherical symmetry of these spacetimes, we were able to characterize the string-loop axially symmetric motion in terms of a single string parameter combining the string angular momentum and tension ($\mu > 0$), and the spacetime parameters. An effective-potential method can be used efficiently by introducing a boundary energy function where the string parameter $J$ plays a role analogical to those of the axial angular momentum in the effective potential of the test-particle motion. The conditions put on the energy boundary function enable to find four types of the motion: a) capture by black hole, b) trapping, c) escape, d) scatter (backscatter) with escape. The minima (maxima) of the energy boundary function determine stable (unstable) equilibrium positions of the string loops that are governed by both the string parameters ($\mu, \JJ$) and spacetime parameters ($M,\Lambda$). This method enables a clear representation of the cosmic-repulsion role in the string-loop motion. 

We have shown that the region of the string-loop trapped states is strongly restricted by the cosmic repulsion and can exist only in the SdS spacetimes with the cosmological parameter $\cosm < \cosm_{\rm trap} = 0.00497$ that is by one order larger than $\cosm_{\rm ms} = 0.000237$ limiting SdS spacetimes admitting stable circular orbits and bound orbits of test particles that may be considered as an analogy of the trapped oscillatory states of string loops. The regions of trapped states is limited by the static radius $\rr_{\rm s} = \cosm^{-\frac{1}{3}}$ from above and by the radius of marginal trapping at the equatorial plane $\xx_{\rm mt} \sim 3.3$ from below. This is very close to the value determined for the \Schw{} spacetime for all relevant SdS spacetimes, and is substantially lower in comparison with the radii of the marginally stable and bound circular orbits of test particles, being on the other hand relatively close to the photon circular orbit located at $\rr_{\rm ph}=3$ independently of the value of the cosmological constant.

Due to the non-integrable and chaotical nature of the string equations of motion, numerical integration was used and string trajectories were obtained for typical initial conditions. We can see that in the SdS spacetimes, the trapped motion of the string loops is possible only in an appropriatelly  restricted region of the spacetime given by the influence of the cosmological constant, contrary to the case of motion in the \Schw{} spacetime where the trapped orbits are not limited from above. At distances large enough (above the static radius), the string-loop motion is fully governed by the cosmic repulsion. 

In the \Schw{} and SdS spacetimes there is one specific center of symmetry related to the gravitating mass and it is evident that the amplitude of the oscillations will not be constant if the string loop will move in the $y$-direction as the gravitational attraction of the center will be changed. On the other hand, in the flat and dS spacetimes where the center can be chosen at any spacetime point, the amplitude of the string-loop oscillations remains constant. Our numerical integrations demonstrated (in agreement with results of \cite{Lar:1994:CLAQG:,Jac-Sot:2009:PHYSR4:}) that the amplitude of the oscillations in the $\xx$-direction can be reduced (increased) while the string loop passes the black hole, being accelerated (deccelerated) in the $\yy$-direction. 

Using the basin boundary method appropriate for chaotic motion, we separated the phase space of initial conditions into the four regions corresponding to four types of the string-loop motion. Again, the role of the cosmological constant is clearly reflected by the properties of these sections, and the role of the static radius is again immediately demonstrated. 

We can summarize the effects of the repulsive cosmological constant in the following way.
\begin{itemize}

\item The cosmological constant (even very small) can completely change character of the string-loop oscillatory motion. 

\item The static radius of the SdS spacetimes, being very relevant for the test-particle motion, plays a crucial role for the string-loop motion too, giving boundary to the trapped states of oscillating strings. A string loop starting above the static radius from the rest in the $\yy$-direction, has to go directly to infinity see Fig. \ref{stringFIG_11c},\ref{stringFIG_12c}.

\item The string loop is accelerated in the $\yy$-direction due to the repulsive cosmological constant influence - see Fig. \ref{stringFIG_08b}.

\item In the SdS (de~Sitter) spacetimes there is, for some combination of string parameters $\EE,\JJ$, an extra ``pathological'' case of the string-loop motion (see points 5 (2) in Fig. \ref{stringFIG_17A} (Fig. \ref{stringFIG_13})). The string-radius is exponentialy growing in the $\xx$-direction due to the cosmic repulsion - an example trajectory can be found in Fig. \ref{stringFIG_01b}.

\end{itemize}

It should be noted that both effects (cosmic repulsion and conversion due to black hole field) related to the acceleration of the string loop in the $\yy$-direction are interesting from the point of view of acceleration of jets in active galactic nuclei and microquasars. Similarly, the inverse process of amplification of the string oscillation amplitude in vicinity of the horizon can be interesting in relation to excitation of quasiperiodic oscillations in the field of black holes.  Nevertheless, a large amount of work is necessary in order to check if the phenomena mentioned above could really be astrophysically relevant.

Finally, we would like to stress that the result presented here for the standard General Relativistic SdS geometry can be relevant in the generalization of the General Relativity given by the $f(R)$ models predicting solution of the SdS type \cite{Noriji-Odintsov:XXX:, Sotiriou-Faraoni:XXX:}

\newpage

\section{Acknowledgements}
The present work was supported by the Czech Grants MSM~4781305903 and LC06014, and by the internal student grant of the Silesian University SGS/2/2010. One of the authors (ZS) would like to express his gratitude to the Czech Committee for Collaboration with CERN for financial support and the Theory Division of CERN for perfect hospitality.




\begin{thebibliography}{99}



 
  

\bibitem{Rie-etal:2004:ASTRJ2:}
  A.~G.~Riess et al.,
  ``Type~Ia Supernova Discoveries at $z>1$ From the Hubble Space Telescope: Evidence for Past Deceleration and Constraints on Dark Energy Evolution,''
  Astrophys.\ J.\ {\bf 607}, 665 (2004)
	[arXiv:astro-ph/0402512].

\bibitem{Spe-etal:2007:ASTJS:}
  D.~N.~Spergel et al.,
  ``Three-Year Wilkinson Microwave Anisotropy Probe (WMAP) Observations: Implications for Cosmology,''
  Astrophys.\ J.\ {\bf 170}, 377 (2007)
	[arXiv:astro-ph/0603449].

\bibitem{Cald-Kami:2009:NATURE:}
  R.~Caldwell and M.~Kamionkowski,
  ``Cosmology: Dark matter and dark energy,''
  Nature {\bf 458}, 587 (2009)
  [arXiv:gr-qc/0312027].


\bibitem{Mis-Tho-Whe:1973:Gra:} C.~W.~Misner, K.~S.~Thorne and J.~A.~Wheeler, {\em Gravitation}, (W. H. Freeman and Co, New York, 1973).

\bibitem{Stu:1983:BULAI:}
  Z.~Stuchl{\'i}k,
  ``The motion of test particles in black-hole backgrounds with non-zero cosmological constant,''
  Bull.\ Astron.\ Inst.\ Czechosl.\ {\bf 34}, 129 (1983).

\bibitem{Stu:1984:BULAI:}
  Z.~Stuchl{\'i}k,
  ``An Einstein--Strauss--de~Sitter model of the universe,''
  Bull.\ Astron.\ Inst.\ Czechosl.\ {\bf 35}, 205 (1984).

\bibitem{Uzan-Ellis-Larena:2010:arXiv:1005.1809v1:}
  J.-P.~Uzan, G.~F.~R.~Ellis and J.~Larena,
  ``A two-mass expanding exact space-time solution,''
  ArXiv e-prints (2010)
  [arXiv:gr-qc/1005.1809].

\bibitem{Grenon-Lake:2010:PHYSR4:}
 	C.~Grenon and K.~Lake,
  ``Generalized Swiss-cheese cosmologies: Mass scales,''
  Phys.\ Rev.\ D {\bf 81}, 023501 (2010)
	[arXiv:astro-ph/0910.0241].

\bibitem{Stu:2005:MODPLA:}
  Z.~Stuchl{\'i}k,
  ``Influence of the Relict Cosmological Constant on Accretion Discs,''
  Mod.\ Phys.\ Lett.\  A {\bf 20}, 561 (2005)
  [arXiv:astro-ph/0804.2266].

\bibitem{Stu-Hle:1999:PHYSR4:}
  Z.~Stuchl{\'i}k and S.~Hled{\'i}k,
  ``Some properties of the Schwarzschild--de~Sitter and Schwarzschild--anti-de~Sitter spacetimes,''
  Phys.\ Rev.\ D {\bf 60}, 044006 (1999).

\bibitem{Stu:2000:APS:}
  Z.~Stuchl{\'i}k,
  ``Spherically Symmetric Static Configurations of Uniform Density in Spacetimes with a Non-Zero Cosmological Constant,''
  Acta Phys.\ Slovaca {\bf 50}, 219 (2000)
  [arXiv:gr-gc/0803.2530].
%

\bibitem{Boh:2004:GENRG:}
  C.~G.~B{\"o}hmer,
  ``Eleven spherically symmetric constant density solutions with cosmological constant,''
  Gen.\ Relativ.\ Gravit.\ {\bf 36}, 1039 (2004)
  [arXiv:gr-qc/0312027].

\bibitem{Kot:1918:}
  F.~Kottler,
  ``{\"U}ber die physikalischen Grundlagen der Einsteinschen Gravitationstheorie,''
  Ann.\ Phys.\ {\bf 361}, 401 (1918).

\bibitem{Car:1973:BlaHol:} B.~Carter, {\em Black Hole Equilibrium States}, in Black Holes pp.57--214 (Gordon and Breach, New York--London--Paris, 1973).

\bibitem{Stu:1990:BULAI:}
  Z.~Stuchl{\'i}k,
  ``Note on the properties of the Schwarzschild--de~Sitter spacetime,''
  Bull.\ Astron.\ Inst.\ Czechosl.\ {\bf 41}, 341 (1990).

\bibitem{Stu-Hle:2000:CLAQG:}
  Z.~Stuchl{\'i}k and S.~Hled{\'i}k,
  ``Equatorial photon motion in the Kerr-Newman spacetimes with a non-zero cosmological constant,''
  Class.\ Quant.\ Grav.\ {\bf 17}, 4541 (2000)
  [arXiv:gr-gc/0803.2539].

\bibitem{Stu-Cal:1991:GENRG2:}
  Z.~Stuchl{\'i}k and M.~Calvani,
  ``Null geodesics in black hole metrics with non-zero cosmological constant,''
  Gen.\ Relativ.\ Gravit.\ {\bf 23}, 507 (1991).

\bibitem{Lake:2002:PHYSR4:}
  K.~Lake,
  ``Bending of light and the cosmological constant,''
  Phys.\ Rev.\ D {\bf 65}, 087301 (2002)
  [arXiv:gr-qc/010305].

\bibitem{Bak-etal:2007:CEJP:}
  P.~Bakala, P.~{\v C}erm{\'a}k, S.~Hled{\'i}k, Z.~Stuchl{\'i}k and K.~Truparov{\'a},
  ``Extreme gravitational lensing in vicinity of Schwarzschild-de~Sitter black holes,''
  Central Eu.\ J.\ of Phys.\ {\bf 5}, 599 (2007)
	[arXiv:astro-ph/0709.4274].

\bibitem{Scha-Zai:2008:}
  T.~Sch{\"u}cker and  N.~Zaimen,
  ``Cosmological constant and time delay,''
	Astronom.\ Astrophys.\ {\bf 484}, 103 (2008)
	[arXiv:astro-ph/0801.3776].

\bibitem{Ser:2008:}
  M.~Sereno,
  ``Influence of the cosmological constant on gravitational lensing in small systems,''
  Phys.\ Rev.\ D {\bf 77}, 043004 (2008)
  [arXiv:astro-ph/0711.1802].

\bibitem{Ser:2009:}
  M.~Sereno,
  ``Role of {$\Lambda$} in the Cosmological Lens Equation,''
  Phys.\ Rev.\ Lett.\ {\bf 102}, 021301 (2009)
  [arXiv:astro-ph/0807.5123].

\bibitem{Mul:2008:}
  T.~M{\"u}ller,
  ``Falling into a Schwarzschild black hole. Geometric aspects,''
  Gen.\ Relativ.\ Gravit.\ {\bf 40}, 2185 (2008).

\bibitem{Stu-Sla:2004:PHYSR4:}
  Z.~Stuchl{\'i}k and P.~Slan{\'y},
  ``Equatorial circular orbits in the Kerr--de~Sitter spacetimes,''
  Phys.\ Rev.\ D {\bf 69}, 064001 (2004)
  [arXiv:gr-gc/0307049].

\bibitem{Kra:2004:CLAQG:}
  G.~V.~Kraniotis,
  ``Precise relativistic orbits in Kerr and Kerr--(anti)-de~Sitter spacetimes,''
  Class.\ Quant.\ Grav.\ {\bf 21}, 4743 (2004)
  [arXiv:gr-gc/0405095].

\bibitem{Kra:2005:CLAQG:}
  G.~V.~Kraniotis,
  ``Frame dragging and bending of light in Kerr and Kerr--(anti)-de~Sitter spacetimes,''
  Class.\ Quant.\ Grav.\ {\bf 22}, 4391 (2005)
  [arXiv:gr-gc/0507056].

\bibitem{Kra:2007:CLAQG:}
  G.~V.~Kraniotis,
  ``Periapsis and gravitomagnetic precessions of stellar orbits in Kerr and Kerr--de~Sitter black hole spacetimes,''
  Class.\ Quant.\ Grav.\ {\bf 24}, 1775 (2007)
  [arXiv:gr-gc/0602056].

\bibitem{Cru-Oli-Vil:2005:CLAQG:}
  N.~Cruz, M.~Olivares and J.~R.~Villanueva,
  ``The geodesic structure of the Schwarzschild--anti-de~Sitter black hole,''
  Class.\ Quant.\ Grav.\ {\bf 22}, 1167 (2005)
  [arXiv:gr-qc/0408016].

\bibitem{Kag-Kun-Lam:2006:PHYSR4:}
  V.~Kagramanova, J.~Kunz and C.~L{\"a}mmerzahl,
  ``Solar system effects in Schwarzschild--de~Sitter space time,''
  Phys.\ Lett.\  B {\bf 634}, 465 (2006)
  [arXiv:gr-qc/0602002].

\bibitem{Ior:2009:NewA:}
  L.~Iorio,
  ``Constraining the cosmological constant and the DGP gravity with the double pulsar PSR J0737-3039,''
  New Astronomy {\bf 14}, 196 (2009)
  [arXiv:gr-qc/0808.0256].

\bibitem{Stu-Hle:2001:PHYSR4:}
  Z.~Stuchl{\'i}k and S.~Hled{\'i}k,
  ``Equilibrium of a charged spinning test particle in Reissner-Nordstr{\"o}m backgrounds with a nonzero cosmological constant,''
  Phys.\ Rev.\ D {\bf 64}, 104016 (2001).

\bibitem{Stu:1999:ACTPS2:}
  Z.~Stuchl{\'i}k,
  ``Equilibrium of spinning test particles in the Schwarzschild--de~Sitter spacetimes,''
  Acta Phys.\ Slovaca {\bf 49}, 319 (1999).

\bibitem{Stu-Kov:2006:CLAQG:}
  Z.~Stuchl{\'i}k and J.~Kov{\'a}{\v r},
  ``Equilibrium conditions of spinning test particles in Kerr--de~Sitter spacetimes,''
  Class.\ Quant.\ Grav.\ {\bf 23}, 3935 (2006)
  [arXiv:gr-gc/0611153].

\bibitem{Mor-Moh:2009:GRG:}
	M.~Mortazavimanesh and M.~Mohseni,
  ``Spinning particles in Schwarzschild--de~Sitter space-time,''
  Gen.\ Relativ.\ Gravitat.\ {\bf 41}, 2697 (2009)
  [arXiv:gr-qc/0904.1263].

\bibitem{Mul-Asch:2007:CLAQG:NonMonoVel}
  A.~M{\"u}ller and B.~Aschenbach,
  ``Non-monotonic orbital velocity profiles around rapidly rotating Kerr--(anti-)de~Sitter black holes,''
  Class.\ Quant.\ Grav.\ {\bf 24}, 2637 (2007)
  [arXiv:gr-qc/0704.3963].
  
\bibitem{Sla-Stu:2008:CLAQG:} 
  P.~{Slan\'{y}} and Z.~{Stuchl\'{i}k}.
  ``Comment on non-monotonic orbital velocity profiles around rapidly rotating Kerr--(anti-)de~Sitter black holes''
  Class\ Quant.\ Grav.\, {\bf 25}, 038001, (2008).

\bibitem{Sla-Stu:2005:CLAQG:}
  P.~Slan{\'y} and Z.~Stuchl{\'i}k,
  ``Relativistic thick discs in the Kerr--de~Sitter backgrounds,''
  Class.\ Quant.\ Grav.\ {\bf 22}, 3623 (2005).

\bibitem{Stu-Sla-Hle:2000:ASTRA:}
  Z.~Stuchl{\'i}k, P.~Slan{\'y} and S.~Hled{\'i}k,
  ``Equilibrium configurations of perfect fluid orbiting Schwarzschild--de~Sitter black holes,''
	Astronom.\ Astrophys.\ {\bf 363}, 425 (2000).

\bibitem{Rez-Zan-Fon:2003:AA:}
  L.~Rezzolla, O.~Zanotti and J.~A.~Font,
  ``Dynamics of thick discs around Schwarzschild--de~Sitter black holes,''
	Astronom.\ Astrophys.\ {\bf 412}, 603 (2003)
	[arXiv:gr-qc/0310045].
	
\bibitem{Asch:2008:}
  B.~Aschenbach,
  ``Measurement of Mass and Spin of Black Holes with QPOs,''
  Chin.\ J.\ Astron.\ Astrophys.\ {\bf 8}, 291 (2008)
  [arXiv:astro-ph/0710.3454].

\bibitem{Stu-Sla-Tor-Abr:2005:PHYSR4:}
  Z.~Stuchl{\'i}k, P.~Slan{\'y}, G.~T{\"o}r{\"o}k and M.~A.~Abramowicz,
  ``Aschenbach effect: Unexpected topology changes in the motion of particles and fluids orbiting rapidly rotating Kerr black holes,''
  Phys.\ Rev.\ D {\bf 71}, 024037 (2005)
  [arXiv:gr-qc/0411091].

\bibitem{Stu-Sla-Kov:2009:CLAQG:}
  Z.~Stuchl{\'i}k, P.~Slan{\'y} and J.~Kov{\'a}{\v r},
  ``Pseudo-Newtonian and general relativistic barotropic tori in Schwarzschild--de~Sitter spacetimes,''
  Class.\ Quant.\ Grav.\  {\bf 26}, 215013 (2009)
  [arXiv:gr-qc/0910.3184].
  
\bibitem{Stu-Kov:2008:INTJMD:}
  Z.~Stuchl{\'i}k and J.~Kov{\'a}{\v r},
  ``Pseudo-Newtonian Gravitational Potential for Schwarzschild--de~Sitter Space-Times,''
  Int.\ J.\ Mod.\ Phys.\  D {\bf 17}, 2089 (2008)
  [arXiv:gr-gc/0803.3641].

\bibitem{Stu-Schee:2010:}
  Z.~Stuchl{\'i}k and J.~Schee,
  ``Influence of the cosmological constant on the motion of Magellanic Clouds in the field of Milky Way,''
  submitted.

\bibitem{Jac-Sot:2009:PHYSR4:}
   T.~Jacobson and T.~P.~Sotiriou, 
   ``String dynamics and ejection along the axis of a spinning black hole,``
   Phys.\ Rev.\ D {\bf 79}, 065029, (2009)
   [arXiv:gr-gc/0812.3996].

\bibitem{Sem-Dya-Pun:2004:Sci:}
  V.~Semenov, S.~Dyadechkin and B.~Punsly,
  ``Simulations of jets driven by black hole rotation,''
  Science {\bf 305}, 978 (2004)
  [arXiv:astro-ph/0408371].

\bibitem{Chri-Hin:1999:PhRvD:}
  M.~Christensson and M.~Hindmarsh,
  ``Magnetic fields in the early universe in the string approach to MHD,''
  Phys.\ Rev.\  D {\bf 60}, 063001 (1999)
  [arXiv:astro-ph/9904358].

\bibitem{Spr:1981:AA:}
	H.~C.~Spruit, 
	``Equations for thin flux tubes in ideal MHD,''
	Astron.\ Astrophys.\ {\bf 102}, 129 (1981).

\bibitem{Lar:1994:CLAQG:}
  A.~L.~Larsen,
  ``Chaotic string capture by black hole,''
  Class.\ Quant.\ Grav.\  {\bf 11}, 1201 (1994)
  [arXiv:hep-th/9309086].
  
\bibitem{Fro-Lar:1999:CLAQG:}
  A.~V.~Frolov and A.~L.~Larsen,
  ``Chaotic scattering and capture of strings by a black hole,''
  Class.\ Quant.\ Grav.\  {\bf 16}, 3717 (1999)
  [arXiv:gr-qc/9908039].

\bibitem{Gu-Cheng:2007:GRG:}
  Z.~Gu and H.~Cheng, 
  ``The circular loop equation of a cosmic string in Kerr--de~Sitter spacetimes,''
  Gen.\ Relativ.\ Gravit.\ {\bf 39}, 1 (2006)
  [arXiv:hep-th/0412297].

\bibitem{Wit:1985:NuclPhysB:}
   E.~Witten,
   ``Superconducting Strings,''
   Nucl.\ Phys.\ B {\bf 249}, 557 (1985).
  
\bibitem{Lar:1993:CLAQG:}
  A.~L.~Larsen,
  ``Dynamics of cosmic strings and springs; a covariant formulation,''
  Class.\ Quant.\ Grav.\  {\bf 10}, 1541 (1993)
  [arXiv:hep-th/9304086].
  
\bibitem{Ziol:2008:CJA:}
  J.~Ziolkowski,
  ``Masses of Black Holes in the Universe,''
  Chin.\ J.\ Astron.\ Astrophys.\ {\bf 8}, 273 (2008)
	[arXiv:astro-ph/0808.0435].

\bibitem{Ott:1993:book:} E.~Ott, {\em Chaos in dynamical systems}, (Cambridge University Press, Cambridge, 1993).

\bibitem{Noriji-Odintsov:XXX:}
 	S.~Nojiri and  S.~D.~Odintsov,
  ``Dark energy, inflation and dark matter from modified F(R) gravity,''
  ArXiv e-prints (2008)
  [arXiv:hep-th/0807.0685].
   
\bibitem{Sotiriou-Faraoni:XXX:}
	T.~P.~Sotiriou and V.~Faraoni,
	``f(R) theories of gravity''
	Rev.\ Mod.\ Phys.\ {\bf 82}, 451 (2010)
	[arXiv:gr-qc/0805.1726].
	
\end{thebibliography}
\end{document}